\documentclass[utf8]{./FrontiersinHarvard} 

\usepackage{url,hyperref,lineno,microtype,subcaption}
\usepackage[onehalfspacing]{setspace}
\usepackage{lmodern}

\def\scriptO{{{\it O}\kern -.42em {\it `}\kern + .20em}}
\def\RR{{{\rm l}\kern - .15em {\rm R} }}
\def\PP{{{\rm l}\kern - .15em {\rm P} }}
\def\L2{{{\sf L}^2}}
\def\H1{{{\sf H}^1}}
\def\PN2{{\PP_{N}-\PP_{N-2}}}

\def\complex{{{\rm C} \kern - .53em {\rm l} \kern + .38em}}

\def\a1{{ | \lambda_{\min} |}}

\def\l1{{   \lambda_{\min}  }}

\def\bu0{{\underline {\bf 0}}}

\def\bu{{\bf u}}

\def\Oh{{\hat \Omega}}

\def\ub{{\underline b}}

\def\ur{{\underline r}}

\def\ux{{\underline x}}

\def\uz{{\underline z}}
\def\u0{{\underline 0}}

\linenumbers

\def\keyFont{\fontsize{8}{11}\helveticabold }
\def\firstAuthorLast{Misun Min {et~al.}} 
\def\Authors{Misun Min\,$^{1,*}$, Yu-Hsiang Lan\,$^{2}$, Paul Fischer\,$^{1,2,3}$, Thilina Rathnayake\,$^{2}$ and John Holmen\,$^{4}$}
\def\Address{$^{1}$Argonne National Laboratory, Mathematics and Computer Science Divison, Lemont, IL, USA\\
$^{2}$University of Illinois Urbana-Champaign, Department of Computer Science, Urbana, IL, USA\\
$^{3}$University of Illinois Urbana-Champaign, Department of Mechanical Science and Engineering, Urbana, IL, USA\\
$^{4}$Oak Ridge National Laboratory, Oak Ridge Leadership Computing Facility, Oak Ridge, TN, USA}
\def\corrAuthor{Corresponding Author}

\def\corrEmail{mmin@mcs.anl.gov}

\begin{document}
\nolinenumbers
\onecolumn
\firstpage{1}

\title {Nek5000/RS Performance on Advanced GPU Architectures} 

\author[\firstAuthorLast ]{\Authors} 
\address{} 
\correspondance{} 

\extraAuth{}

\maketitle

\begin{abstract} 
We demonstrate NekRS performance results on various advanced GPU architectures.
NekRS is a GPU-accelerated version of Nek5000 that targets high performance on 
exascale platforms. It is being developed in DOE's Center of Efficient Exascale 
Discretizations, which is one of the co-design centers under the Exascale 
Computing Project.  In this paper,
we consider Frontier, Crusher, Spock, Polaris, Perlmutter, ThetaGPU, and Summit.
Simulations are performed with 17$\times$17 rod-bundle geometries from small 
modular reactor applications. We discuss strong-scaling performance and analysis. 


 \keyFont{\section{Keywords:} 
  Nek5000/RS, exascale, strong scaling, small modular reactor, rod-bundle} 
\end{abstract}

\section{Introduction}

\begin{figure*}[t]
  \begin{center}
     \includegraphics[width=0.9\textwidth]{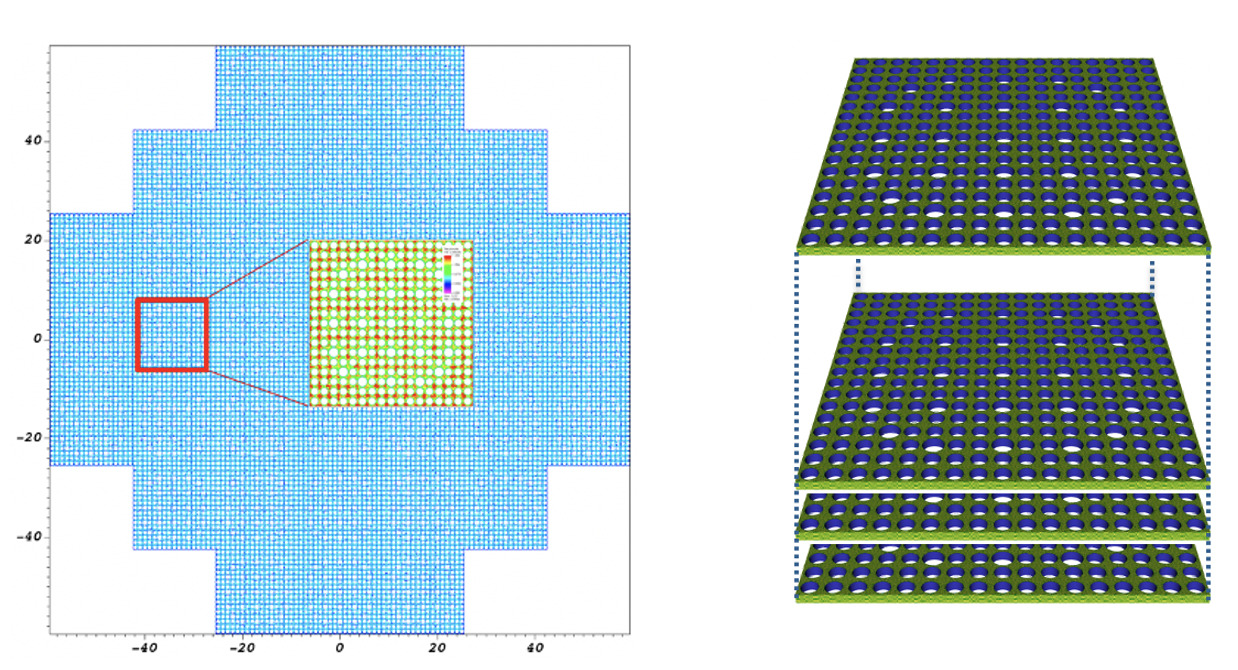 }
  \end{center}
\caption{\label{rod1717}
Full-core configuration on the left and a single 17$\times$17 rod bundle on the right.}
\end{figure*}

%
%
%
%
%
%
%
%

As part of its Exascale Computing Project, the U.S. Department of Energy
leadership computing facilities deploy platforms capable of reaching $>$ 1
$\approx 10^3$--$10^4$ nodes, each equipped with powerful CPUs and
anywhere from 4 to 8 accelerators (i.e., GPUs), which provide the bulk of the
compute power.  For reasons of efficiency, a favored programming model for
these architectures is to assign a single process (i.e., MPI rank) to each GPU
(or GPU processing unit, such as a GCD on the AMD MI250X or a tile on the Intel
PVC)
and execute across the GPUs using a private distributed-memory programming
model.  With $P=10^3$--$10^5$ MPI ranks, this approach affords a significant
amount of internode parallelism and contention-free bandwidth with no
increase in memory-access latency, save for the relatively sparse internode
communication that is handled by MPI.

In this paper we describe performance results for the open source thermal-fluids
simulation code Nek5000/RS \cite{nek5000,nekrs,nekrs1} on serveral of DOE's
recently installed high-performance computing (HPC) platforms.  
%
%
NekRS is a GPU-oriented version of Nek5000 that was
developed under DOE's Center for Efficient Exascale Discretizations (CEED).
%
%
%
For portability, all the GPU kernels are written in OCCA \cite{occa,occa-repo}.
Many of the high-performance kernels came out of the libParanumal
library \cite{libp}.

\subsection{Performance Metrics}

Of great interest to computational scientists is the speed that can be realized
for a particular application for a given architecture. (Here, an {\em
application} is a particular {\em problem} that uses Nek5000/RS, which is an {\em
application code}.)  For example, one frequently needs to estimate the number
of node-hours and number of wall-clock hours that might be required for a large
simulation campaign.  A common metric, which is very much case-specific, is the
number of degrees of freedom (dofs) per second that can be realized on a
platform, or perhaps the number of dofs per second per accelerator (i.e., per
MPI rank.\footnote{We prefer dofs per second per rank because AMD's MI250X has
two compute units (GCDs) per GPU and Aurora's Intel has two tiles per
PVC---users view these as two processors.})  In the sequel, we will assign
MDOFS to the quantity ``millions of dofs per second per rank.''  The
case-specificity aspect of MDOFS is that one can realize a much larger MDOFS
value for, say, linear solution of $A\ux = \ub$ than would be possible for an
incompressible Navier--Stokes solver.   

Despite the large variance in MDOFS from one problem class to the next, it is
nonetheless a worthy metric when making platform-to-platform comparisons.   A
related metric is the time-to-solution or, in the case of a timestepping
simulation code that could take an aribitrarily large number of steps, the
time per step, $t_{step},$\footnote{We typically take the average time over several 
hundreds or thousands timesteps.} which we measure in seconds.  
Even for a given code and architecture this latter quantity is subject to
significant variability because some problems or computational meshes are more
ill-conditioned than others, which leads to higher iteration counts in the
linear solvers (e.g., in the pressure solve for an incompressible flow
simulation) and hence a longer time per step.  

MDOFS and $t_{step}$ are dependent, or output, parameters.  For a fixed
platform, code, and problem, users still have two independent parameters at
their disposala: $n$, the problem size or number of dofs,\footnote{For fluid flow
simulations, some authors set $n$ to be 4 times the number of grid points
because there are typically 3 velocity components and one pressure unknown at
each grid point.  We prefer to take $n$ to be the number of grid points.   The
problems are generally large enough that we do not need to distinguish between
interior points where the solution is unknown and surface points where boundary
data is prescribed.  We simply set $n$ to represent the union of these sets
given that some quanities might have Neumann conditions one the boundary while
others have Dirichlet.} and $P$, the number of ranks (here, accelerator
devices) to use.  For a fixed problem size, $n$ (which is what a user typically
has), there is only one variable, namely, $P$.  A user who is contemplating a
simulation campaign will often be interested in predicting performance
over a range of $n$.  Under the given conditions, we will see that the most
important quantitity in predicting performance is
\begin{eqnarray} \label{eq:np}
 \frac{n}{P} &=& \mbox{the number of grid points per device},
\end{eqnarray}
where we reiterate that we are assigning a single MPI rank to each device.

Two other dependent quantities of interest are parallel efficiency, $\eta^{}_P$,
and realized FLOPS, the measurable number of 64-bit floating-point operations
per second.  We typically will report FLOPS per rank, which is more universal
than aggregate FLOPS.  Also, in cases where mixed precision is used (e.g., 
when 32-bit arithmetic is used in a preconditioner), we count the FP32 flops
as a half-flop each.   A typical definition of (strong-scaling) parallel efficiency is
\begin{eqnarray} \label{eq:eff1}
 \eta^{}_P &:=& \frac{P_0 \, t_{step}(P_0)}{P \, t_{step}(P)} 
\;=\; \frac{\mbox{MDOFS}(P)}{\mbox{MDOFS}(P_0)},
\end{eqnarray}
where $t_{step}(P)$ is the time per step when running a fixed problem of fixed
size, $n$, on $P$ ranks and MDOFS$(P)$ is the corresponding number of mega-dofs
per second per rank.  Here, $P_0$ is the smallest number of ranks that is able
to hold the problem.  On some architectures, the amount of memory per GPU is
relatively small, which prevents extensive strong-scaling studies.  (From a
users perspective, however, this is potentially a happy circumstance since
there is ``just enough'' memory and not a lot of idle memory that incurs
unnecessary capital and power overhead.)

An alternative definition of parallel efficiency is given by the relationship
\begin{eqnarray} \label{eq:eff2}
   S_P &=& \eta^{}_P \, P \, S_1.
\end{eqnarray}
Here, $S_P$ is the {\em speed} on $P$ processors, 
which could be measured in total (not per rank)
FLOPS or MDOFS.  This definition is equivalent to (\ref{eq:eff1}) when $P_0=1$.  
The utility of this definition is that one can consider it for either weak-
(fixed $n/P$) or strong- (fixed $n$) scaling studies.  If FLOPS are used, it is
relatively easy to get FLOPS on one rank for a smaller version of the application
problem (although that might not be a useful starting point in the exascale era
given that no exascale problem comes anywhere close to fitting on
one rank).   

What (\ref{eq:eff2}) tells us is that the speed on $P$ ranks should be $\approx
P$ times the speed on 1 rank, if only we can sustain close to unity efficiency,
$\eta_P^{} \approx 1$.  We remark that HPC users generally want to run as fast
as possible, particularly for large campaigns, so they want $P \gg 1$.
However, they also need to efficiently use their allocation, which implies
$\eta^{}_P \approx 1$.  This latter condition places a constraint on
time-to-solution that is generally
stronger than the unconstrained ``min-time-to-solution'' result.  In our studies
we will assume that the user is willing to run at 80\% parallel efficiency,
$\eta^{}_P = 0.8$.  Of course, other values are possible, and a user can change
$P$ and, hence, $\eta^{}_P$ on a submission-by-submission basis for each case
run in a given campaign.
However, $\eta^{}_P = 0.8$ is a reasonable starting point for analysis.

The first analysis question we address is, {\em For a fixed problem size $n$,
how many ranks can we use before $\eta^{}_P < 0.8$?}  An accompanying question
is, {\em What is the time per step at that value?}    The key to this first
question is that parallel efficiency typically drops as the local amount of
work, $n/P$ tends towards zero.  So, fixing $\eta_P^{} = 0.8$ implies
$n/P$ is a fixed value (for a given fixed-sized problem with $P$ varying).
We denote this value as $n_{0.8}$.  It is the number of points per rank
where the application realizes 80\% efficiency, which is where we anticipate
that users will typically run.  With this definition, we can address
the question of the expected $t_{step}$ under these conditions.
Assume that a given problem requires a certain amount of work, $W$, 
that is measured in total number of floating-point operations (FLOPS).
Usually, $W \sim C n$, where $C$ is an $n$-independent  constant, which implies
that, to leading order, the amount of works scales with $n$.  On $P$
processors, we therefore can expect that
\begin{eqnarray} \label{eq:tstep}
   t_{step} &=& \frac{W}{S_P} \;=\;
      \frac{C\,n}{\eta^{}_P \, P \, S_1}.
\end{eqnarray}
We define $t_{0.8}$ to be the value of $t_{step}$ at 80\% efficiency, 
\begin{eqnarray} \label{eq:t08}
   t_{0.8} &=& \frac{C}{0.8} \frac{n/P}{S_1}.
      \;=\; \left( \frac{C}{0.8}\right) \frac{n_{0.8}}{S_1}.
\end{eqnarray}
We note that (\ref{eq:tstep})--(\ref{eq:t08}) are predicated on $\eta_P^{}$
being strongly dependent on $(n/P)$ with no direct $P$ dependence.  There are
times when there is a weak $P$ dependence, particularly for $P > 10^4$.  In
this case, one can simply modify the analysis to have a $P$-dependent
$n_{0.8}$.  

We see from (\ref{eq:t08}) that the time per step is governed by the speed on a
single rank, $S_1$ (larger is better), and the amount of work on a single rank,
$n_{0.8}$
(smaller is better), where 80\% efficiency is realized.  If a new platform
comes out with a 2$\times$ increase in $S_1$ but a 4$\times$ increase in
$n_{0.8}$, then the net time-to-solution {\em increases by} $2\times$.  In HPC,
it is the {\em ratio}, $n_{0.8}/S_1,$ that is critical to fast time-to-solution.
Much of this analysis can be found in \cite{fischer15,ceed_bp_paper_2020}.
Communication overhead on GPU-based architectures is discussed in 
\cite{bienz20}.

A typical use case for (\ref{eq:t08}) is that a user knows $n$,
which is the number of gridpoints required to resolve a given simulation,
and wants to know how many processors will be required to efficiently
solver this problem and how long it will take to execute.  
The user also knows $n_{0.8}$ from scaling studies
of the type provided here.  From that, the user can determine
\begin{eqnarray} \label{eq:p08}
  P_{0.8} & = & \frac{n}{n_{0.8}},
\end{eqnarray}
which is the maximum number of ranks that can be employed while sustaining
80\% efficiency.
The time per step will be $t_{0.8}$, and the total required node-hours 
will be 
\begin{eqnarray} 
\mbox{node hours} &\approx& 
  \frac{P_{0.8}}{\mbox{ranks-per-node}} \; \times \;
  \frac{N_{steps} \, t_{0.8}}{3600 \mbox{ s/hour}},
\end{eqnarray}
where $N_{steps}$ is the estimated number of timesteps.

\subsection{Test Cases}

In the following sections we characterize these relevant parameters for NekRS
across several of DOE's pre-exascale and exascale platforms, including
Frontier, Crusher, Spock, Polaris, Perlmutter, ThetaGPU, and Summit.
Simulations are performed using ExaSMR's 17$\times$17 rod-bundle geometry,
illustrated in Fig.~\ref{rod1717}.  This geometry is periodic in the axial
(vertical) flow direction, which allows us to weak-scale the problem by
adding more layers of elements in the $z$ direction.  (The model problem
is essentially homogeneous in $z$.) Each case starts with a pseudo-turbulent
initial condition so that the iterative solvers, which compute only the change
in the solution on each step, are not working on void solutions.  Most of the
cases are run under precisely the same conditions of timestep size, iteration
tolerances, and averaging procedures, which are provided case by case in
the sequel.

We remark that the following performance summaries are for full Navier--Stokes
solution times.  We present a few plots that reflect work in salient kernels,
	 such as the advection operator, which is largely communication-free, and the
pressure-Poisson coarse-grid solve, which is highly communication-intensive.
Detailed kernel-by-kernel breakdowns are presented in \cite{nekrs,min-sc22} and
are available in every logfile generated by NekRS.  Further, we
note that NekRS supports multiple versions of all of its high-intensity kernels
and communication utilities.  These are determined at runtime by running a
small suite of tests for each invocation of the given utility.  Thus, the
performance is optimized under the loading conditions of that particular kernel
for the particular platform for the particular application.  An example of 
these outputs, along with the kernel-by-kernel breakdown, is presented
in Section \ref{sec:disc}.

\section{NekRS Performance on a Single GPU} \label{sec:single}

Table~\ref{test01} demonstrates NekRS performance for ExaSMR's single-rod
simulation on a single GPU. Simulations are performed for 500 steps; and the
average time per step, $t_{\rm step}$, is measured in seconds for the last 400
steps. For a given system, the speedup is the inverse ratio of $t_{\rm step}$ 
to that of Summit.
$v_i$ and $p_i$ represent the average iteration counts per step of the
velocity components and pressure. Timestepping is based on Nek5000's second-order
characteristics method with one substep~\cite{maparo90,patel18}, 
and the timestep size is $\Delta t=1.2$e-3 (CFL=1.82).  Pressure
preconditioning is based on $p$-multigrid with Chebyshev and additive Schwarz 
method (CHEBYSHEV+ASM) smoothing and hypre
AMG (algebraic multigrid) for the  coarse-grid solve~\cite{nekrs}. Tolerances for pressure and velocity
are 1e-4 and 1e-6, respectively.  
We note that this test case has been explored in the context of NekRS kernel
and algorithm development on other architectures in earlier work, including
\cite{ceed_special_issue1,ceed_special_issue2,nekrs1}.

The single-device results of Table~\ref{test01} show that, for the current
version of NekRS,\footnote{NekRS version 22.0 is used.} 
a single GCD of the MI250X on Crusher realizes a 1.32$\times$
gain in Navier--Stokes solution performance over a single V100 on Summit.
Similarly, the A100s are realizing $\approx$ 1.6-fold gain over the V100.

 \begin{table*}
  \footnotesize
  \begin{center}
  \begin{tabular}{|l|r|l|l|r|r|r|c|}
  \hline
  \multicolumn{8}{|c|}{{\bf GPU Performance on a Single GPU}: singlerod, $E=7168$, $n=2,458,624$, $N=7$}\\
  \hline
       System   & GPU  &  Device & API  & $v_i$ & $p_i$  & $t_{\rm step}$ (sec)  &  Speedup    \\
  \hline
       Summit   & 1 & 16GB V100 GPU       & CUDA &  4    &  1  & 7.98e-02  & 1.00  \\
       Spock    & 1 & 32GB MI100 GPU      & HIP  &  4    &  1  & 4.17e-02  & 0.84  \\
       Crusher  & 1 & 64GB MI250X (1 GCD) & HIP  &  4    &  1  & 6.02e-02  & 1.32  \\
       ThetaGPU & 1 & 40GB A100 GPU       & CUDA &  4    &  1  & 6.78e-02  & 1.57  \\
       Perlmutter & 1 & 40GB A100 GPU     & CUDA &  4    &  1  & 4.16e-02  & 1.62  \\
       Polaris  & 1 & 40GB A100 GPU       & CUDA &  4    &  1  & 4.31e-02  & 1.62  \\
  \hline
  \end{tabular}
  \end{center}
  \caption{\label{test01}NekRS performance on various architectures using a single GPU.}
 \end{table*}
\section{NekRS Performance on Frontier vs. Crusher}
\label{nek-frontier}

On Frontier,
{\tt rocm/5.1.0} and {\tt cray-mpich/8.1.17} were used. On Crusher,
simulations were performed with variation of versions such as {\tt rocm/5.1.0},
{\tt rocm/5.2.0}, {\tt cray-mpich/8.1.16}, and {\tt cray-mpich/8.1.19}. On
Crusher, {\tt rocm/5.1.0} is 2\%--5\% faster than {\tt rocm/5.2.0}. 
We observe that the performance on Frontier is better than that on Crusher.

We consider ExaSMR's 17$\times$17 rod-bundle geometry and extend the domain in
streamwise direction with 10, 17, and 170 layers, keeping the mesh density
the same, which correspond to 277 thousand spectral elements of order $N=7$, for a
total of $n=.27$M $\times 7^3 = 95$M grid points, 471 thousand spectral
elements of order $N=7$, for a total of $n=.47$M $\times 7^3 = 161$M grid
points, and 4.7 million spectral elements of order $N=7$, for a total of
$n=4.7$M $\times 7^3 = 1.6$B grid points, respectively.  Table~\ref{cases}
summarizes the configuration of the testing cases.

 \begin{table*}[t]
  \footnotesize
  \begin{center}
  \begin{tabular}{|r|r|r|r|}
  \hline
  \multicolumn{4}{|c|}{{\bf Strong Scaling Test Sets}}\\
  \hline
         &         $E$ &   $n$  &  rank, $P$  \\
  \hline
  Case 1 &     277000  &   95M  &  8--64      \\
  Case 2 &     470900  &  161M  &  14-128     \\
  Case 3 &    4709000  &  1.6B  &  128--16320  \\
  \hline
  \end{tabular}
\end{center}
\caption{\label{cases}Problem setup for strong-/weak-scaling studies. }
\end{table*}

\begin{figure*}
  \begin{center}
     \includegraphics[width=0.5\textwidth]{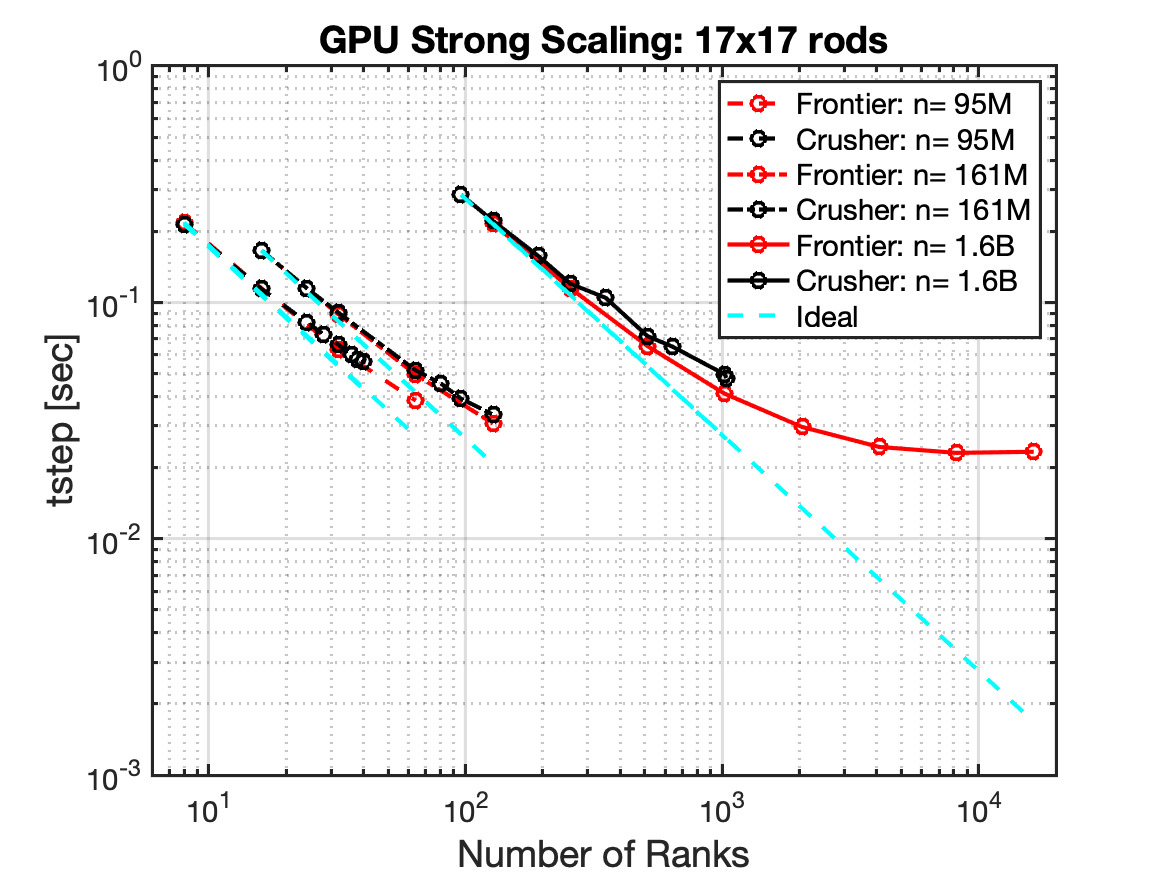 }
     \hspace{-2em}
     \includegraphics[width=0.5\textwidth]{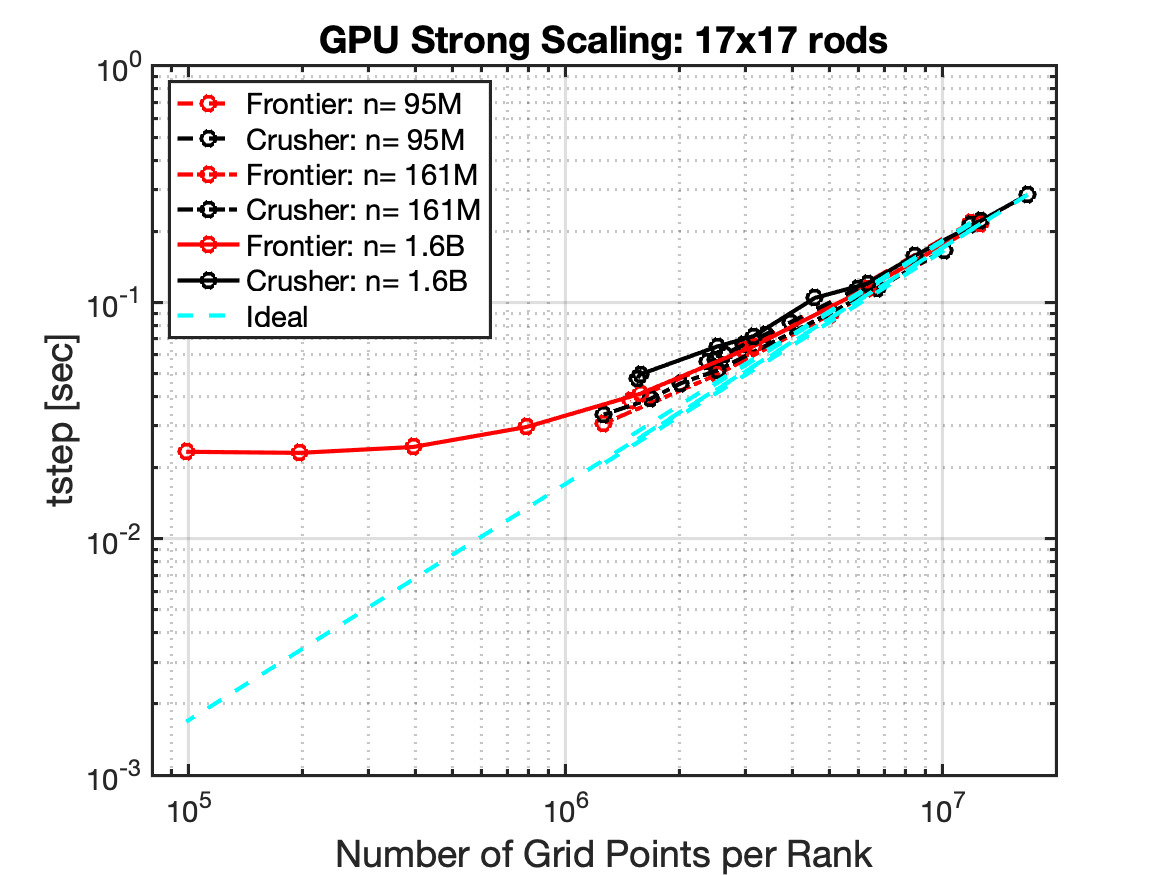 }
  \end{center}
\caption{\label{perf-frontier}Strong scaling on Frontier and Crusher for $17\times17$ rod bundles 
with 10, 17, and 170 layers with total number of grid points of $n=$ 95M, 161M, and 1.6B.  
Average time per step vs. rank, $P$ (left), and average time per step vs. $n/P$ (right).  
Frontier is set with ({\small \tt cray-mpich/8.1.17}, {\small\tt rocm/5.1.0}) and 
Crusher with ({\small\tt cray-mpich/8.1.19}, {\small\tt rocm/5.2.0}).
}
\end{figure*}

Figure~\ref{perf-frontier} compares the scaling performance of Frontier with that
of Crusher.  Simulations are performed for 2,000 steps; and the average
time-per-step, $t_{\rm step}$, is measured in seconds for the last 1,000 steps.
The third-order backward-difference formula (BDF3) combined with the
third-order extrapolation (EXT3)~\cite{dfm02} is used for timestepping. and the
timestep size is $\Delta t$ = 3.0e-04  (CFL=0.82).

Figure~\ref{perf-frontier}, left, shows the classic strong scaling for the
problem sizes of $n$= 95M, 161M, and 1.6B, demonstrating the average
time per step vs. the number of MPI ranks, $P$.  We run a single MPI rank per
GCD, and there are 8 GCDs per node.  The dashed lines in sky-blue color represent ideal
strong-scale profiles for each case.  The solid lines in red are for Frontier,
and the solid lines in black are for Crusher.  We observe that Frontier is
consistently slightly faster than Crusher for these three problem sizes.   For
larger problem sizes and processor counts, the Frontier advantage is increased.

Figure~\ref{perf-frontier}, right, shows the average time per step vs.  the
number of points per MPI rank, $n/P$, where $n$ is the total number of grid
points.  $t_{\rm step}$ based on $n/P$ is independent of the problem
size, $n$.  This is the metric illustrating that the strong-scaling performance
is primarily a function of $(n/P)$ and only weakly dependent on $n$ or $P$
individually, which is in accord with the extensive studies presented
in~\cite{ceed_bp_paper_2020}.  Based on this metric, we can determine a
reasonable number value of $(n/P)$ for a given parallel efficiency and, from
there, determine the number of MPI ranks required for a problem of size $n$ to
meet that expected efficiency.  We provide more detailed performance behaviors
depending on problem sizes in
Figures~\ref{perf10-all-frontier}--\ref{perf170-all-frontier}.


\begin{figure*}
  \begin{center}
     \includegraphics[width=0.5\textwidth]{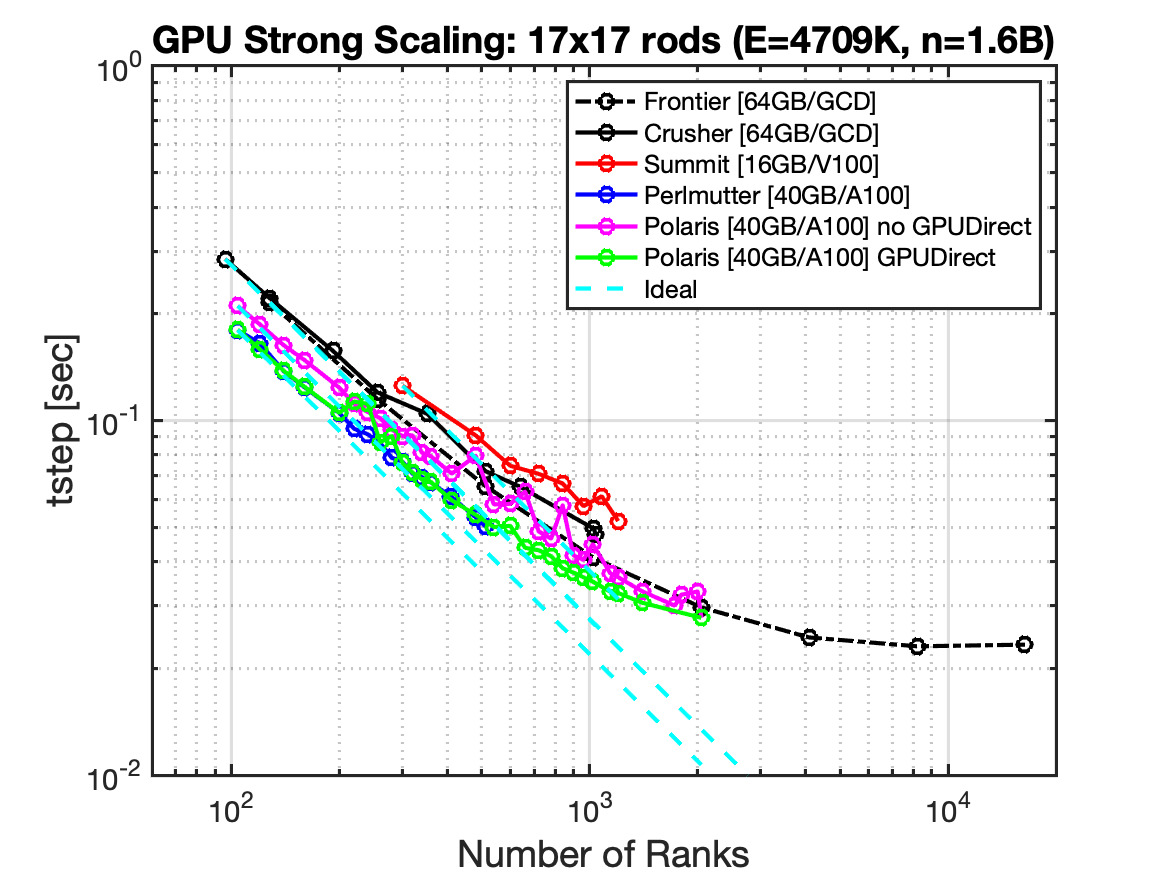 }
     \hspace{-2em}
    \includegraphics[width=0.5\textwidth]{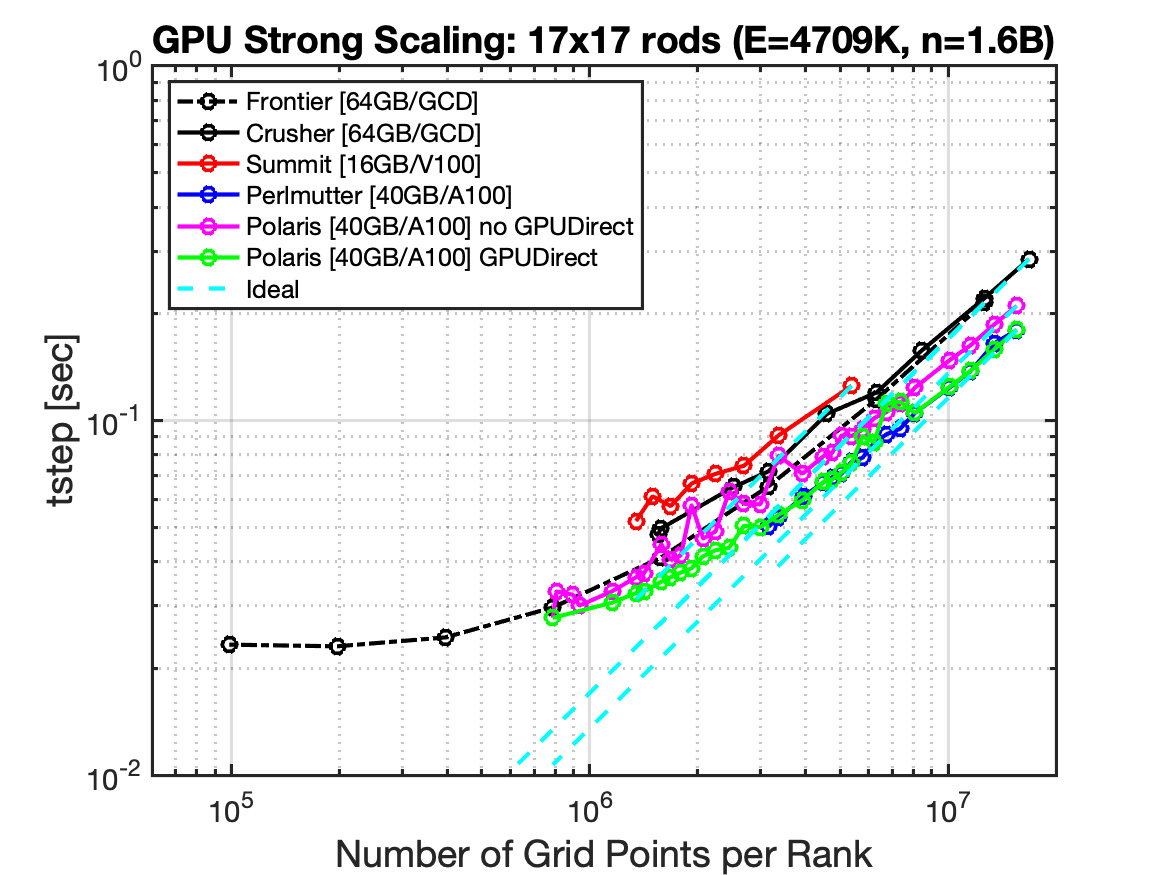 }
  \end{center}
\caption{\label{perf170-all}Strong scaling on  Frontier (MI250X), Crusher (MI250X), Perlmutter (A100),
Polaris (A100), and Summit (V100) for $17\times17$ rod bundles with 170 layers with total number of grid
points of $1.6B$.  Average time per step vs. rank, $P$ (left) and average time per step vs. $n/P$ (right).
}
\end{figure*}

Figure~\ref{perf170-all}, left and right, shows performance for a 17$\times$17
rod bundle with 170 layers ($n=$ 1.6B).  Here we extend our discussion to
other NVIDIA-based GPU architectures such as Summit (V100) at OLCF, Perlmutter
(A100) at NERSC, and Polaris (A100) at ALCF, and we  compare those with Frontier
and Crusher.  While we observe that Frontier is faster than Crusher in
Figure~\ref{perf-frontier}, we see that Crusher is faster than Summit, but not
quite as fast as the A100-based Perlmutter (NERSC) and Polaris (ALCF)
platforms.  
We provide more detailed performance behavior as a function of $P$ in
Figures~\ref{perf10-all-frontier}-\ref{perf170-all-frontier}.
Anomalous behavior for several of these architectures is also discussed
in Section \ref{sec:disc}.

\begin{figure*}
  \begin{center}
     \includegraphics[width=0.44\textwidth]{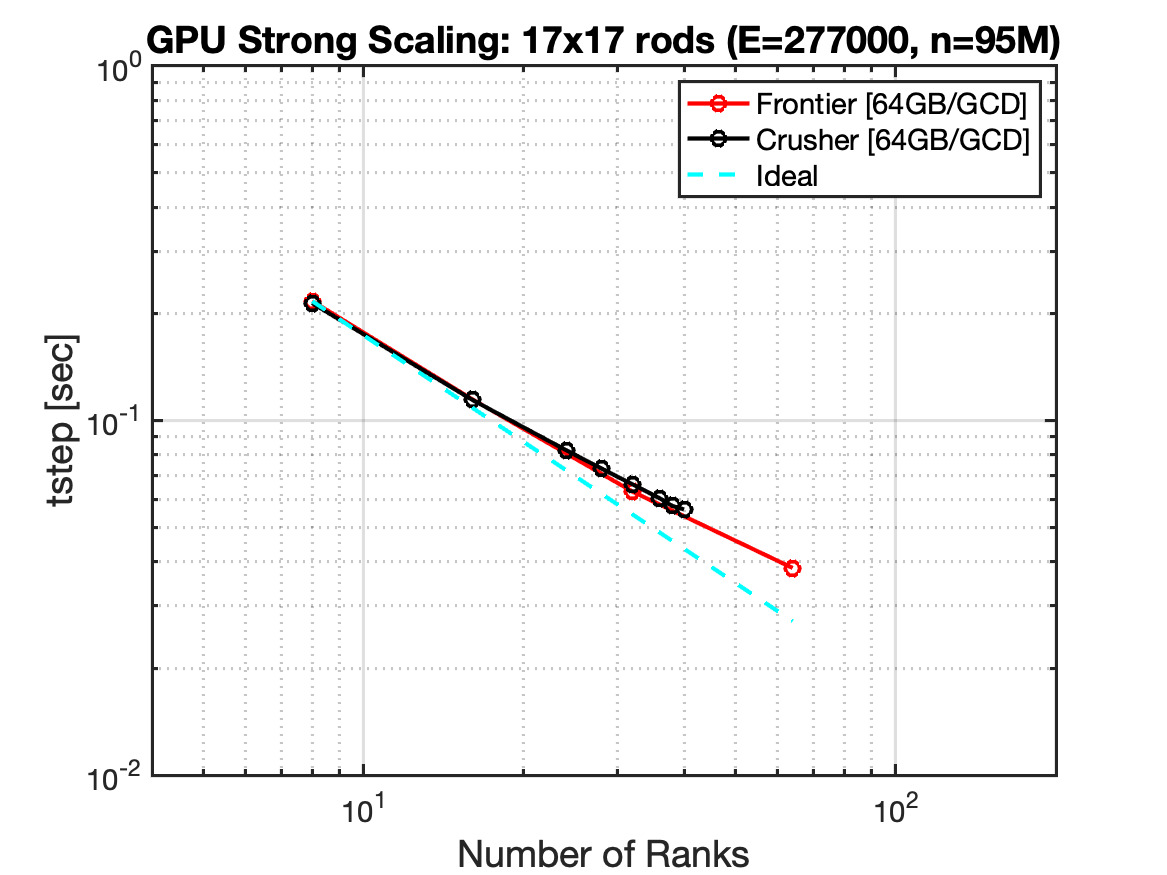 }
     \includegraphics[width=0.44\textwidth]{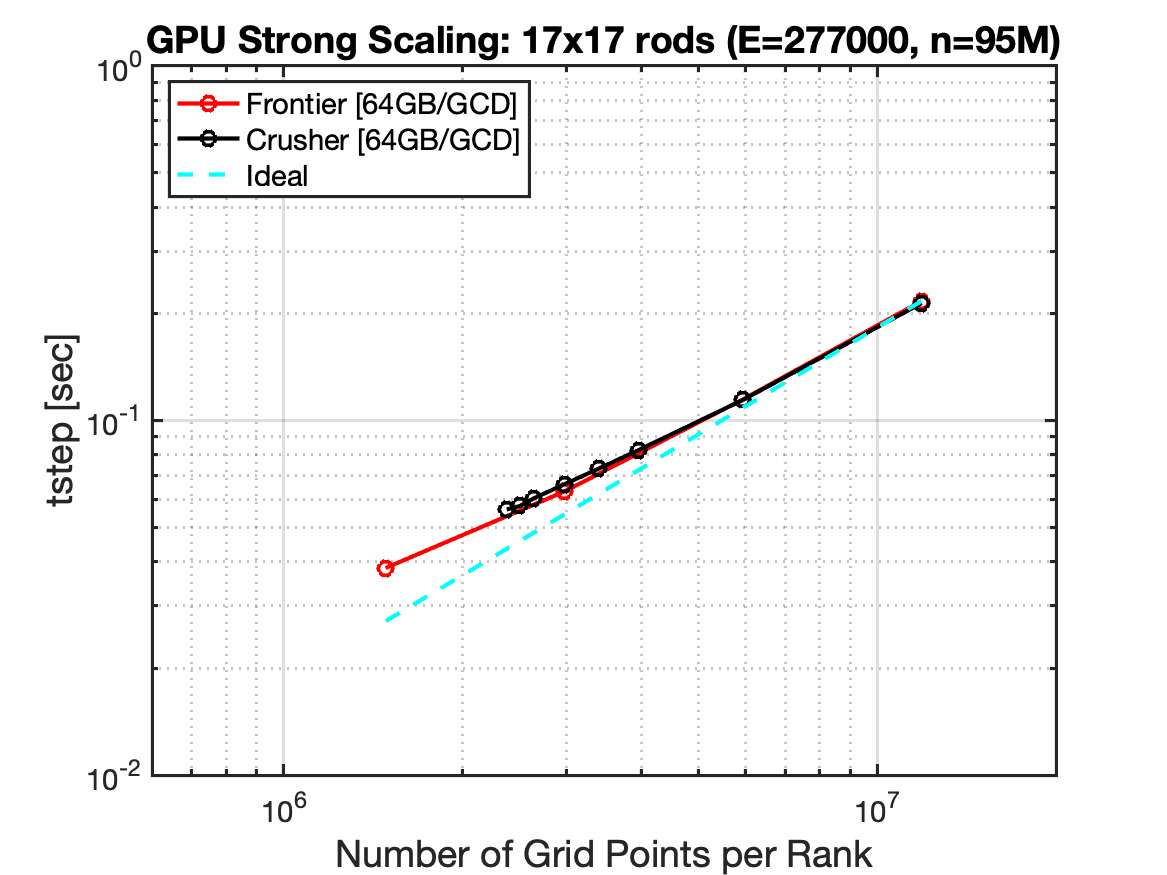 }
     \\
     \includegraphics[width=0.44\textwidth]{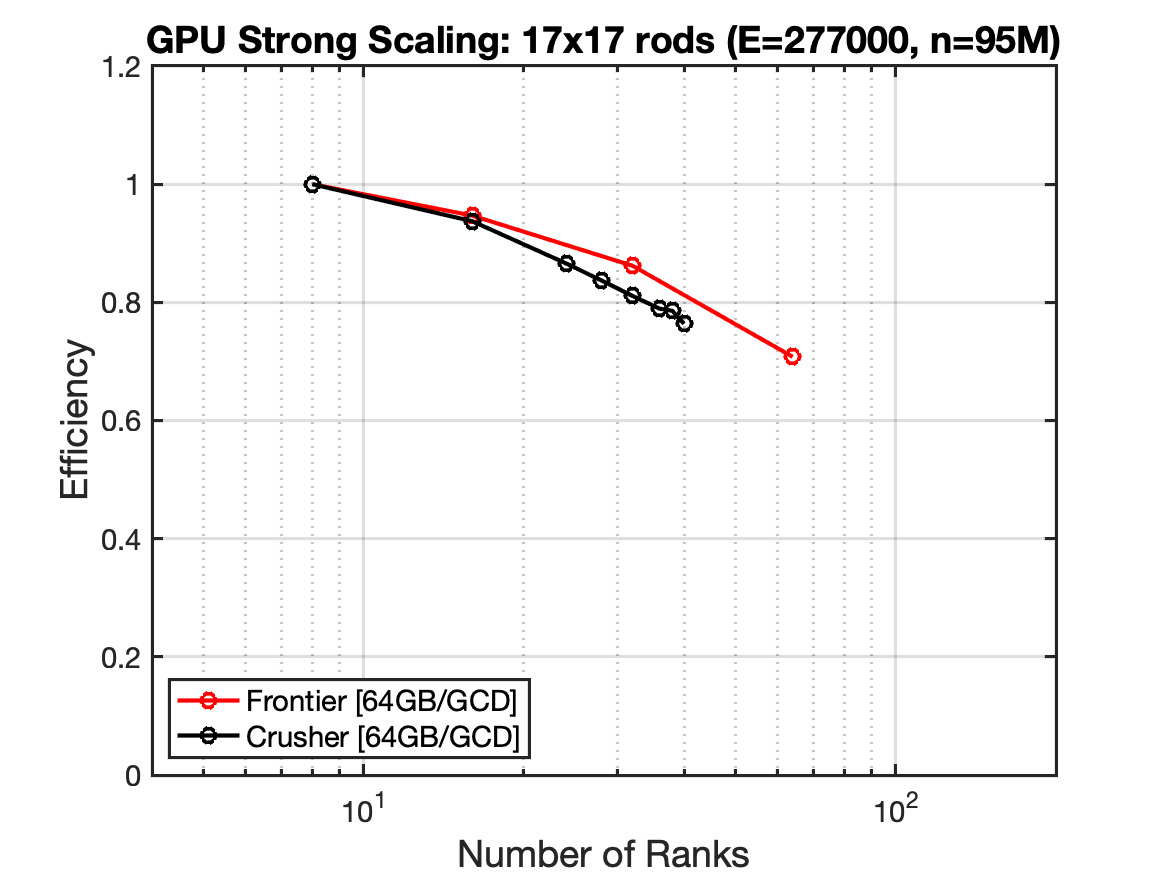 }
     \includegraphics[width=0.44\textwidth]{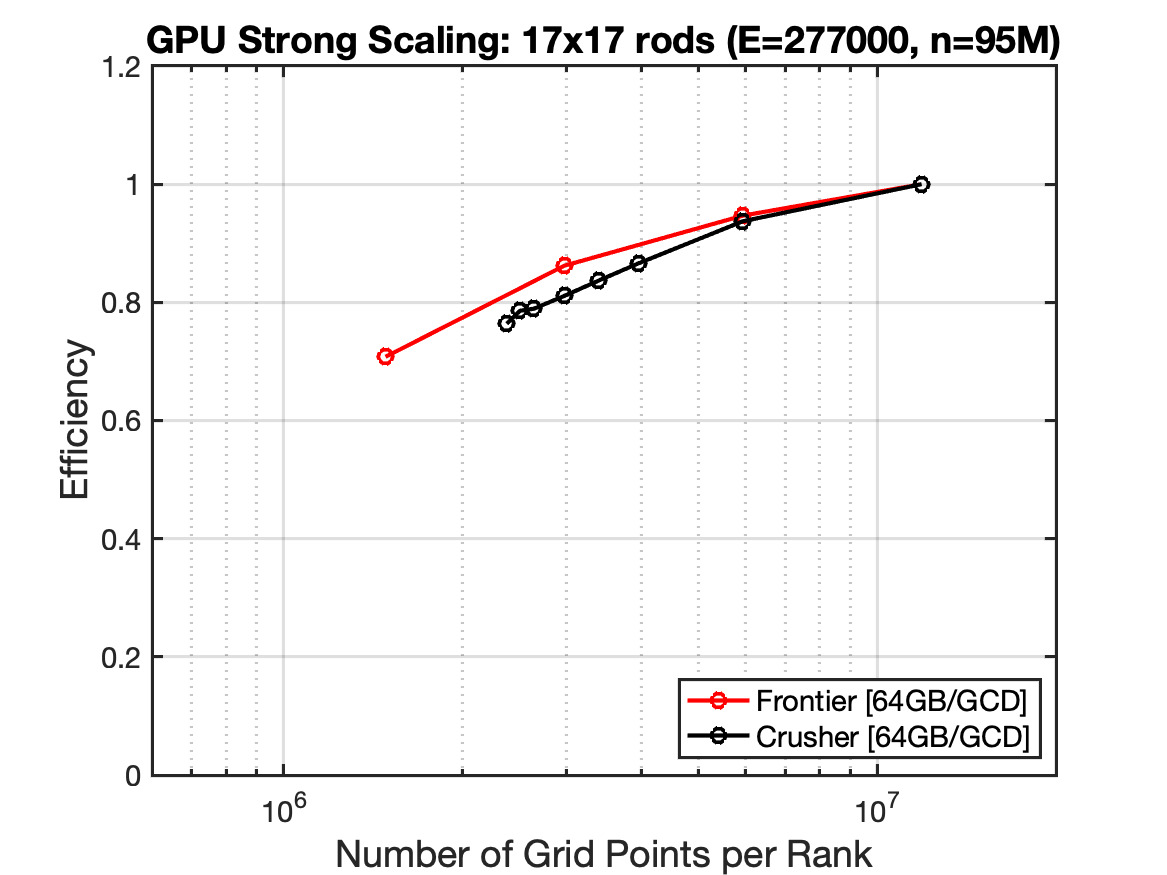 }
     \\
     \includegraphics[width=0.44\textwidth]{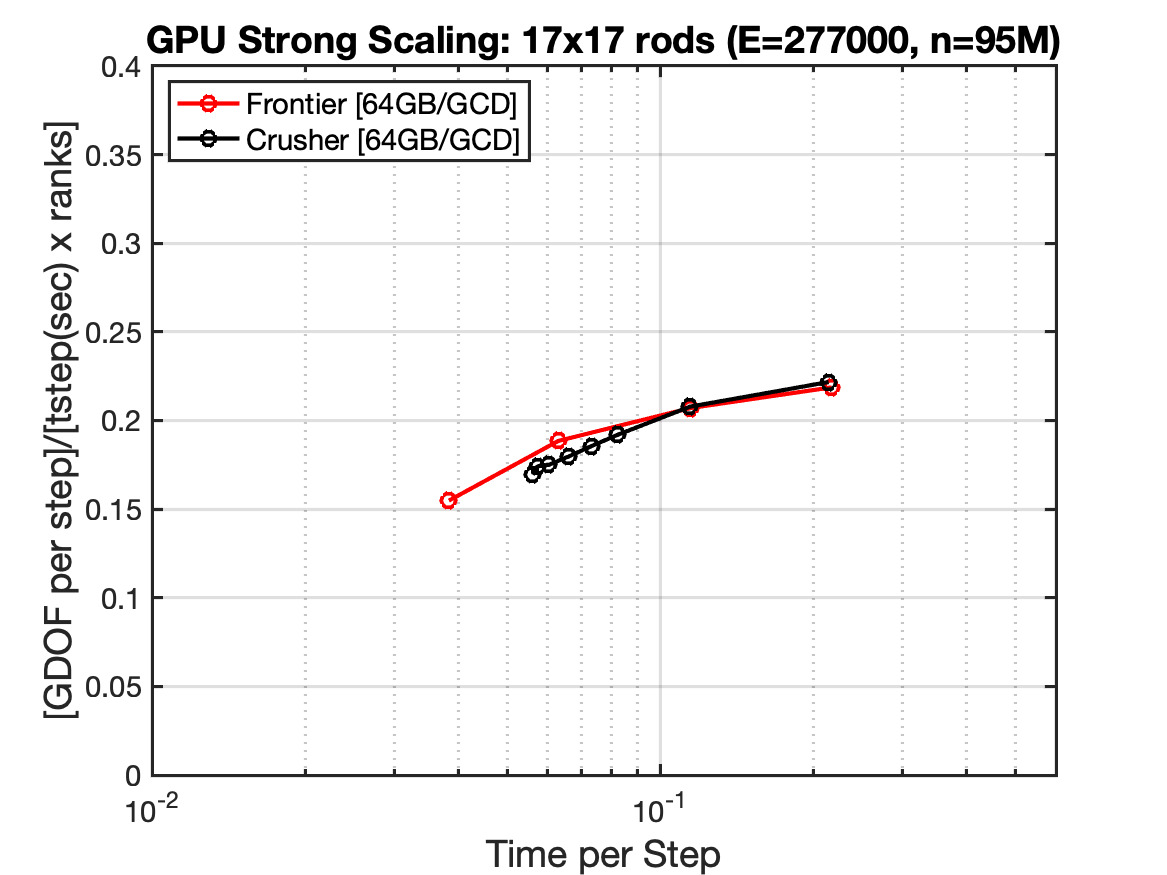 }
     \includegraphics[width=0.44\textwidth]{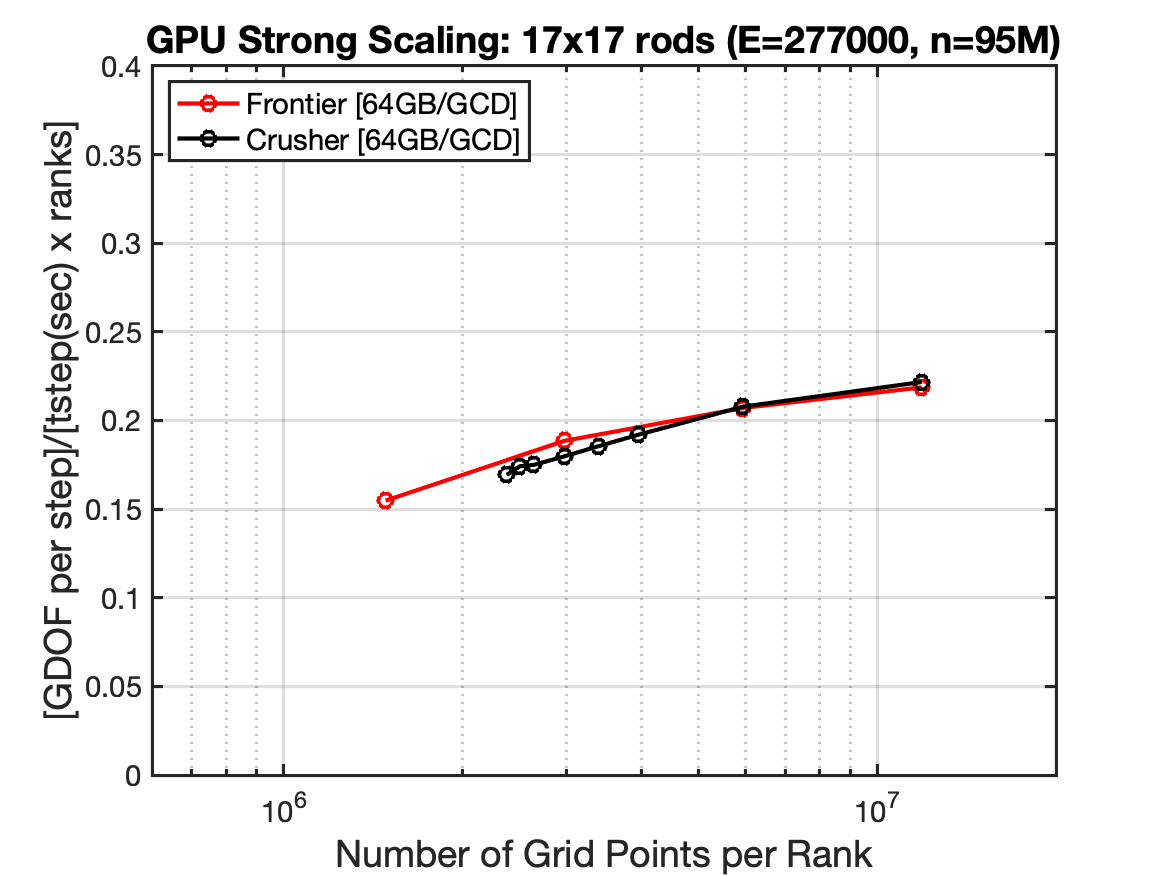 }
     \\
     \includegraphics[width=0.44\textwidth]{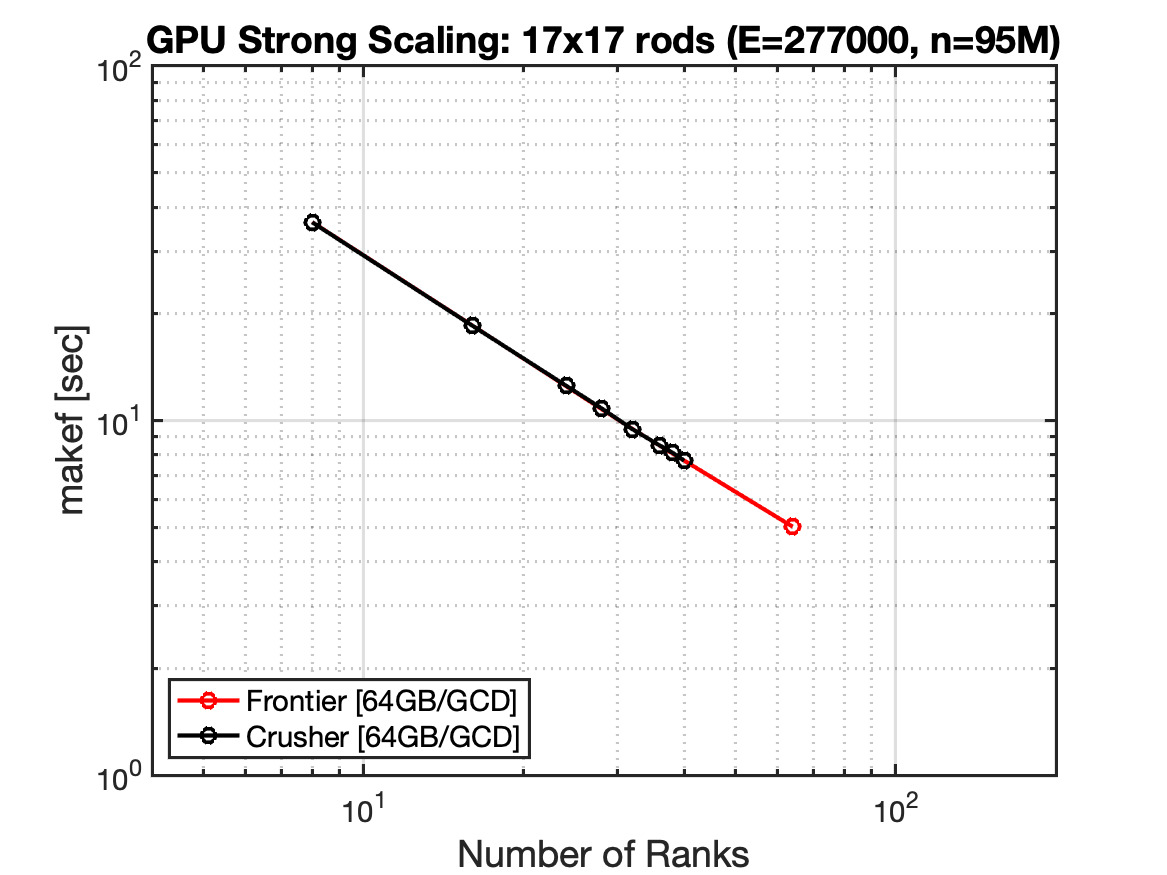 }
     \includegraphics[width=0.44\textwidth]{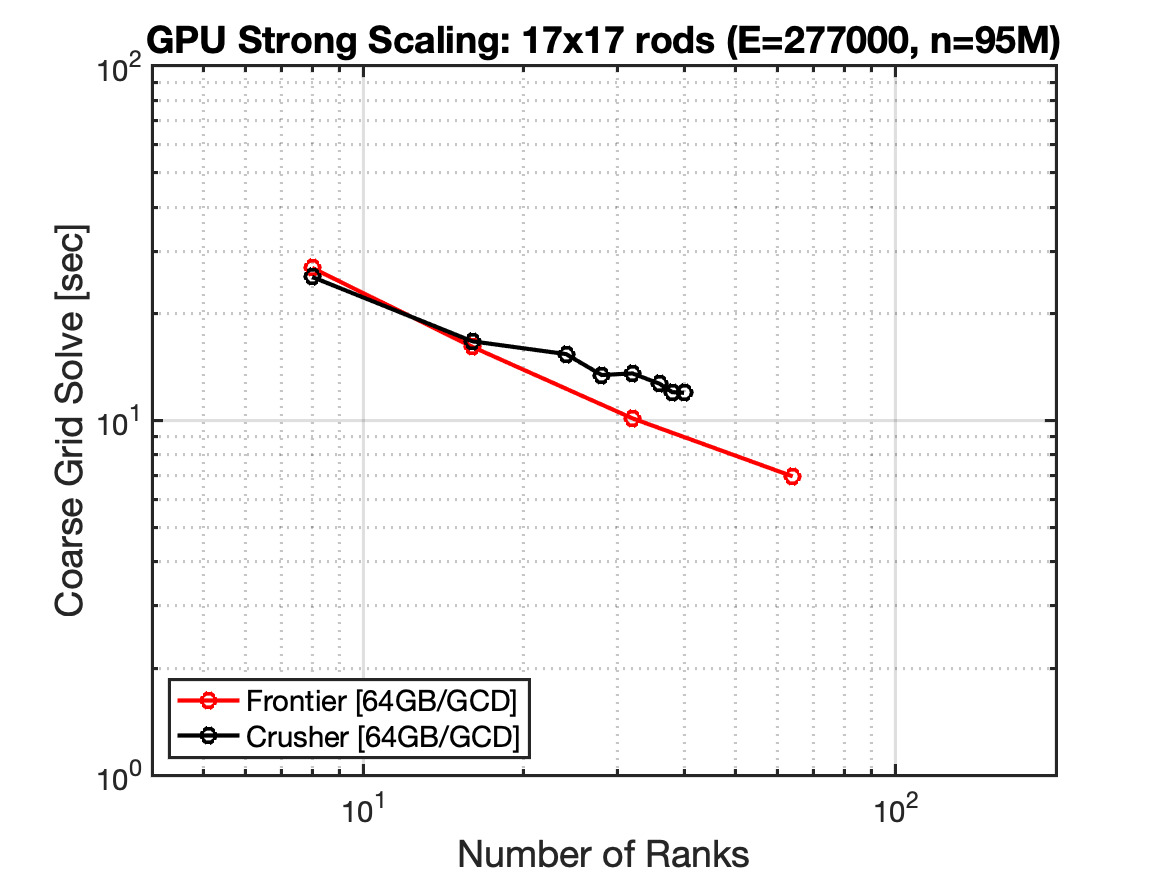 }
   \caption{\label{perf10-frontier}Strong-scaling on Frontier vs. Crusher for 17$\times$17 rod bundle with 10 layers.}
  \end{center}
\end{figure*}
\begin{figure*}
  \begin{center}
     \includegraphics[width=0.44\textwidth]{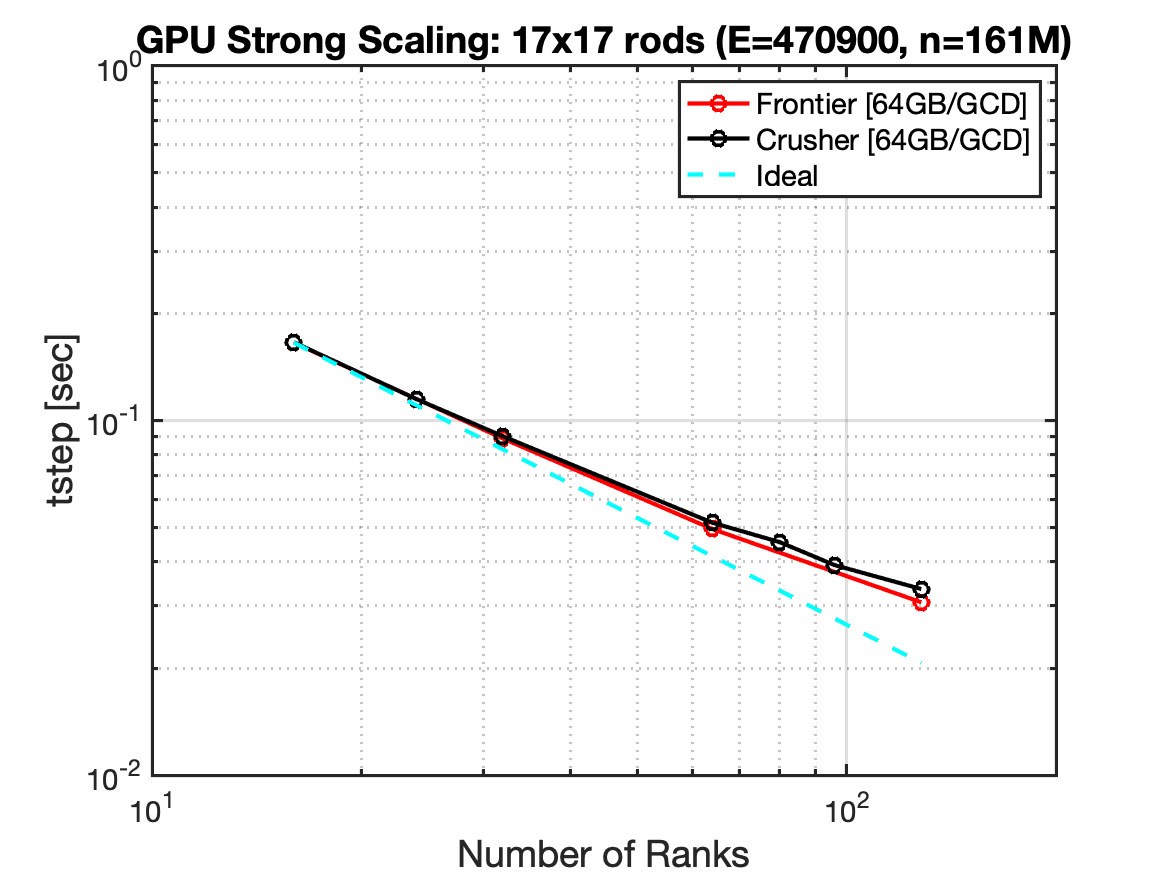 }
     \includegraphics[width=0.44\textwidth]{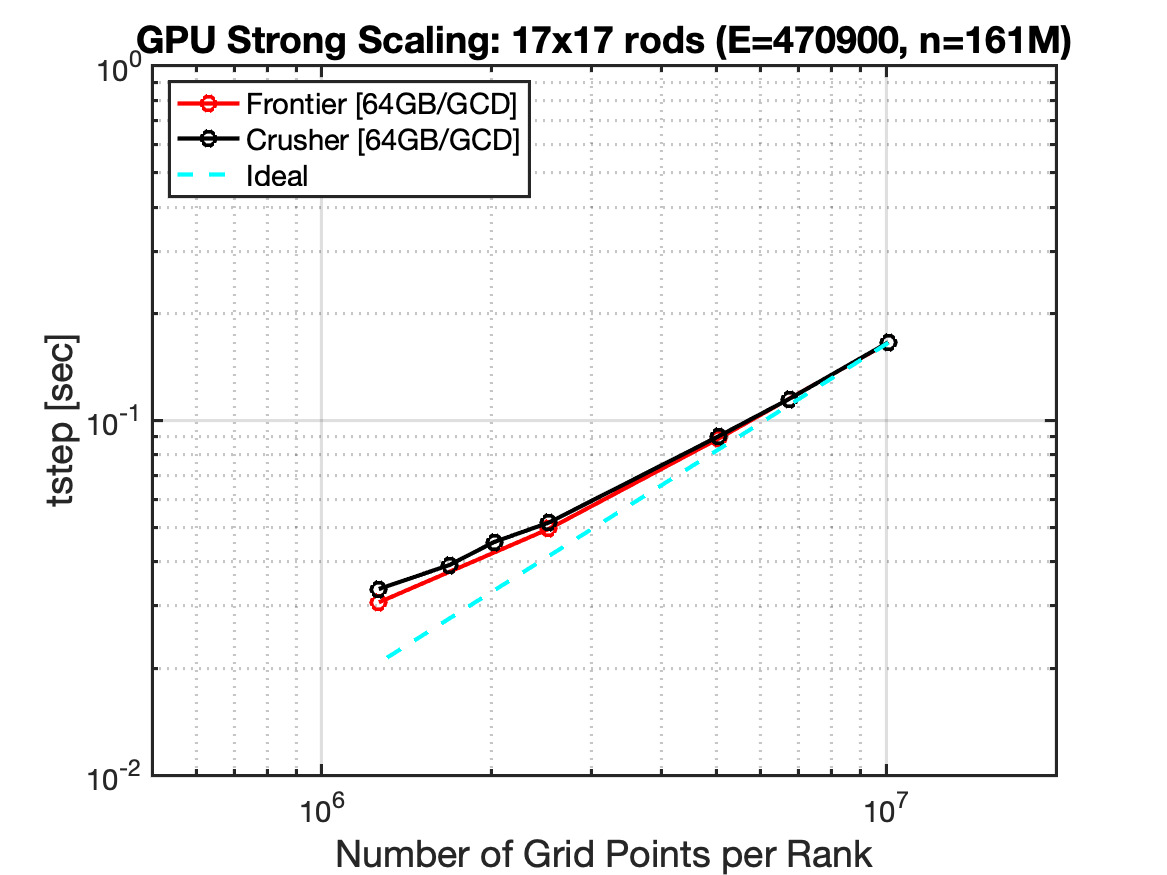 }
     \\
     \includegraphics[width=0.44\textwidth]{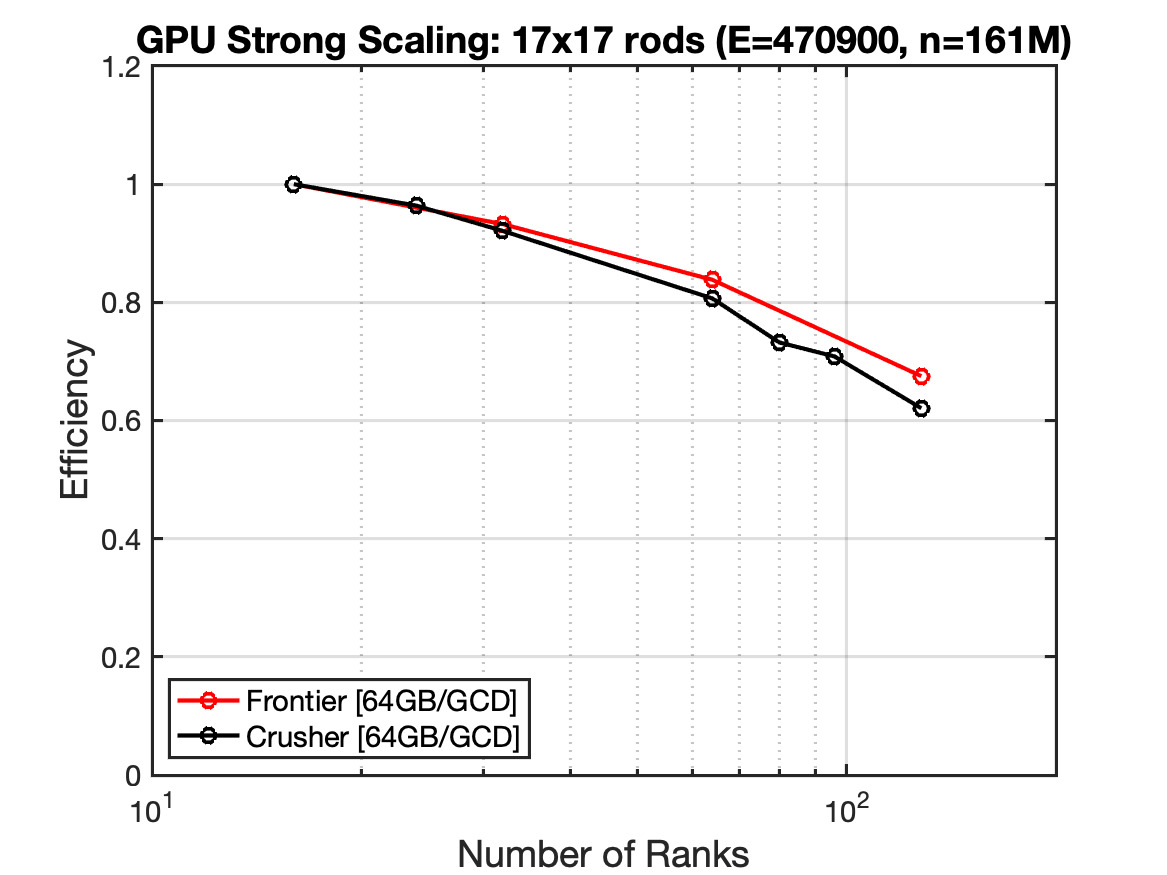 }
     \includegraphics[width=0.44\textwidth]{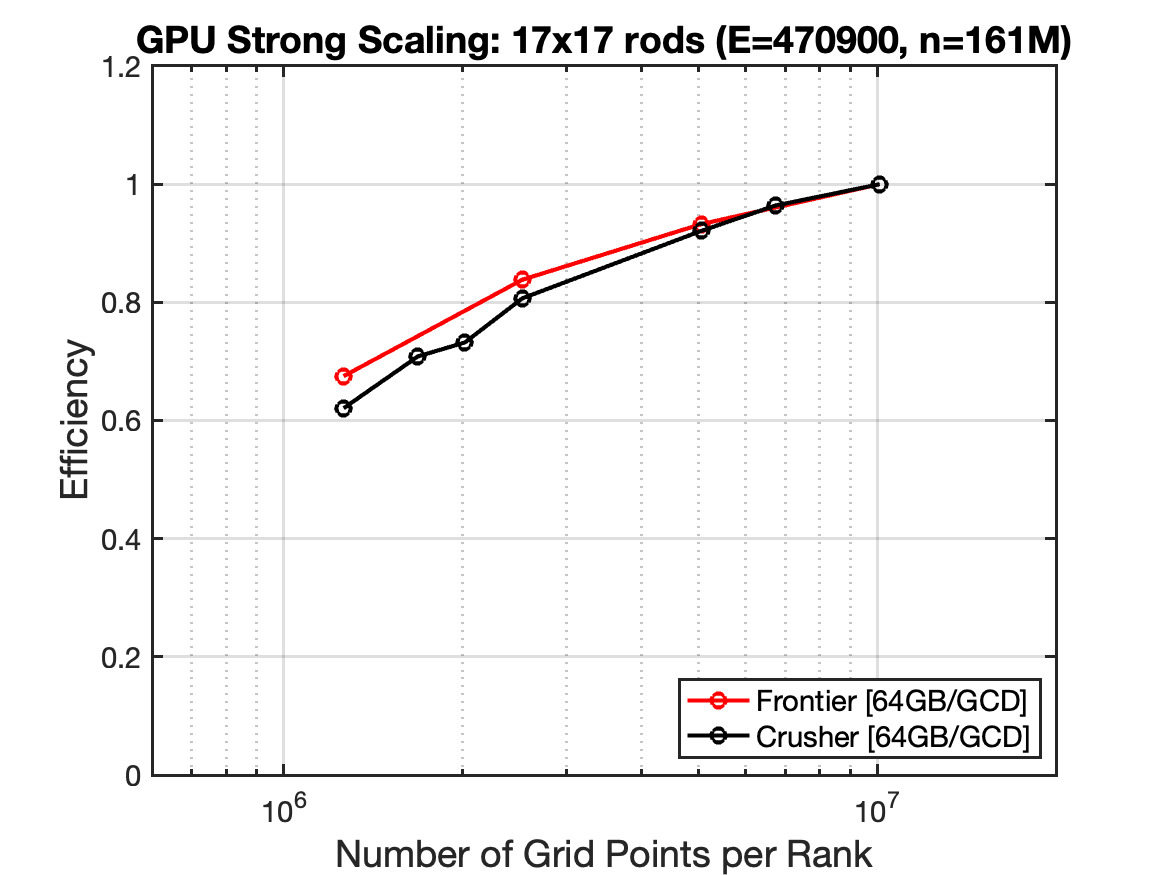 }
     \\
     \includegraphics[width=0.44\textwidth]{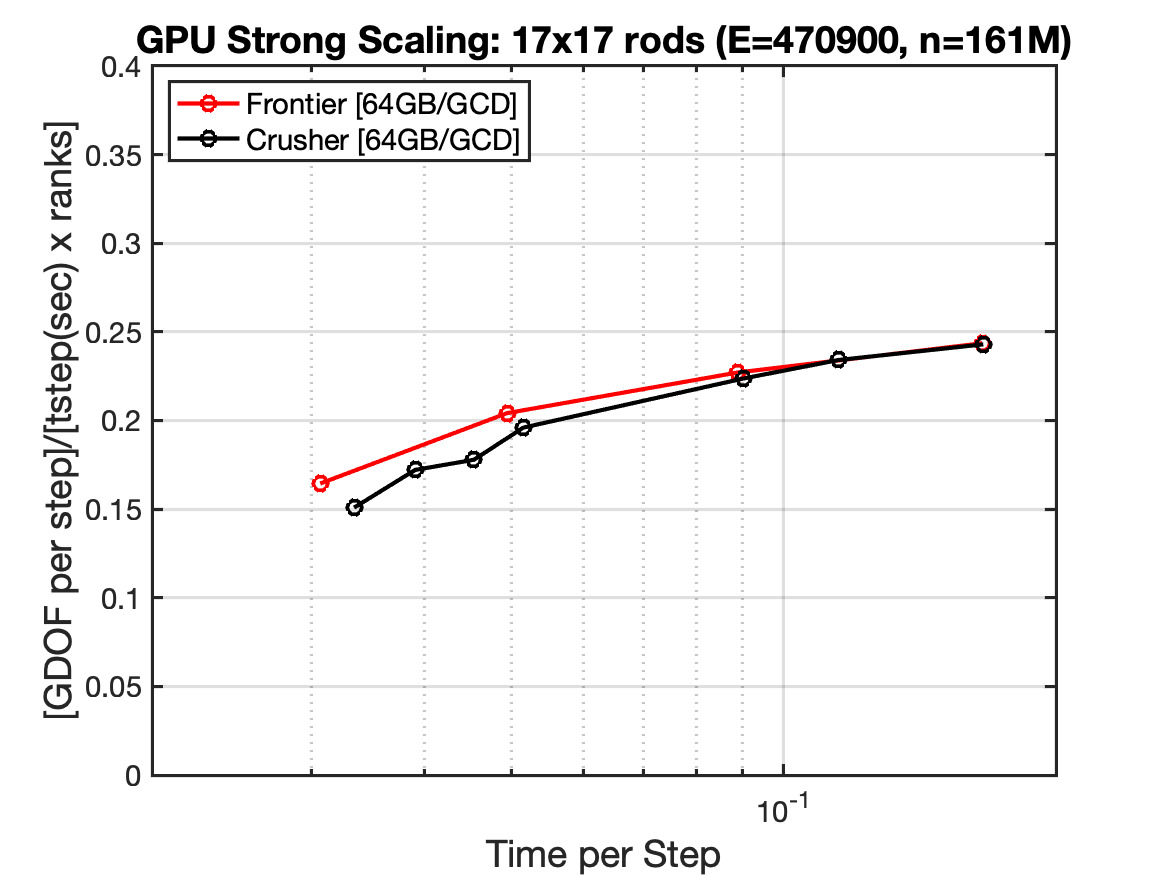 }
     \includegraphics[width=0.44\textwidth]{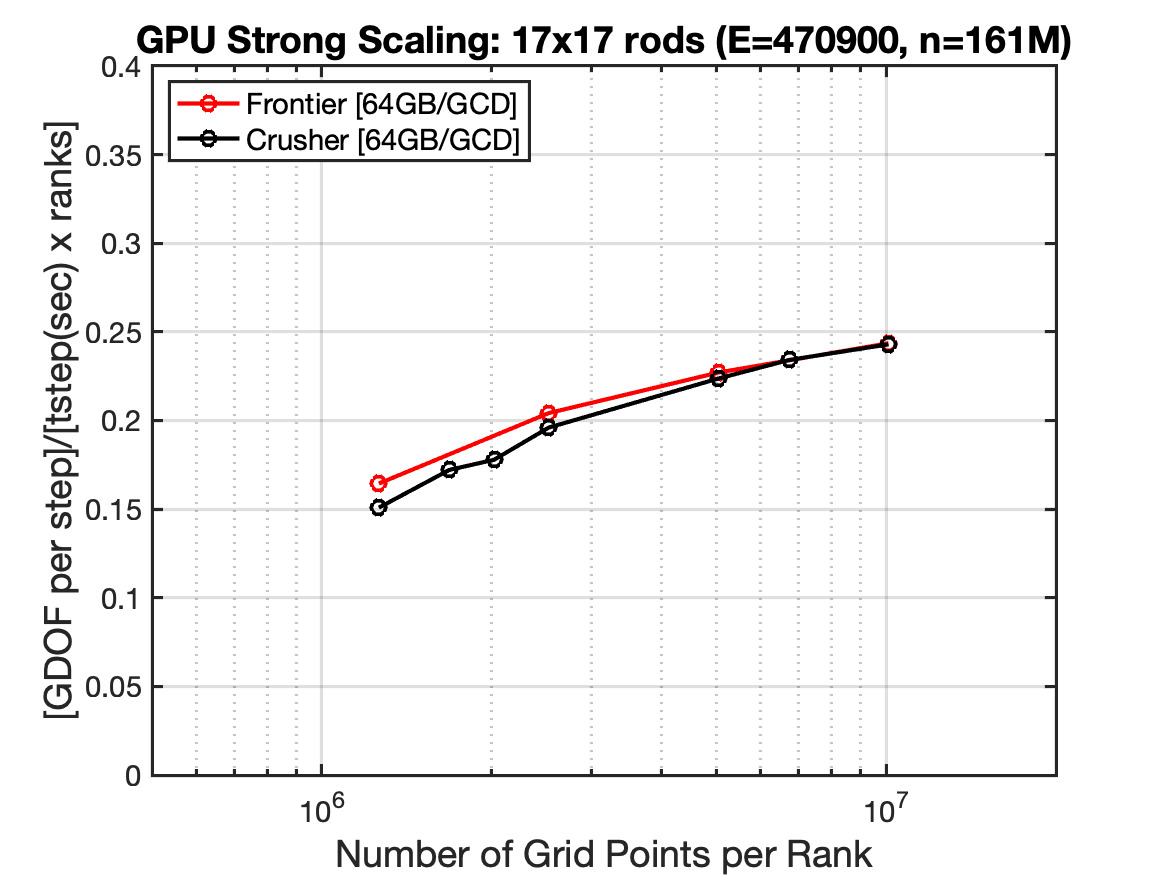 }
     \\
     \includegraphics[width=0.44\textwidth]{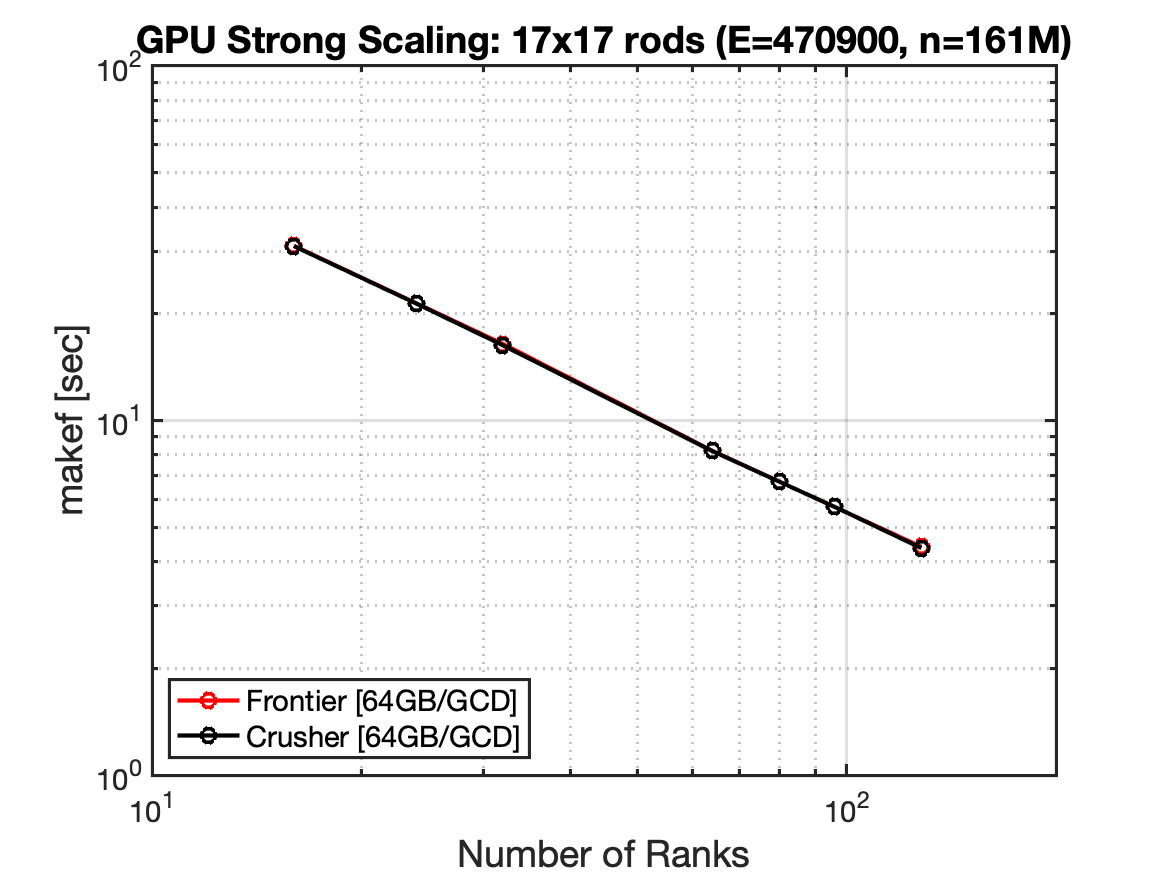 }
     \includegraphics[width=0.44\textwidth]{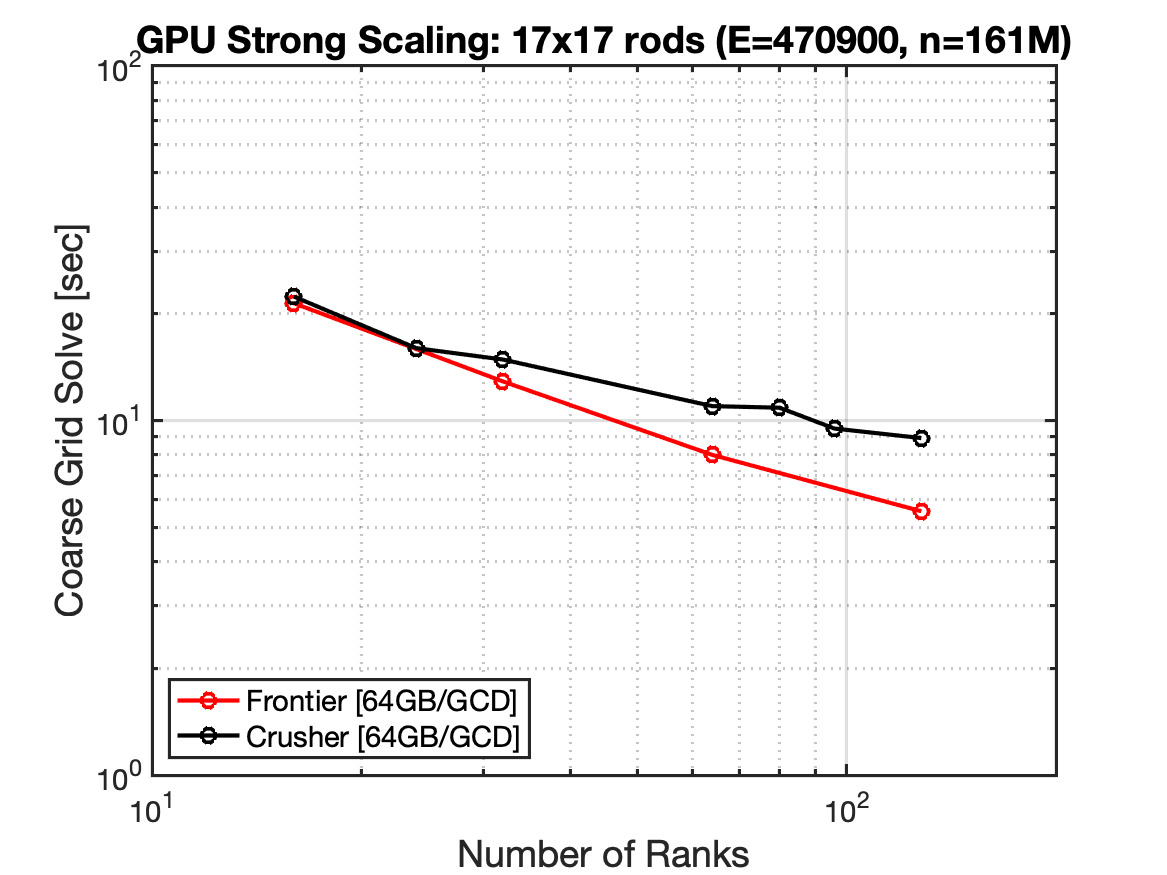 }
   \caption{\label{perf17-frontier}Strong-scaling on Frontier vs. Crusher for 17$\times$17 rod bundle with 17 layers.}
  \end{center}
\end{figure*}
\begin{figure*}
  \begin{center}
     \includegraphics[width=0.44\textwidth]{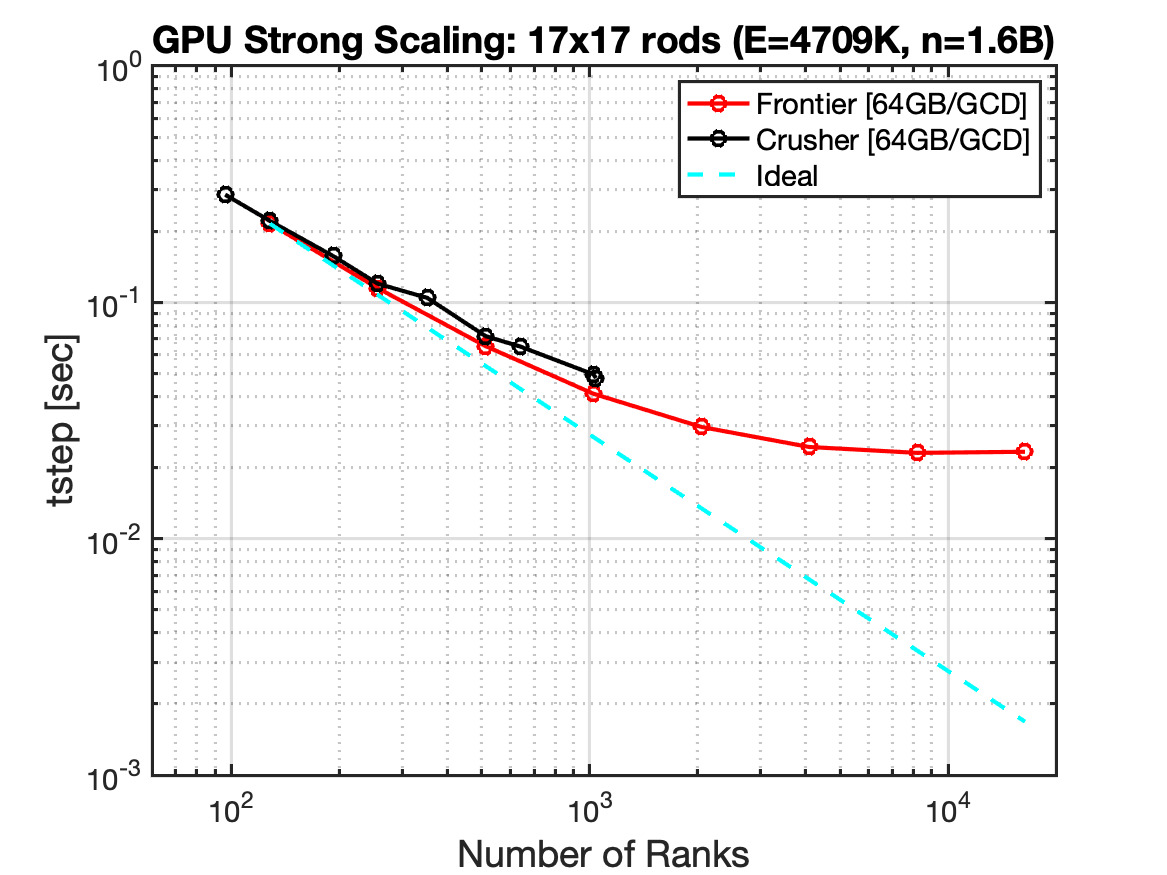 }
     \includegraphics[width=0.44\textwidth]{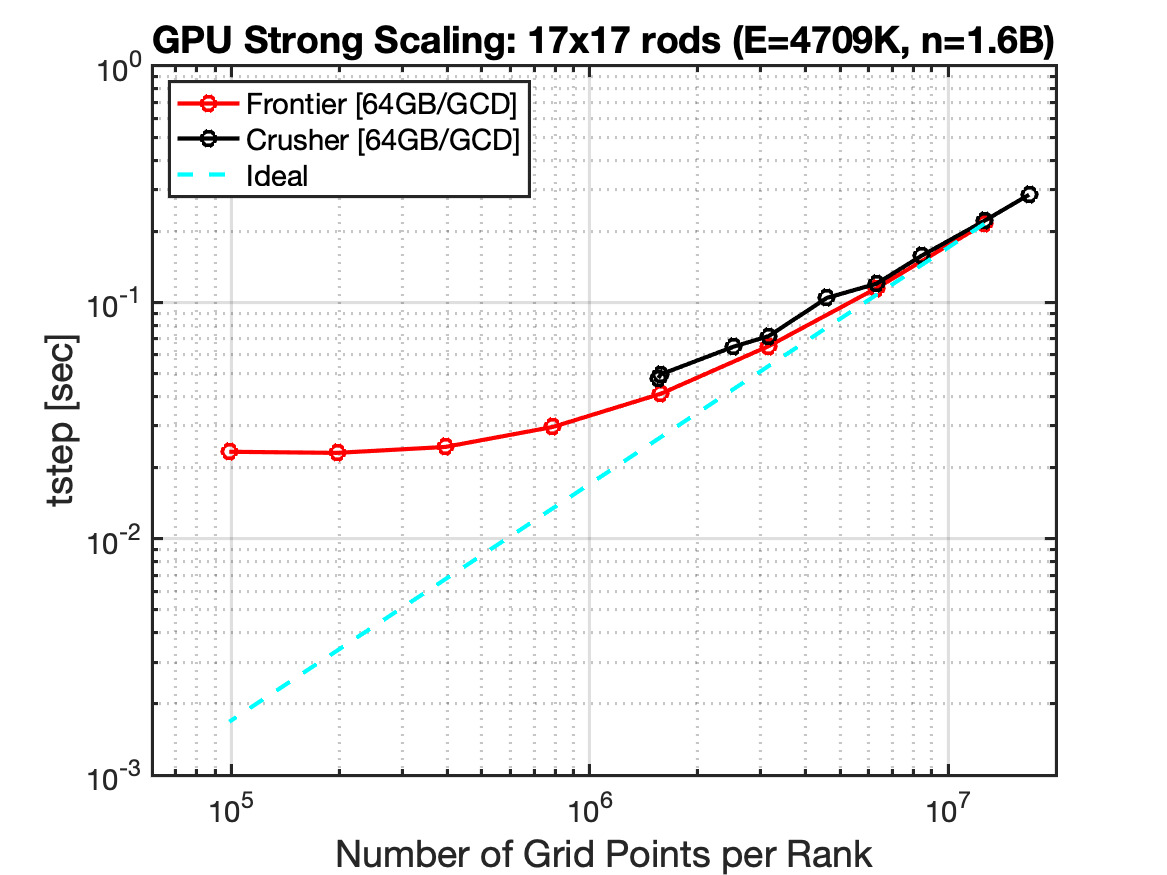 }
     \\
     \includegraphics[width=0.44\textwidth]{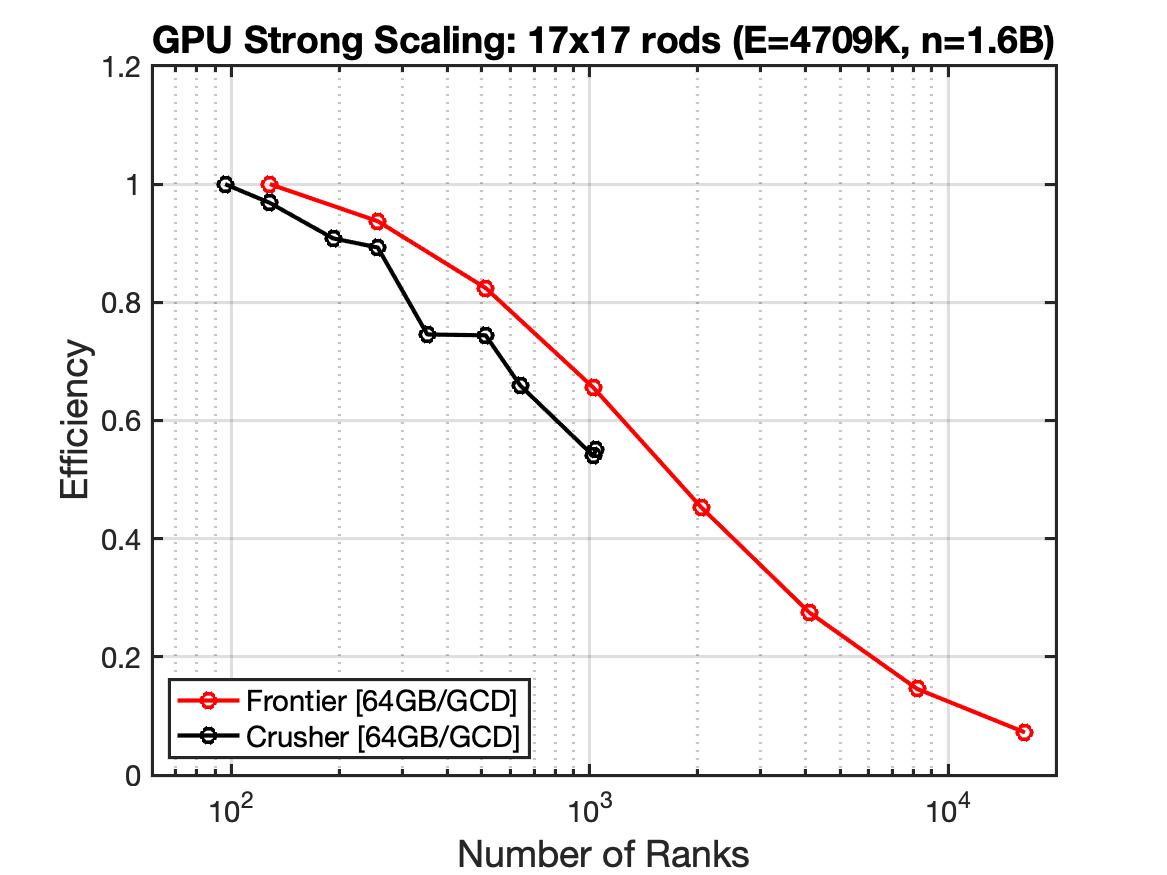 }
     \includegraphics[width=0.44\textwidth]{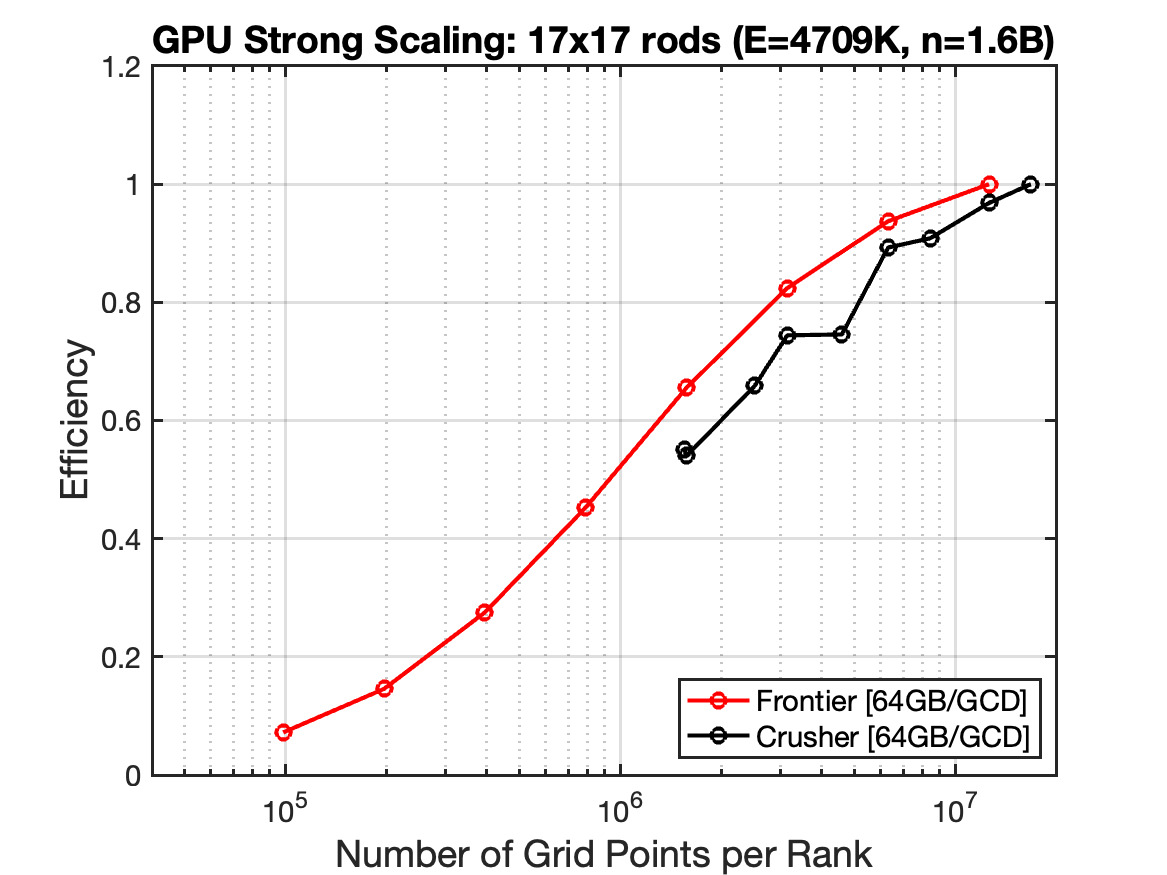 }
     \\
     \includegraphics[width=0.44\textwidth]{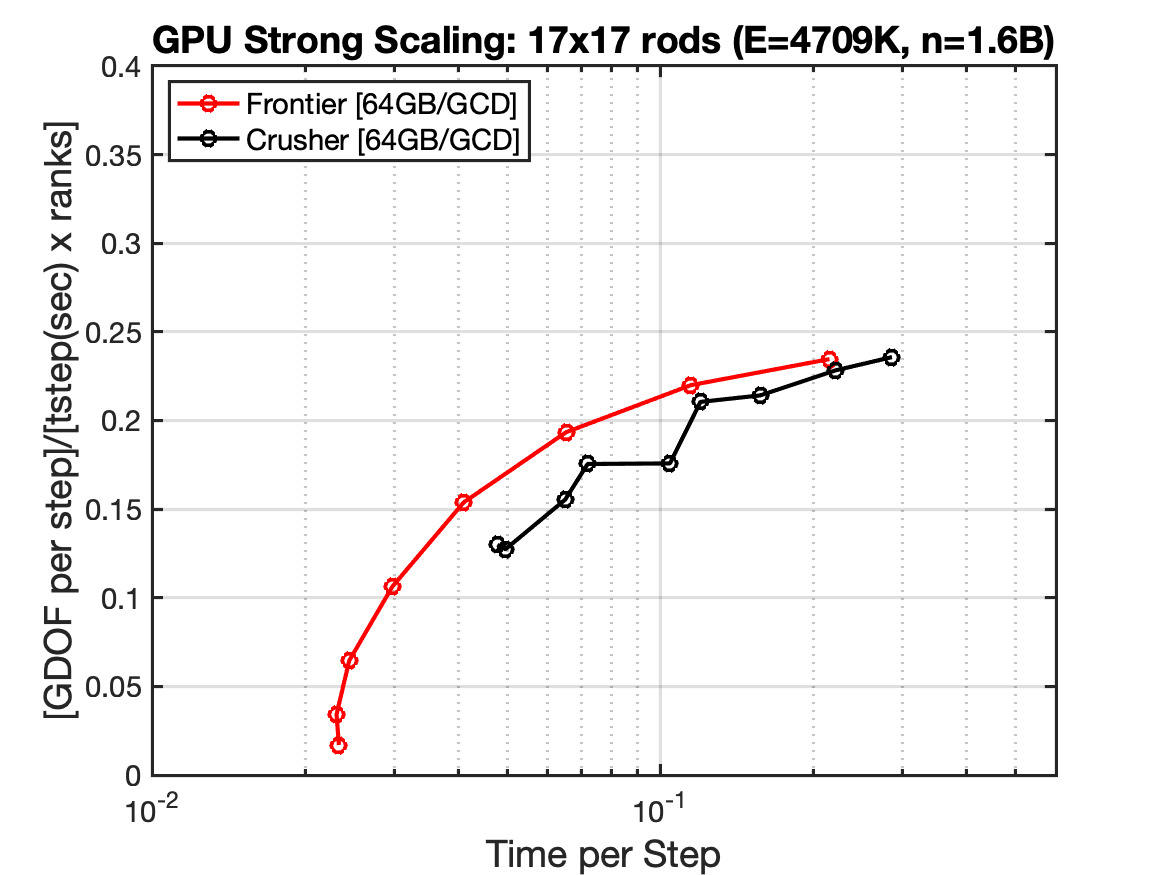 }
     \includegraphics[width=0.44\textwidth]{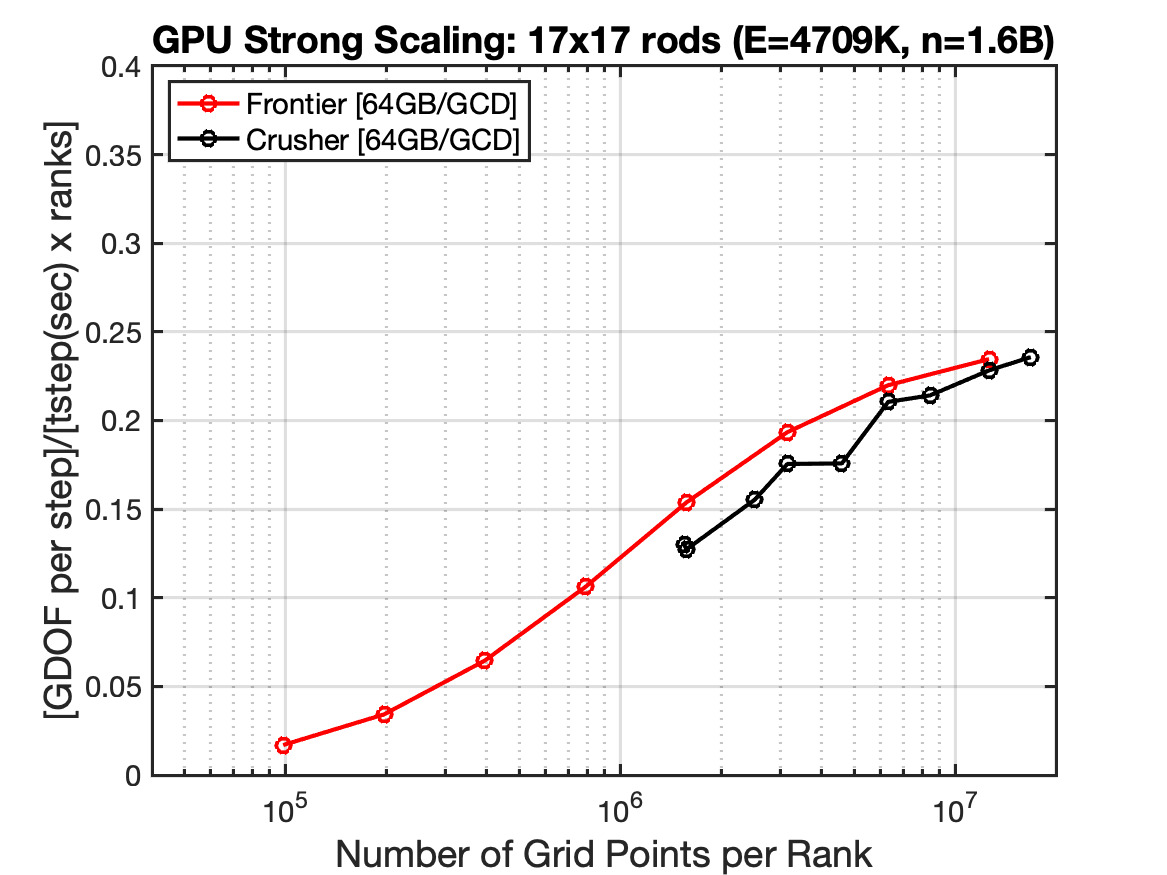 }
     \\
     \includegraphics[width=0.44\textwidth]{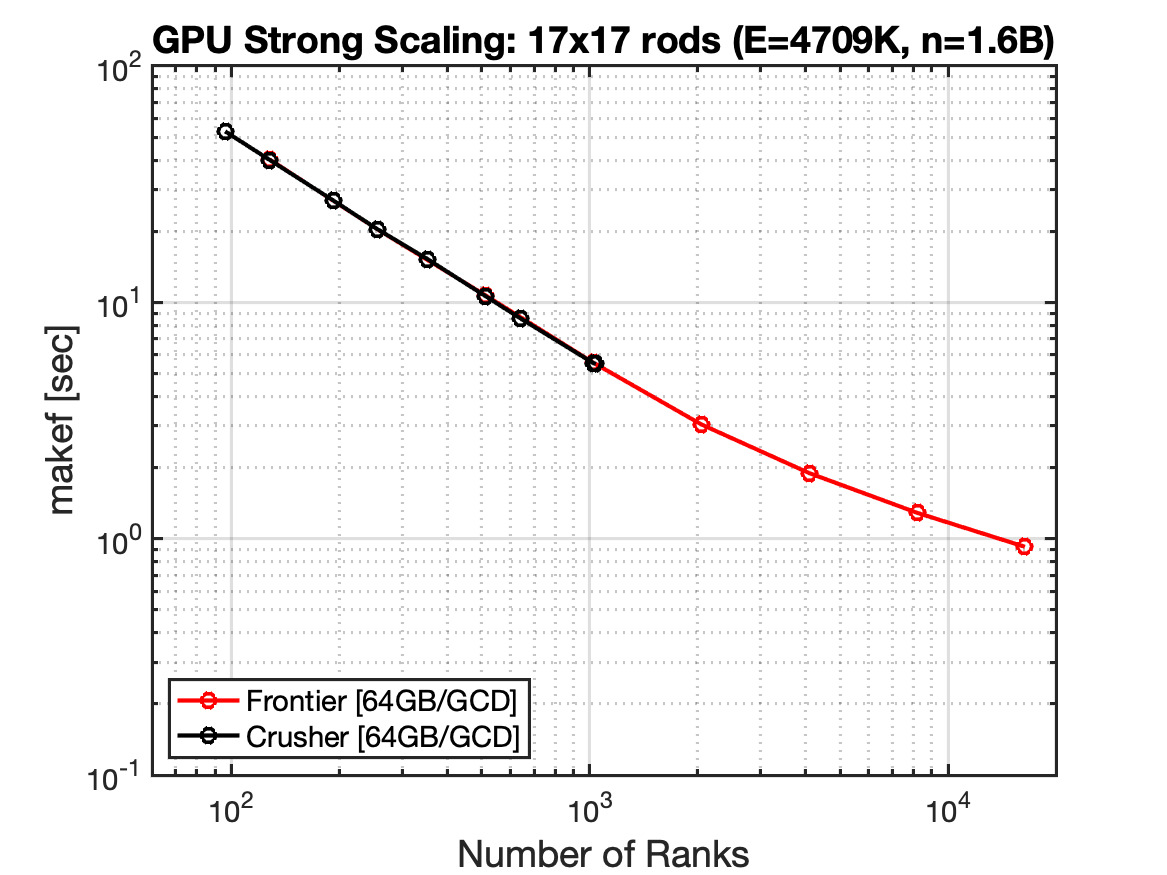 }
     \includegraphics[width=0.44\textwidth]{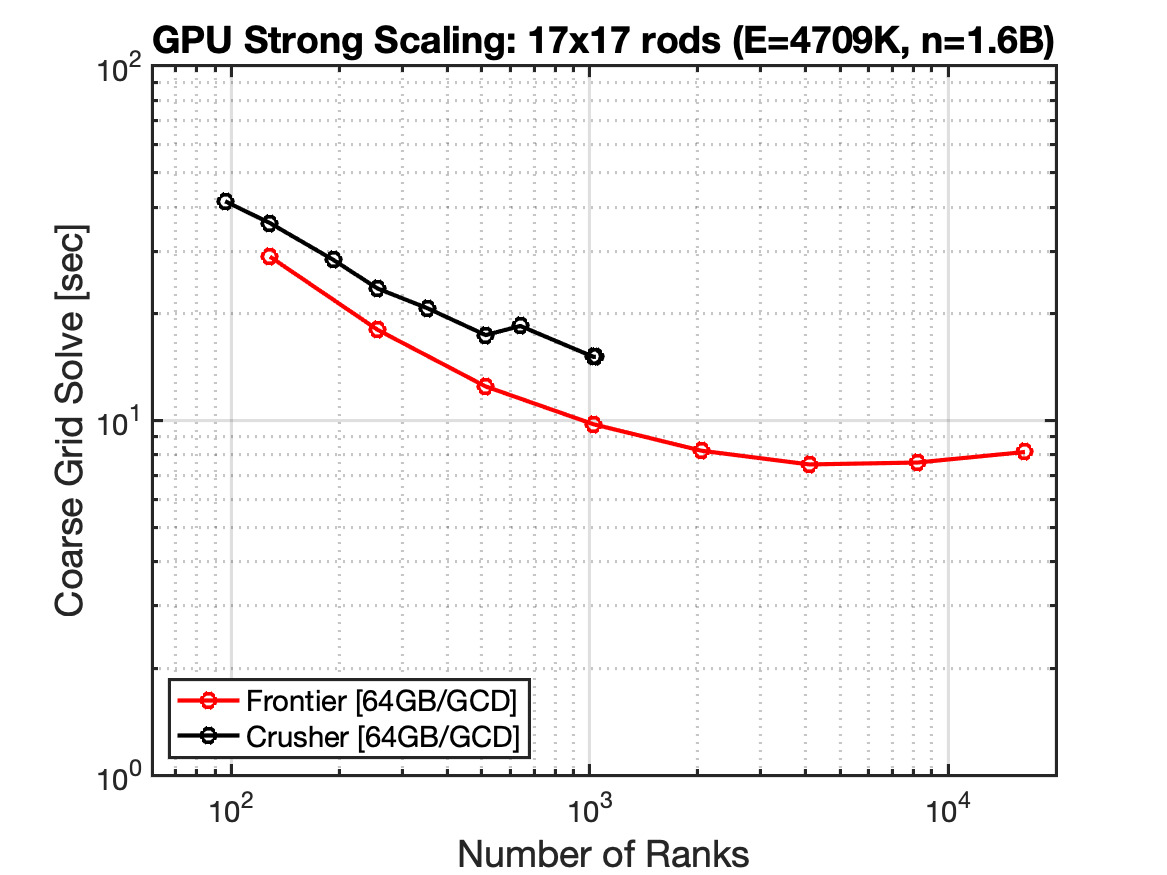 }
   \caption{\label{perf170-frontier}Strong-scaling on Frontier vs. Crusher for 17$\times$17 rod bundle with 170 layers.}
  \end{center}
\end{figure*}

Returning to the results for Frontier vs. Crusher,
Figures~\ref{perf10-frontier}-\ref{perf170-frontier} show detailed information
for each problem size, including timings, parallel efficiency, dofs, and
runtime statistics for the advection kernel {\tt makef} and for the {\tt
coarse-grid-solve} for the pressure Poisson problem.  The bottom left plots in
Figures~\ref{perf10-frontier}--\ref{perf170-frontier} show that Frontier and
Crusher deliver the same performance on the compute-intensive {\tt makef}
kernel, which evaluates the advection term for the Navier--Stokes equations. By
contrast, the bottom right plots in
Figures~\ref{perf10-frontier}-\ref{perf170-frontier} show Crusher is a bit
slower for the communication-intensive coarse-grid solve, which is part of the
multigrid preconditioner for the pressure-Poisson problem. The solve is
performed on the host CPUs using algebraic multigrid with {\tt Hypre}. 
(These lower plots show the times for the full 2,000 steps, not the
time per step.)

For smaller problem sizes with $n=$ 95M and 161M, the second row plots on the
left and right in Figures~\ref{perf10-frontier}--\ref{perf17-frontier} reveal
that the 80\% parallel efficiency is achieved with $n/P$= 2.2M on Frontier and
$n/P$= 2.5M on Crusher.  
The same plots in Figure~\ref{perf170-frontier}
demonstrate that Frontier realizes 80\% parallel efficiency at $n/P\approx 3$M
points per rank while Crusher at $n/P\approx 5$ million points per rank, which
would normally be viewed as the strong-scale limit for NekRS simulations on
these platforms.  This somewhat disappointing increase in $n_{0.8}$ implies
that NekRS is not weak-scaling well on Frontier/Crusher.  As noted in
\cite{fischer15}, weak-scaling performance is generally affected by the
coarse-grid solve, which is one of the few terms for which the cost grows with
$P$ (in this case, as $O(\log P)$), instead scaling as $n/P$.\footnote{Vector
reductions also exhibit $O(\log P)$ growth, but these can be mitigated by
hardware support
\cite{fischer15}.}  Indeed, inspection of the lower right graph in
Figures~\ref{perf17-frontier}--\ref{perf170-frontier} shows that, for the same
$n/P=$ 1.6M, the total coarse-grid solve time (for 2,000 steps) is 1 s for
$P=1000$ and only 0.65 s for $P=100$, which corresponds precisely to the ratio
$\log_2(1000)/\log_2(100)$.  We note that Crusher suffers even more in the
(host-driven) coarse-grid solve.  By contrast, Crusher and Frontier have
identical behavior on the communication-minimal and compute-intensive nonlinear
advection evaluation, as seen in the lower left graph in
Figure~\ref{perf170-frontier}.


The third-row left and right plots in Figures~\ref{perf10-frontier}--\ref{perf170-frontier}
show GDOF per step per $t_{\rm step}$ per rank vs. time per step and 
     GDOF per step per $t_{\rm step}$ per rank vs. $n/P$, where GDOF is defined by the 
{\em total} number of degree of freedom, which is $4n$
(three velocity vectors and one pressure field).
Frontier consistently shows faster performance than does Crusher, particularly
for the larger problem sizes (and, hence, higher processor counts), which
we can primarily attribute to the speed of the coarse-grid solve on Crusher.

Figure~\ref{perf170-frontier} in particular shows the performance as $n/P$ gets
as low as 0.1M, at which point the performance curve flattens and the device
does not speed up further.  There are two reasons for stalled performance:
communication overhead and insufficient work on the device (even in the 
absence of communication), as shown in an earlier study for NekCEM on
ORNL's Titan, which features NVIDIA K20X GPUs~\cite{min2015a} and in
the single-GPU studies on the NVIDIA V100 in the CEED benchmark 
paper ~\cite{ceed_bp_paper_2020}.

\section{NekRS performance on Summit, ThetaGPU, Perlmutter, Polaris, Crusher, and Frontier}
\label{nek-polaris}

\begin{figure*}
  \begin{center}
     \includegraphics[width=0.44\textwidth]{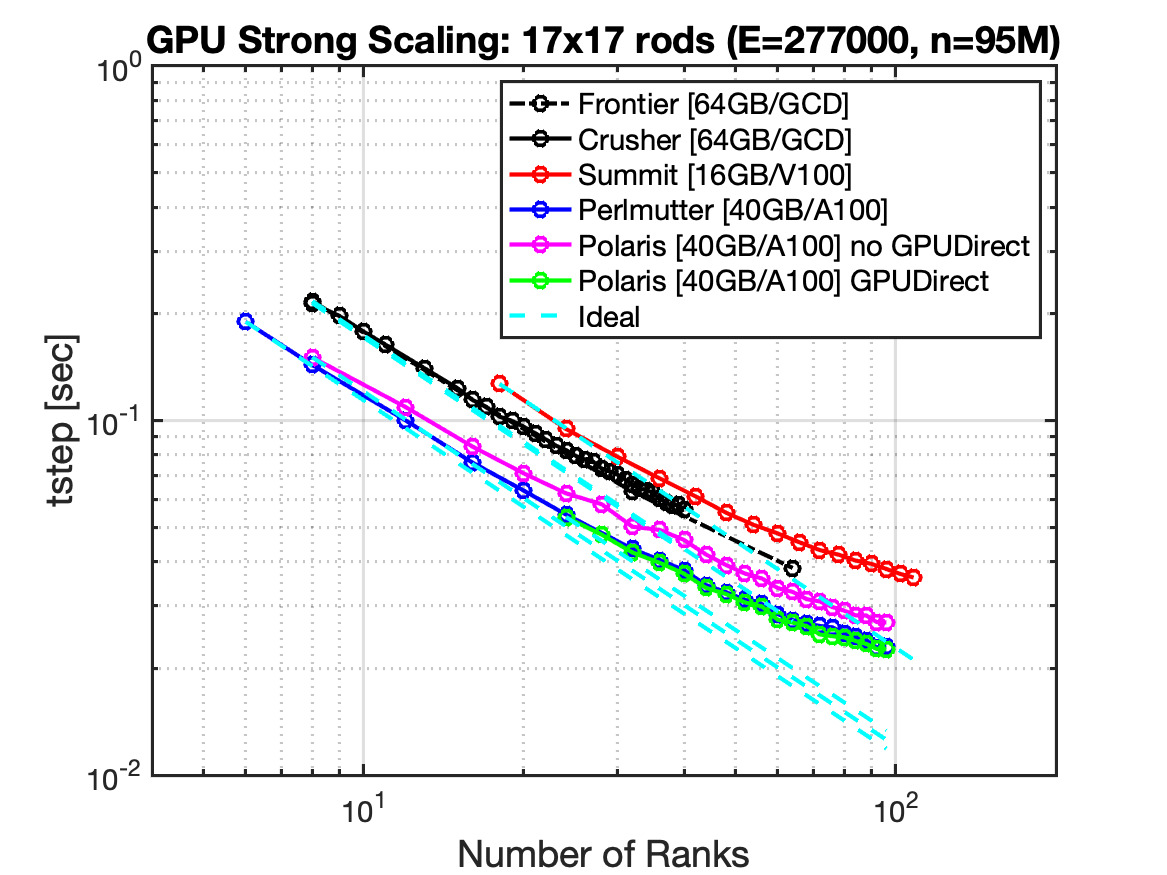 }
     \includegraphics[width=0.44\textwidth]{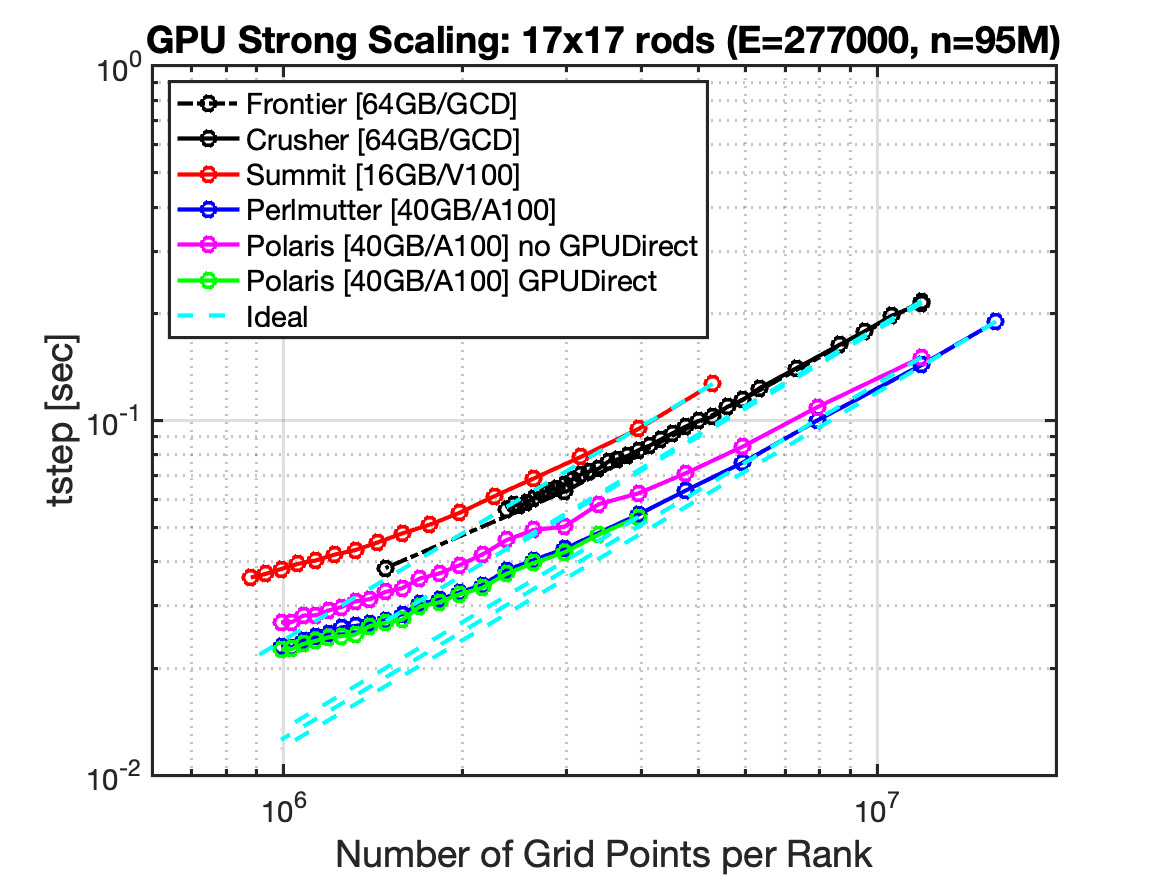 }
     \\
     \includegraphics[width=0.44\textwidth]{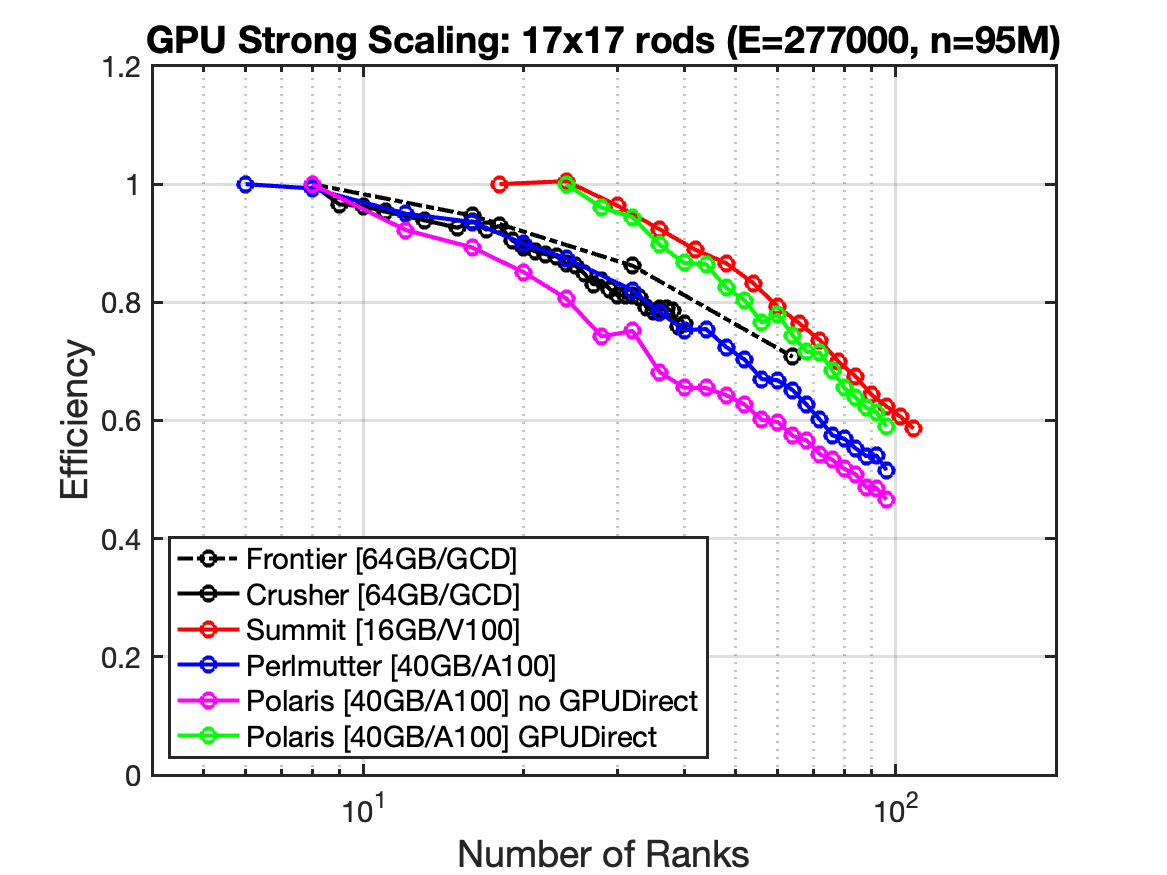 }
     \includegraphics[width=0.44\textwidth]{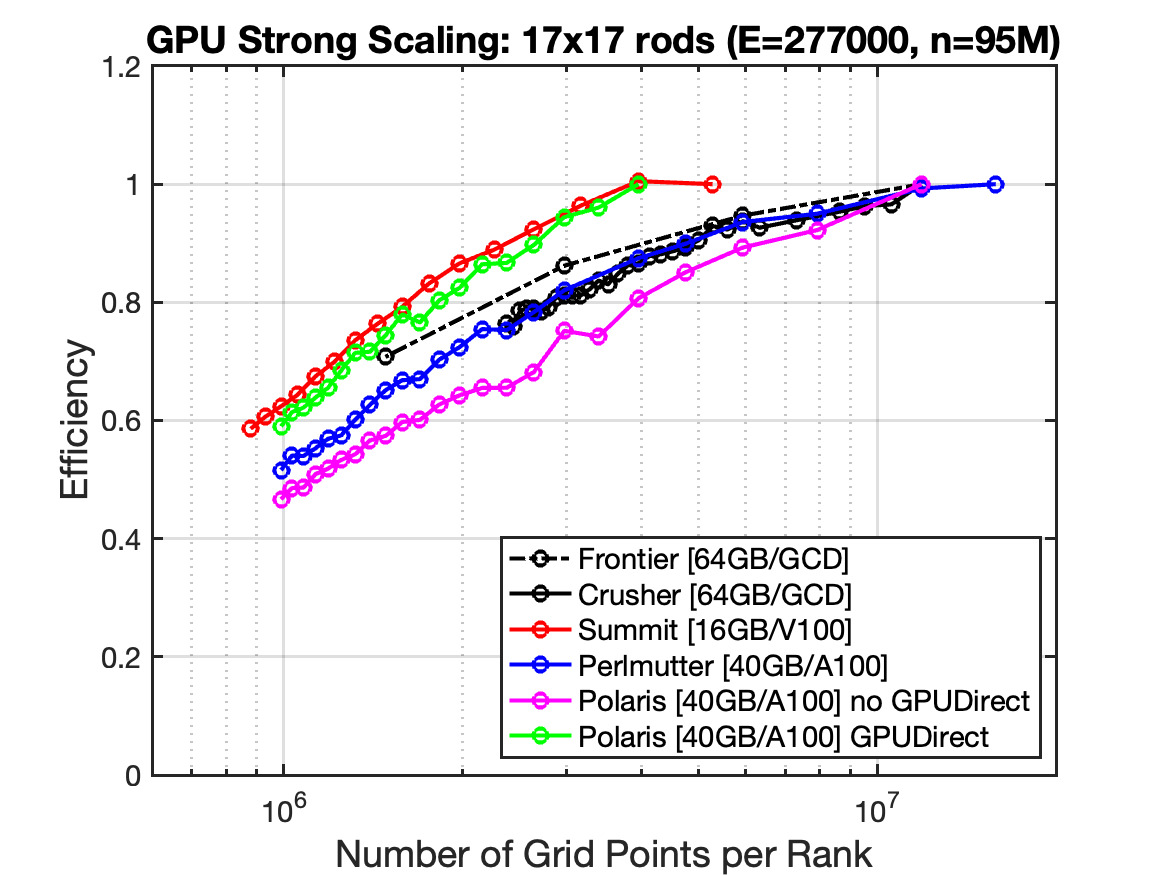 }
     \\
     \includegraphics[width=0.44\textwidth]{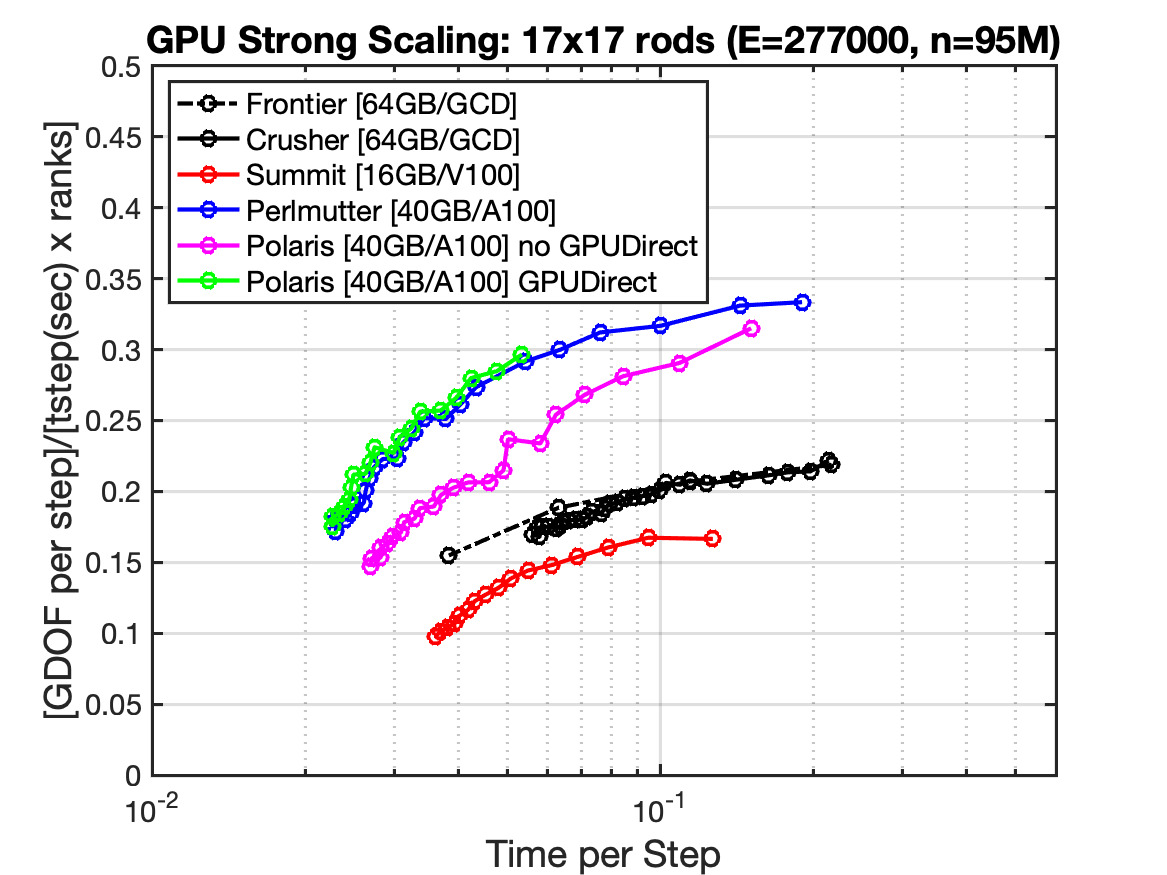 }
     \includegraphics[width=0.44\textwidth]{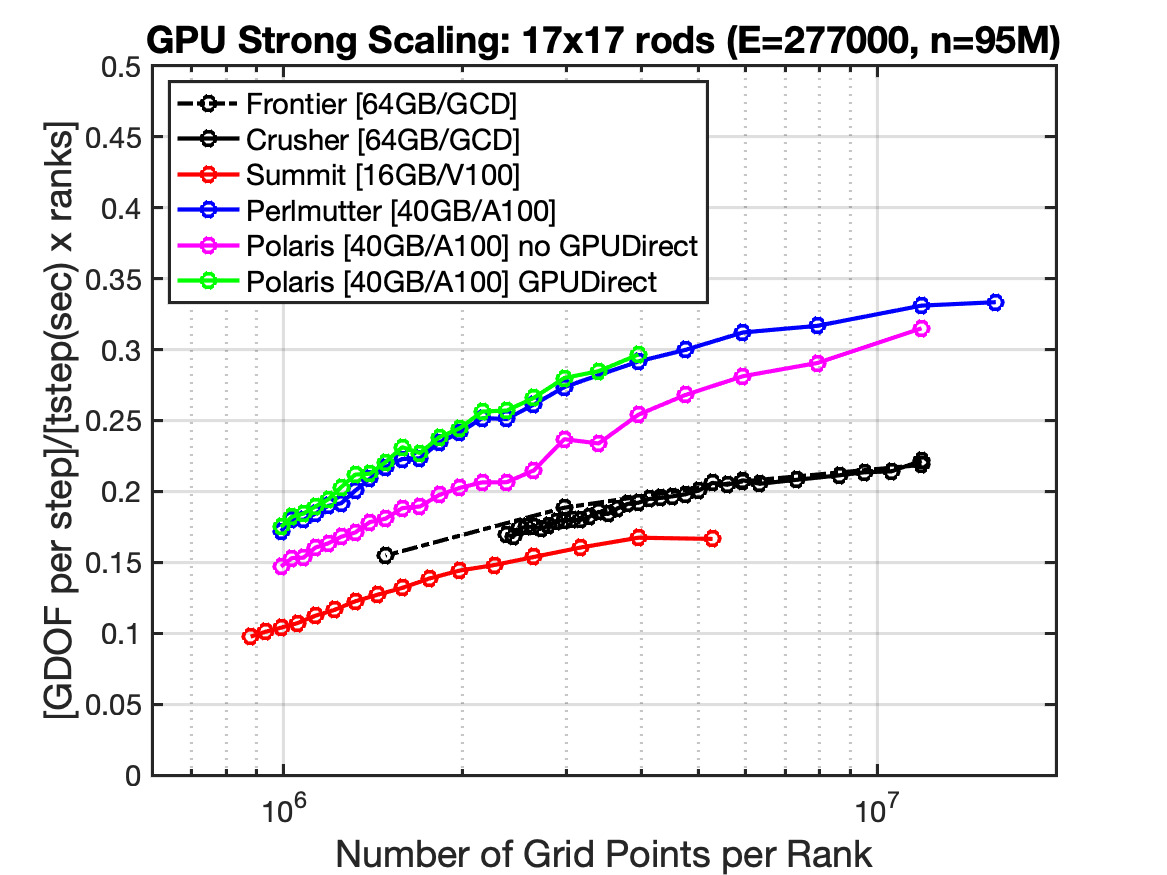 }
     \\
     \includegraphics[width=0.44\textwidth]{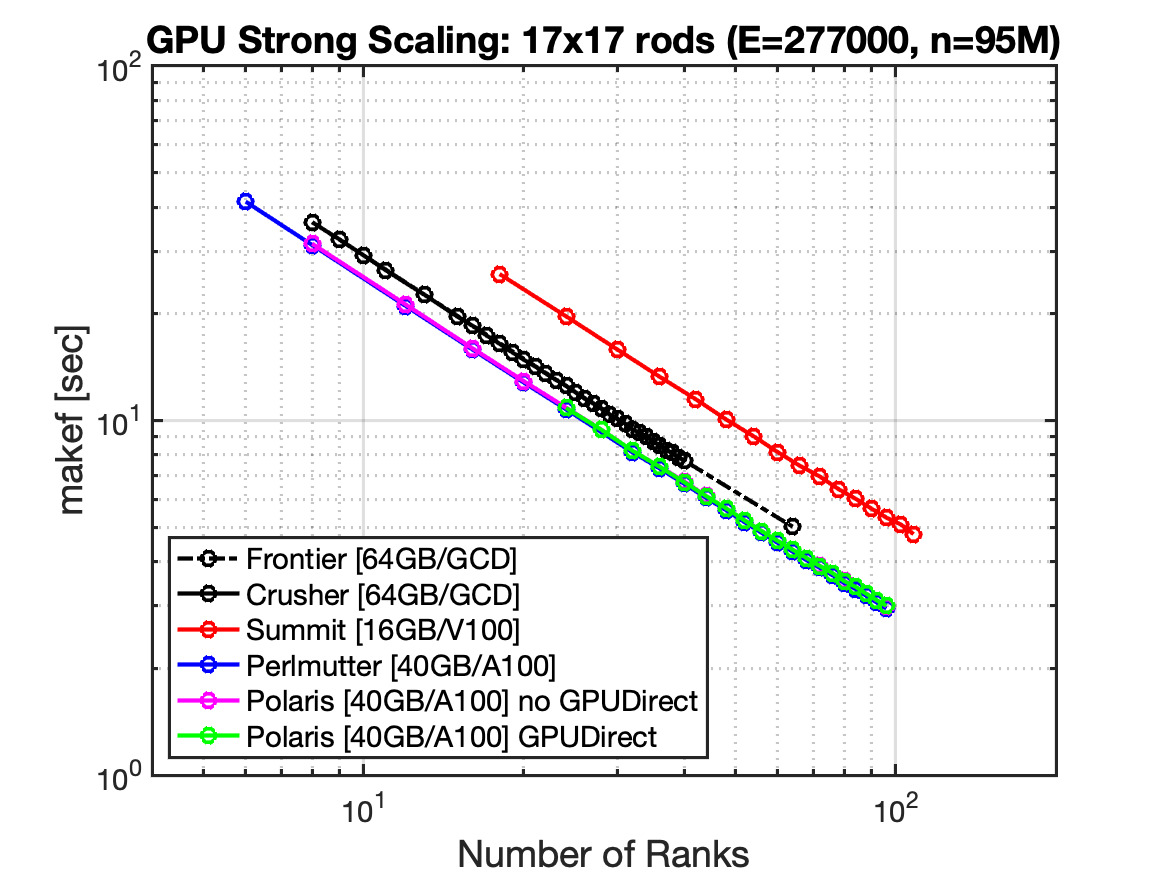 }
     \includegraphics[width=0.44\textwidth]{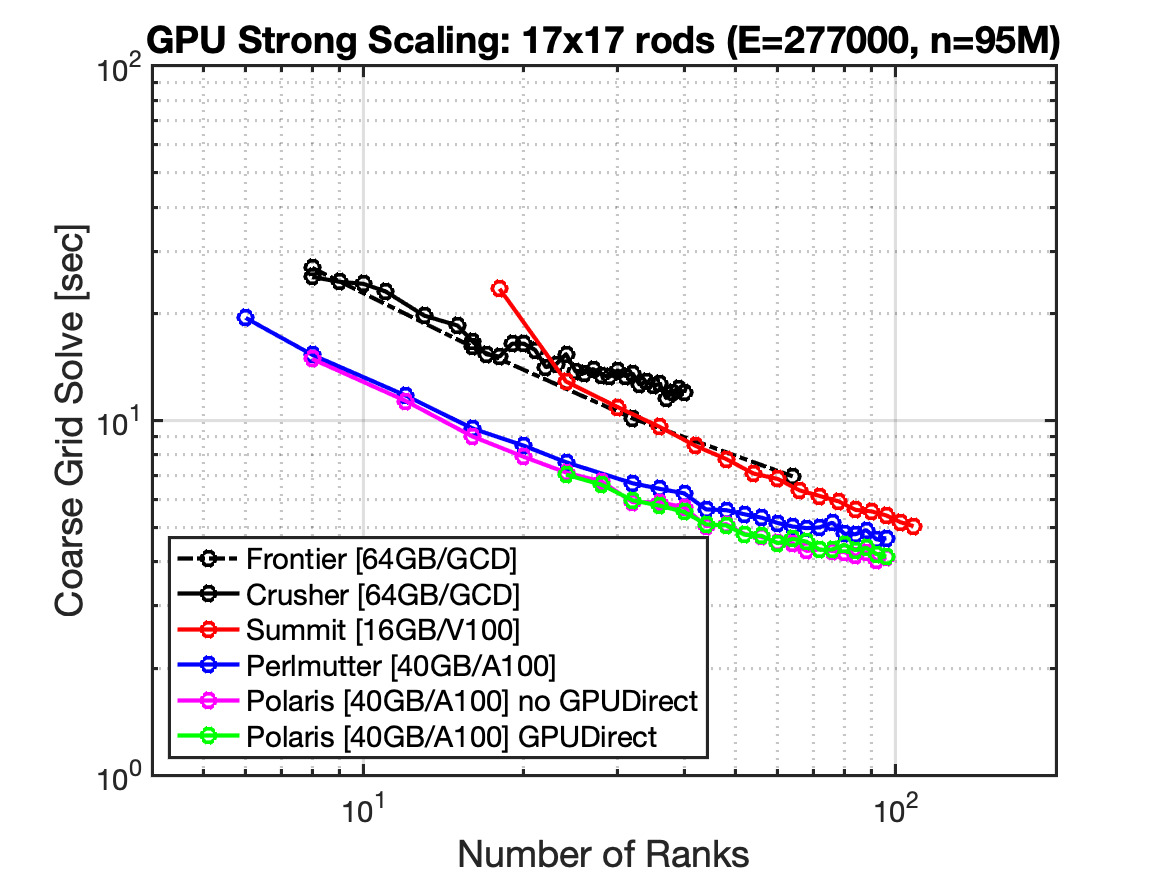 }
   \caption{\label{perf10-all-frontier}Strong-scaling on various GPU architectures  for 17$\times$17 rod bundle with 10 layers.}
  \end{center}
\end{figure*}

\begin{figure*}
  \begin{center}
     \includegraphics[width=0.44\textwidth]{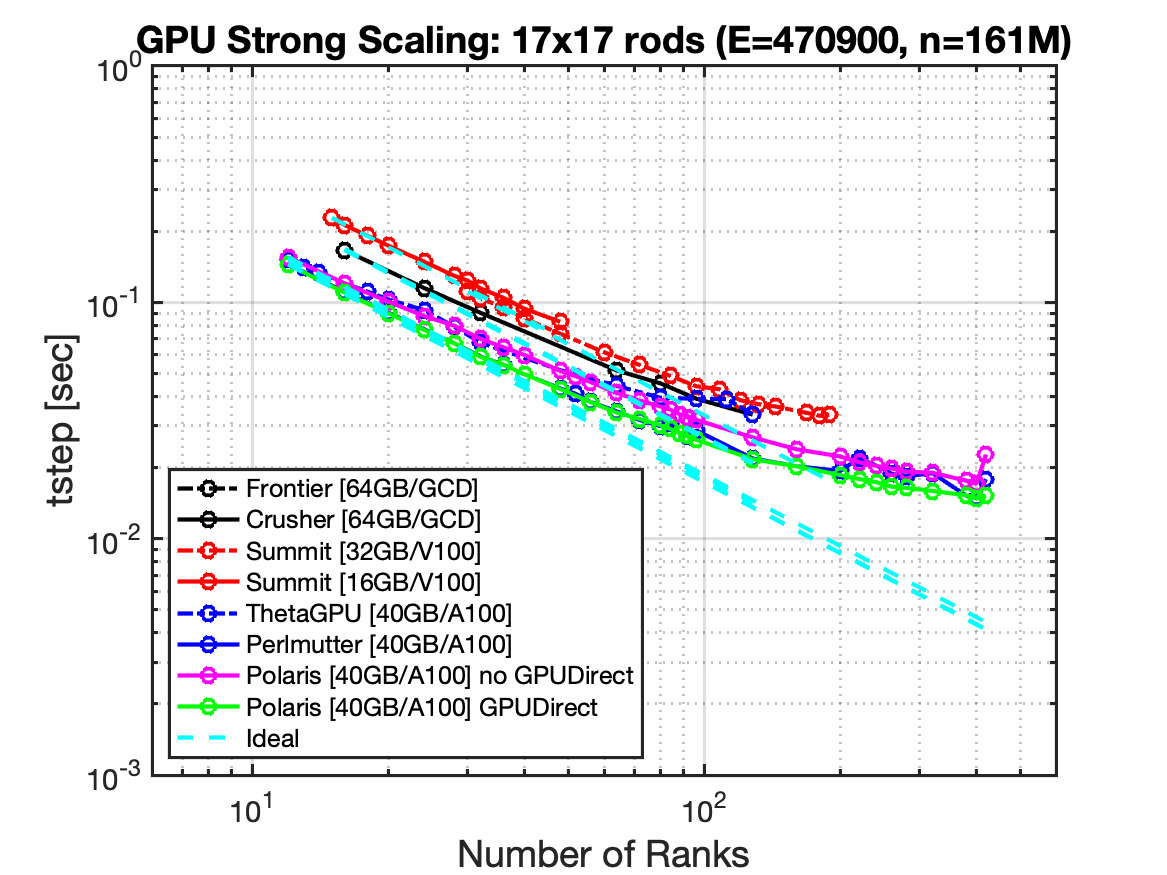 }
     \includegraphics[width=0.44\textwidth]{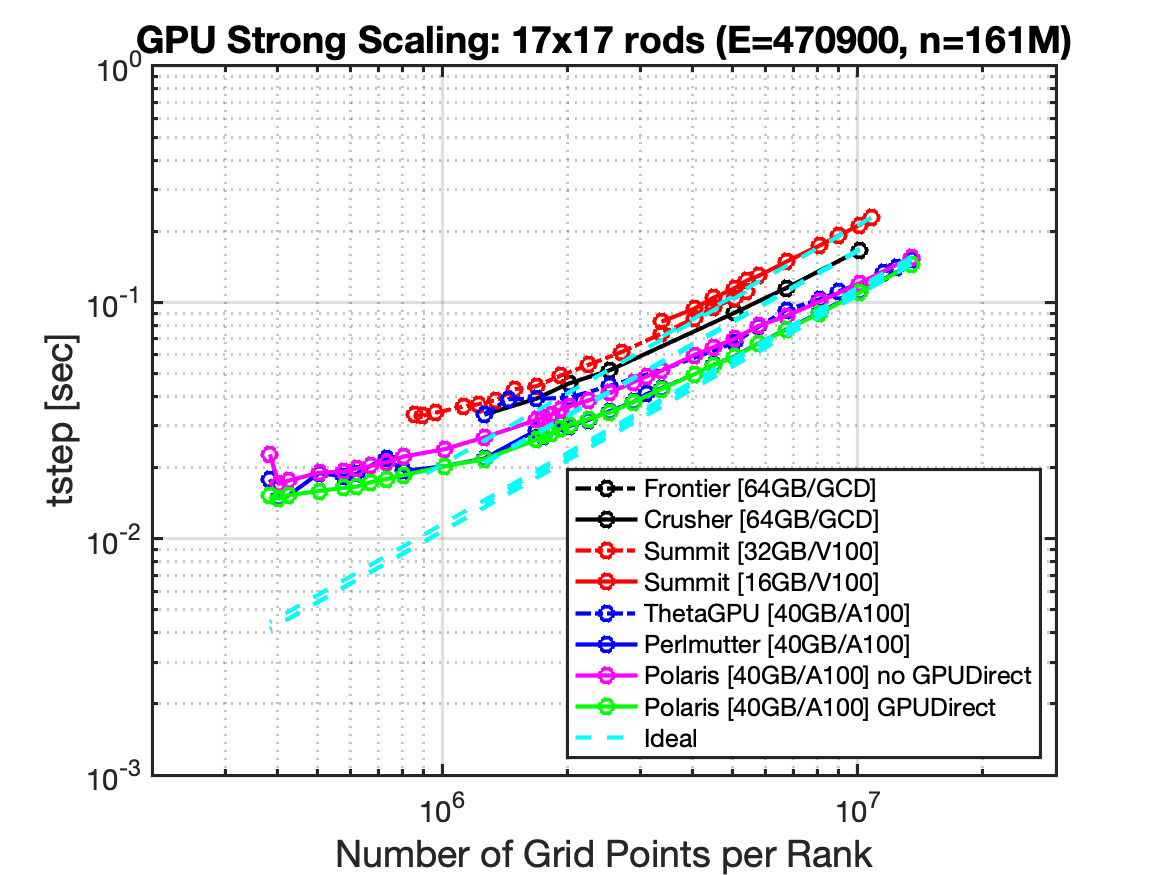 }
     \\
     \includegraphics[width=0.44\textwidth]{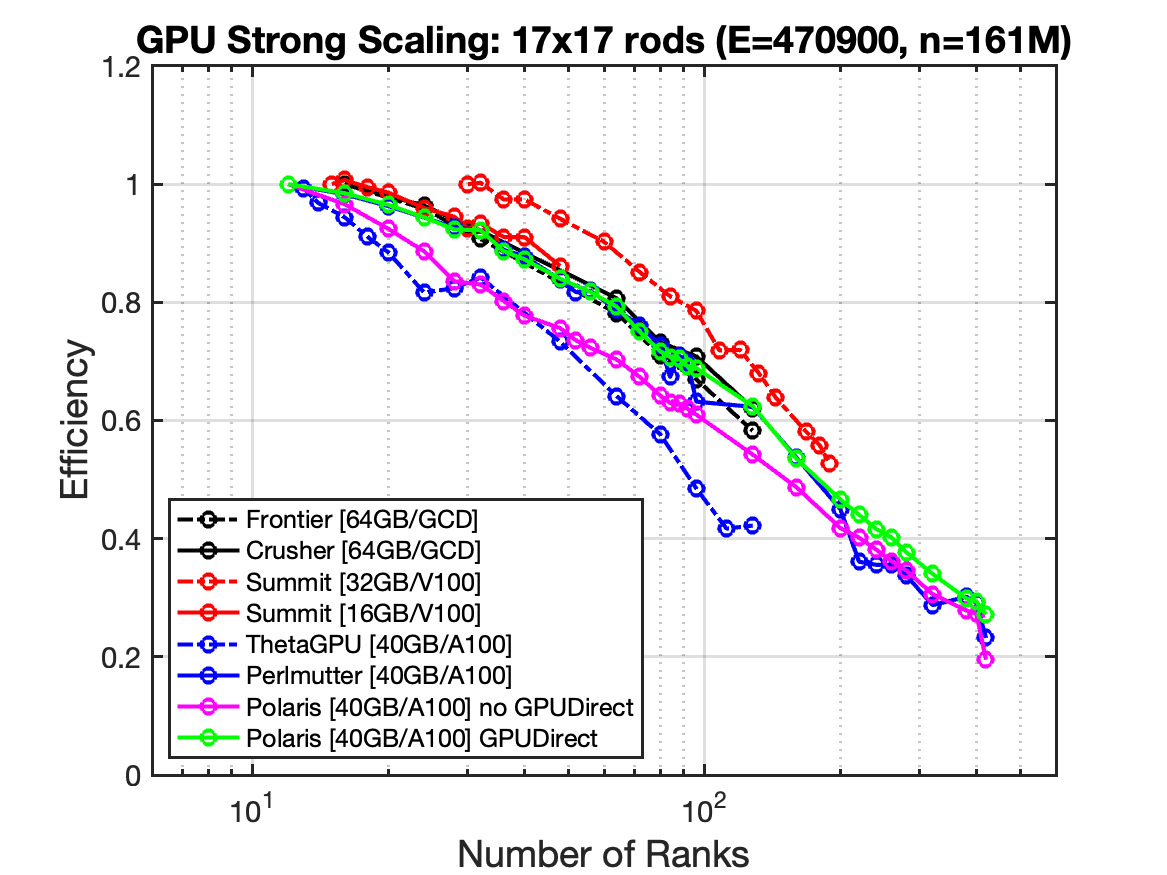 }
     \includegraphics[width=0.44\textwidth]{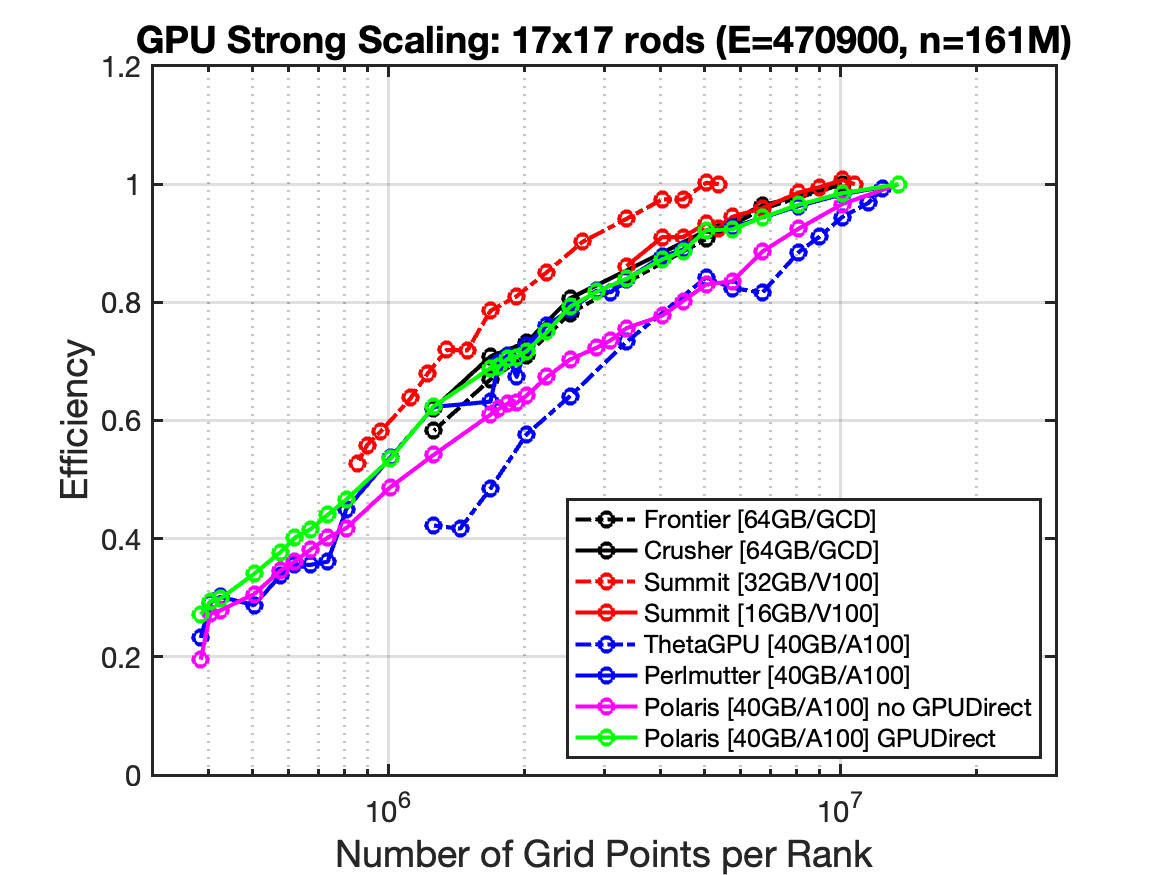 }
     \\
     \includegraphics[width=0.44\textwidth]{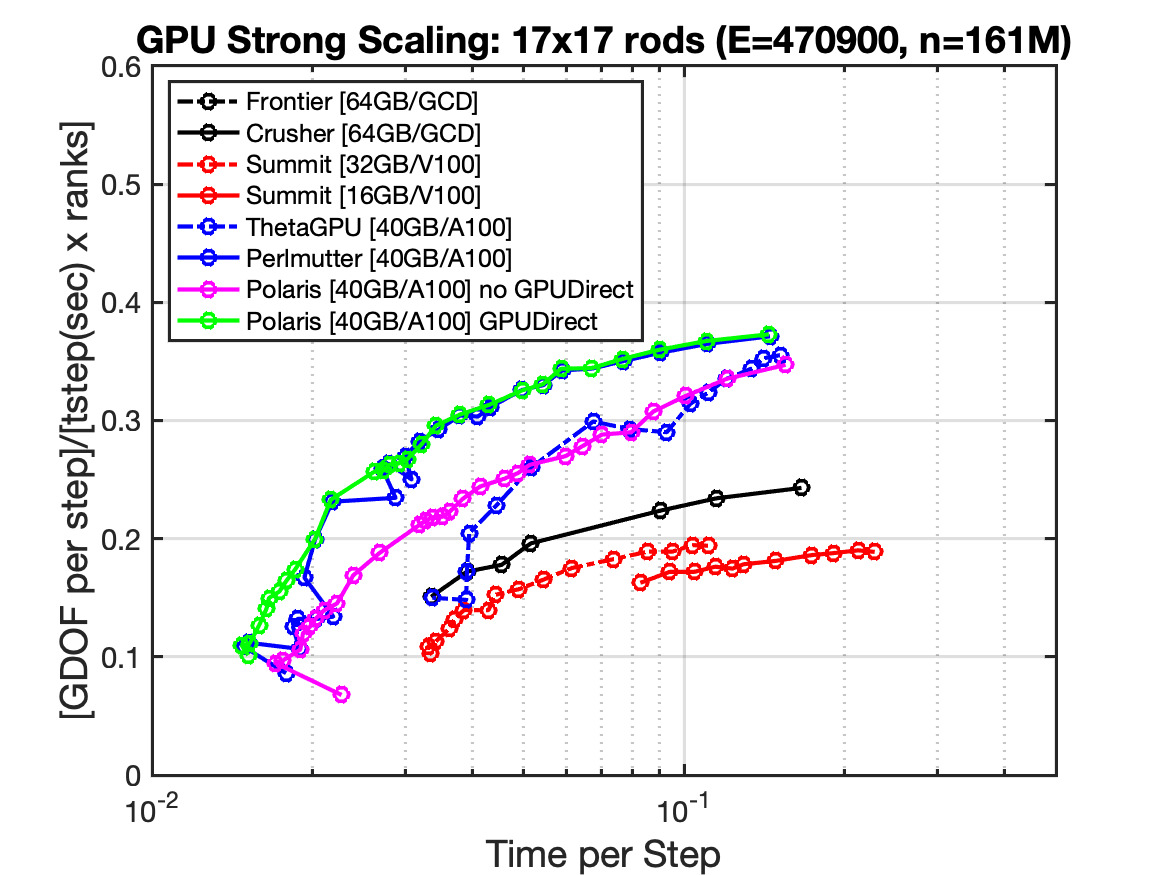 }
     \includegraphics[width=0.44\textwidth]{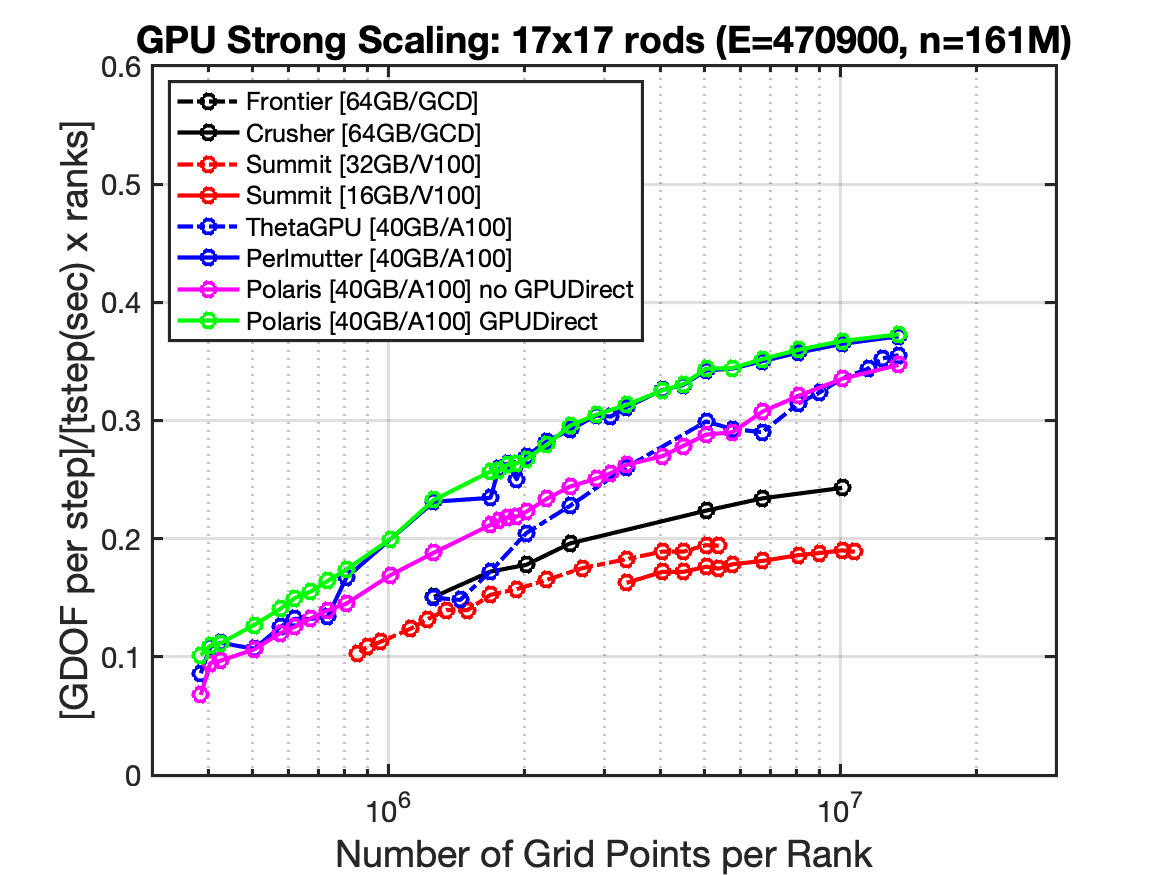 }
     \\
     \includegraphics[width=0.44\textwidth]{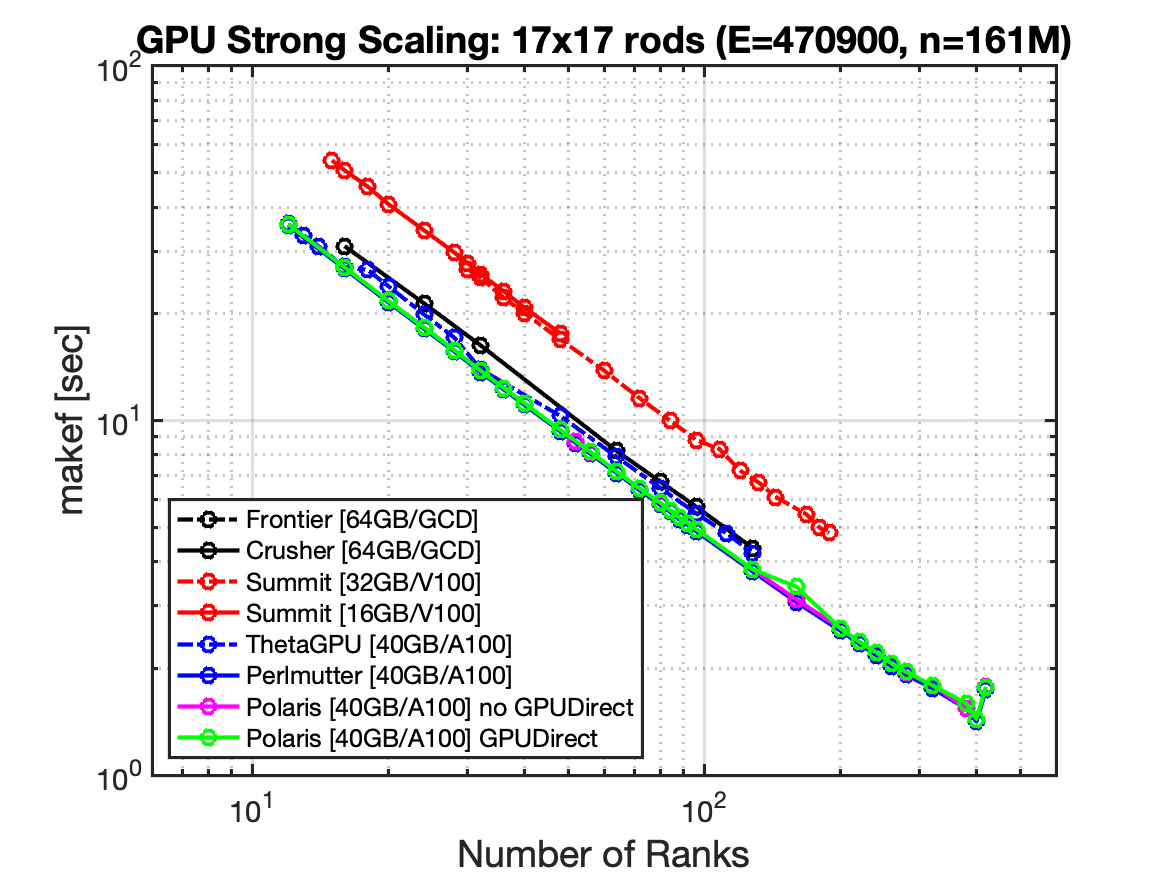 }
     \includegraphics[width=0.44\textwidth]{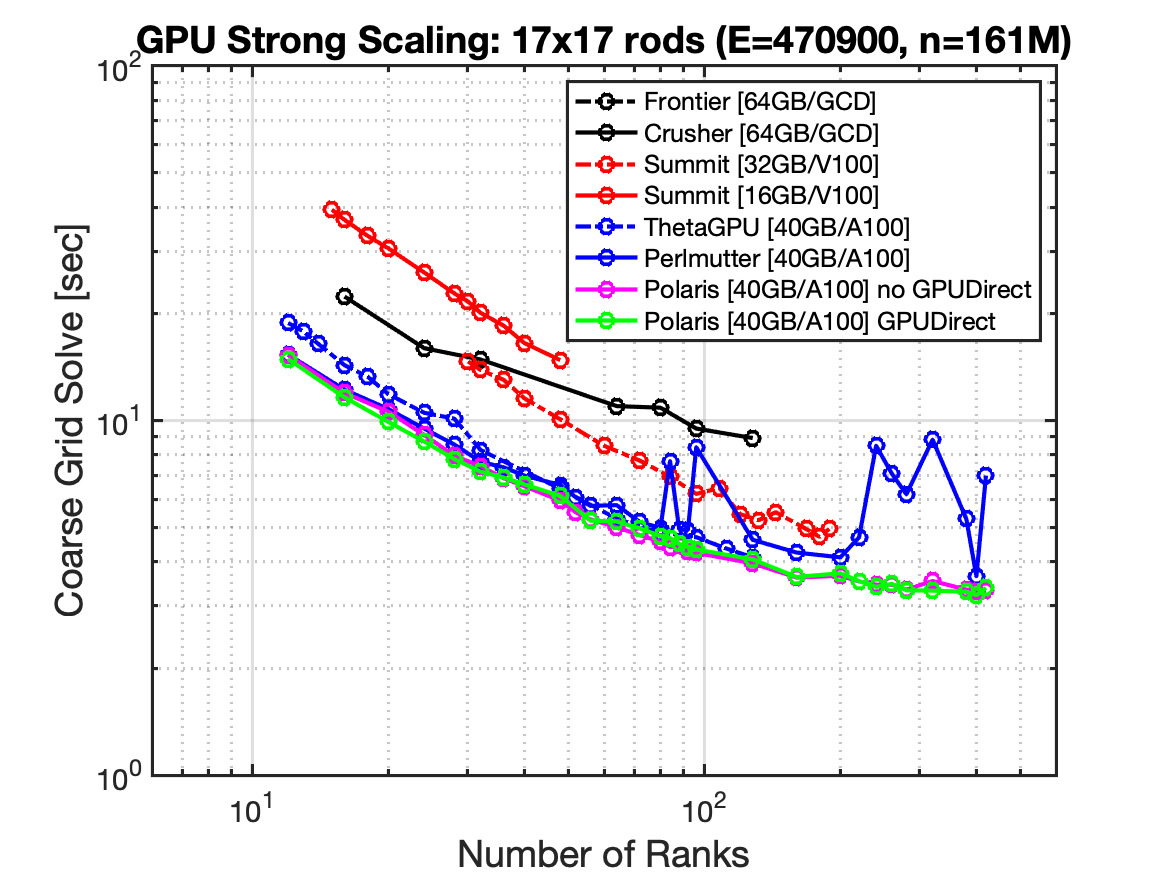 }
   \caption{\label{perf17-all-frontier}Strong-scaling on various GPU architectures  for 17$\times$17 rod bundle with 17 layers.}
  \end{center}
\end{figure*}

\begin{figure*}
  \begin{center}
     \includegraphics[width=0.44\textwidth]{./strong-scale-rod1717-170-p-rank-gmpi-all-frontier.jpg }
     \includegraphics[width=0.44\textwidth]{./strong-scale-rod1717-170-n-p-rank-gmpi-all-frontier.jpg }
     \\
     \includegraphics[width=0.44\textwidth]{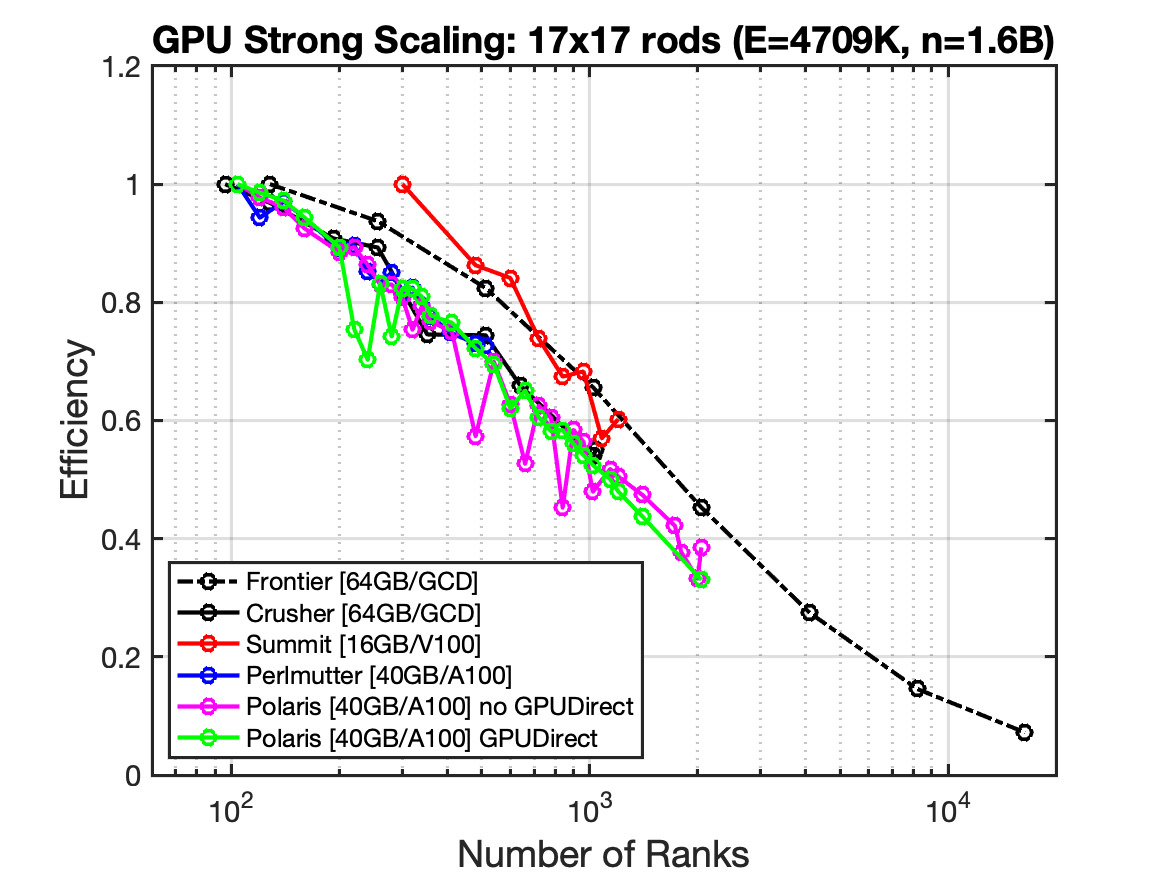 }
     \includegraphics[width=0.44\textwidth]{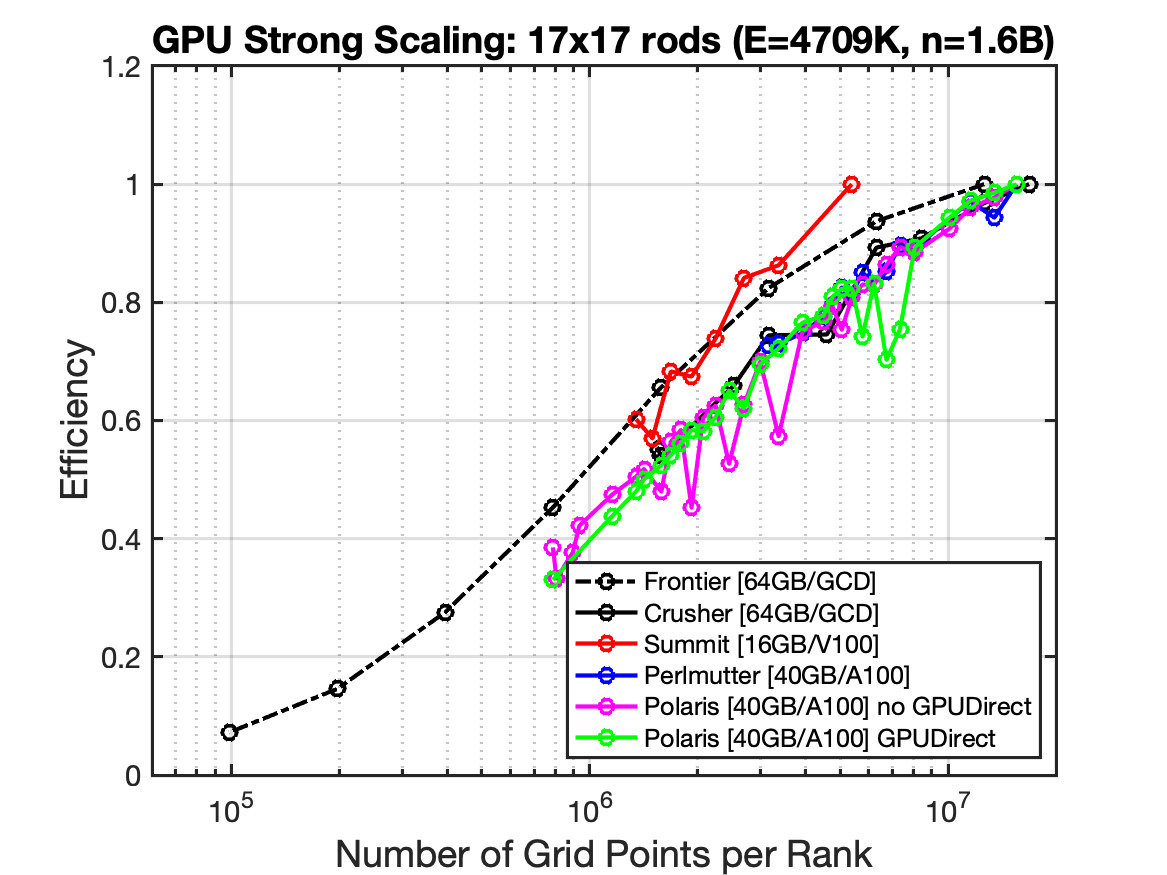 }
     \\
     \includegraphics[width=0.44\textwidth]{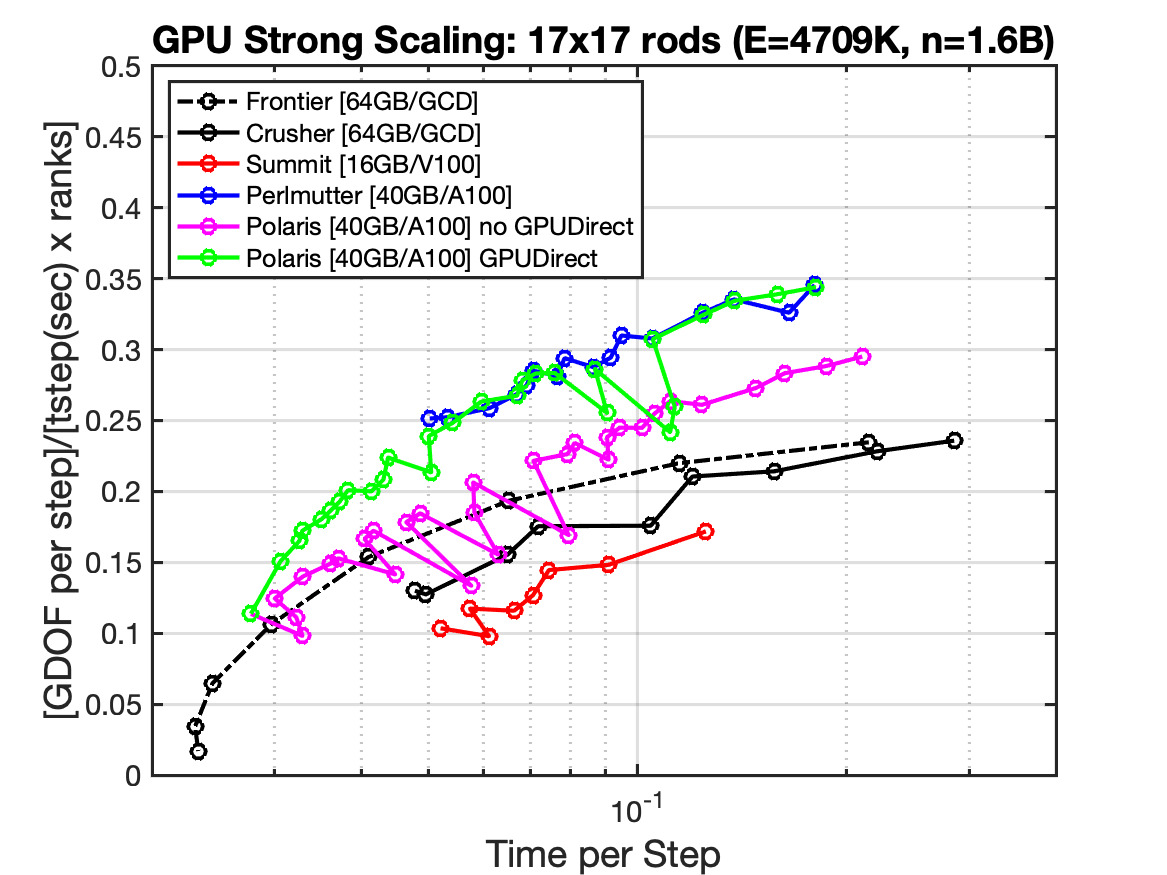 }
     \includegraphics[width=0.44\textwidth]{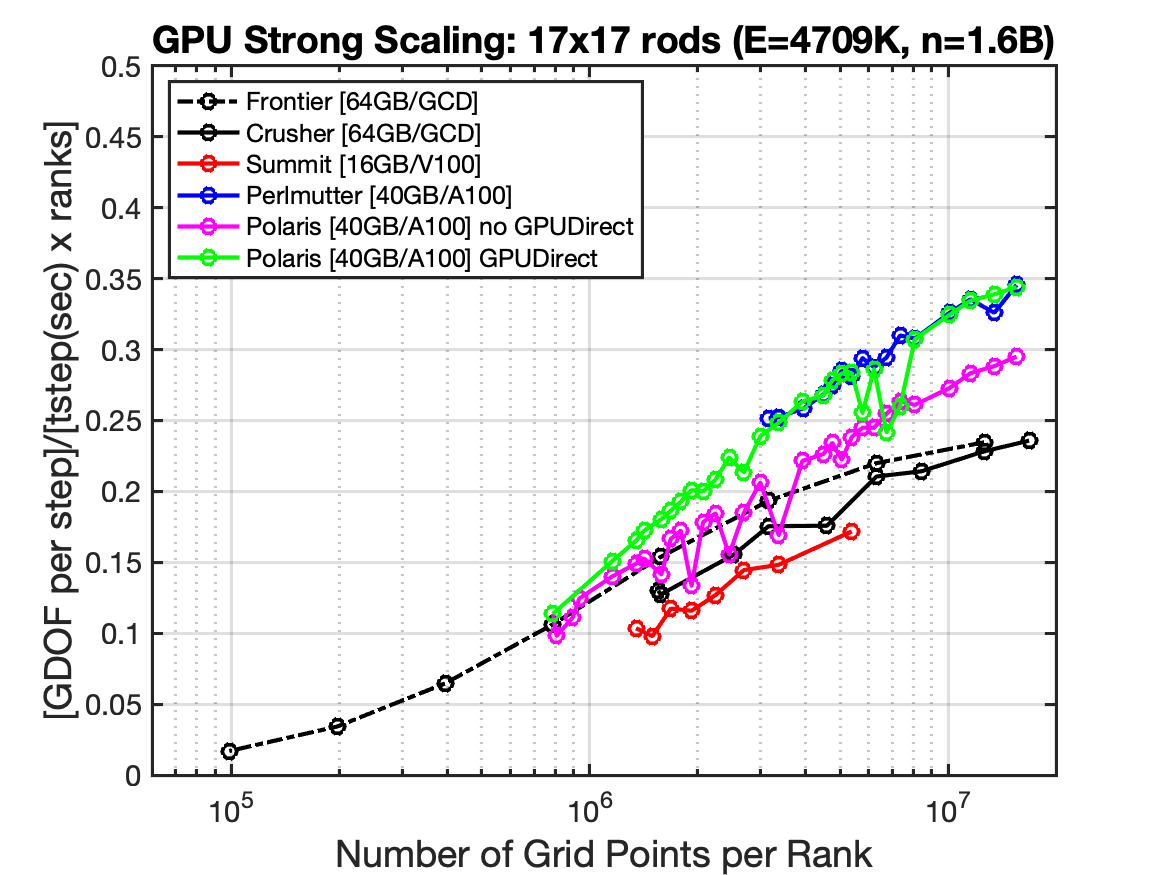 }
     \\
     \includegraphics[width=0.44\textwidth]{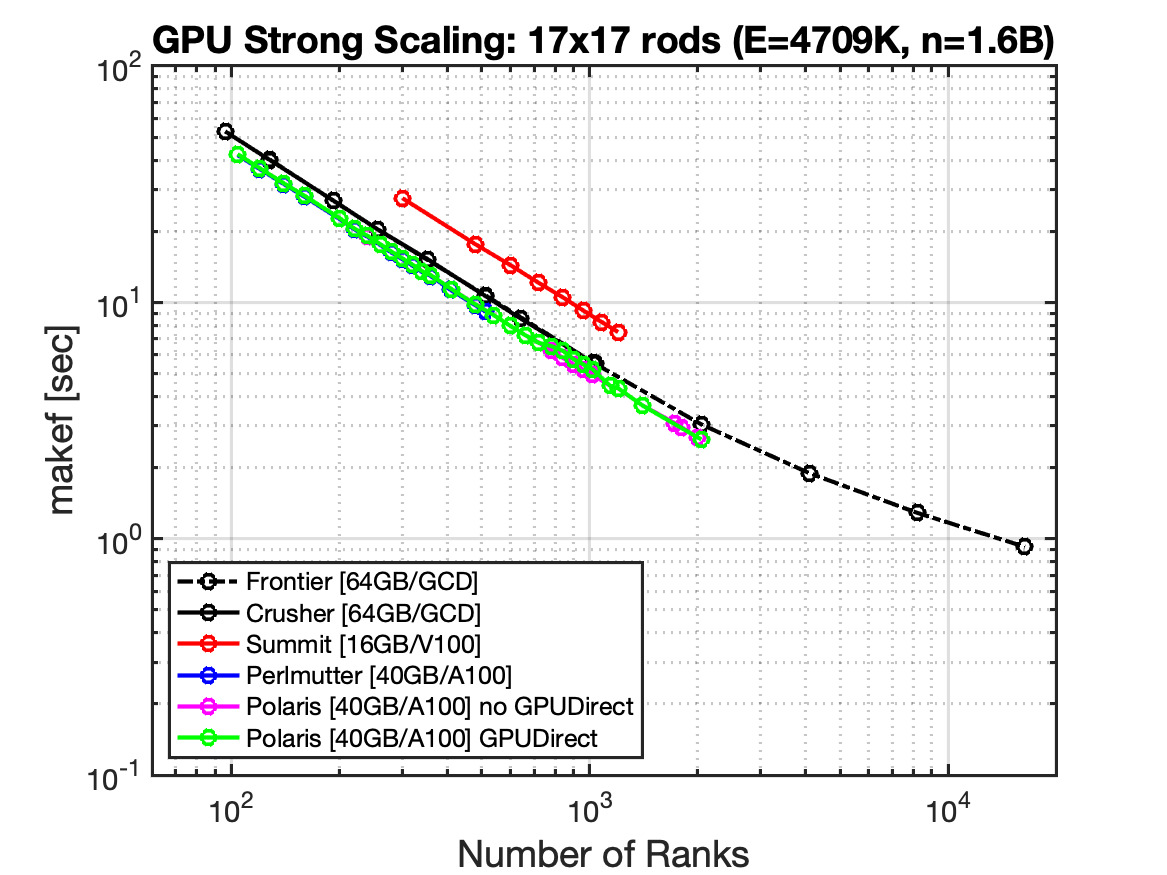 }
     \includegraphics[width=0.44\textwidth]{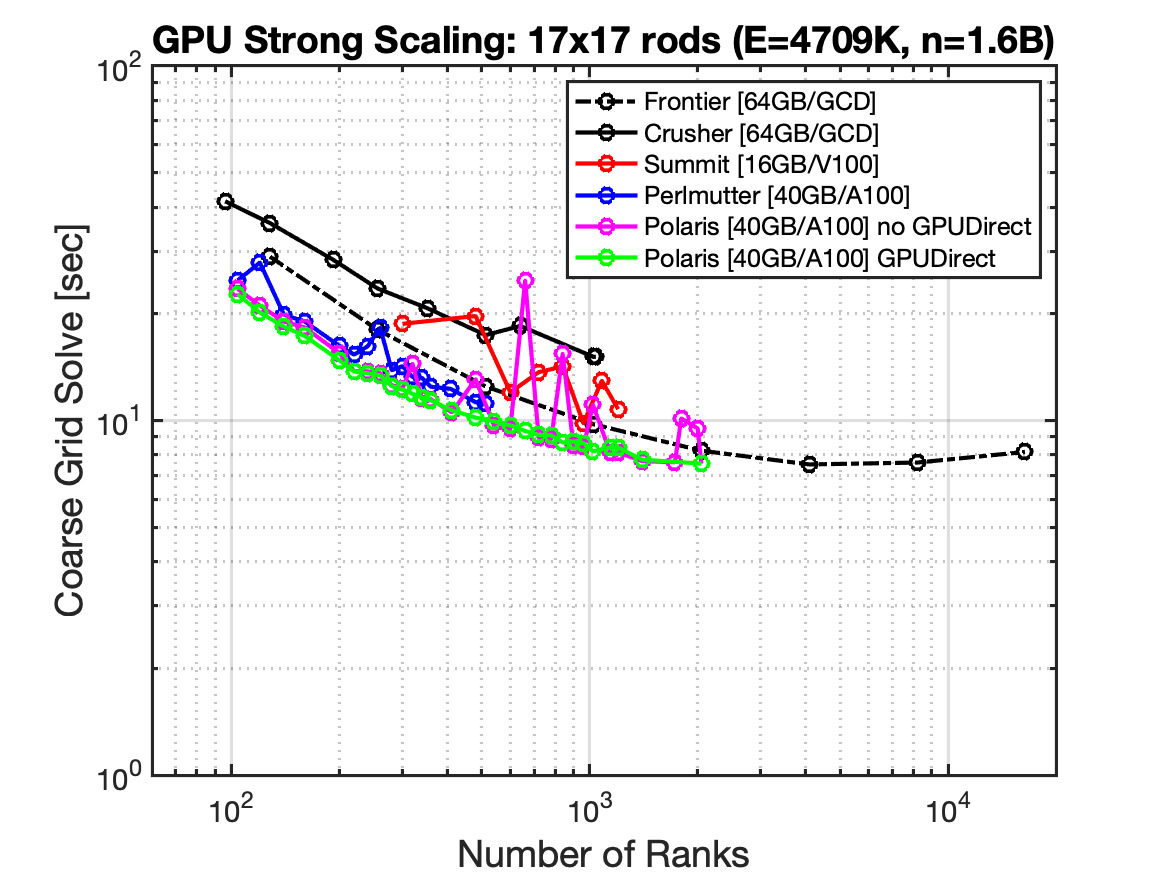 }
   \caption{\label{perf170-all-frontier}Strong-scaling on various GPU architectures for 17$\times$17 rod bundle with 170 layers.}
  \end{center}
\end{figure*}

In this section we continue with scaling studies on the 17$\times$17
rod bundle simulations for the
NVIDIA-based GPU platforms, Summit (V100), ThetaGPU (A100),  Perlmutter (A100), and Polaris
(A100), compared with the AMD MI-250X platforms, Frontier and Crusher.  
Recall that  in Figure~\ref{perf170-all} we showed the 170-layer
case across the platforms of Frontier, Crusher, Summit, Perlmutter, and Polaris.
Here, we discuss the performance in more detail in
Figures~\ref{perf10-all-frontier}--\ref{perf170-all-frontier}.  
In Figure~\ref{perf17-all-frontier} we include performance on ThetaGPU.
We run one MPI rank per V100 or A100 on the NVIDIA-based nodes and one MPI rank
per GCD on the AMD MI250X nodes.

Figures~\ref{perf10-all-frontier}--\ref{perf170-all-frontier} show the same
metrics as for the Crusher-Frontier plots of
Figures~\ref{perf10-frontier}--\ref{perf170-frontier}.
We
 point out that these strong-scaling plots start from a high
level of performance.  NekRS currently leverages extensive tuning of several
key FP64 and FP32 kernels in libParanumal, including the standard spectral
element Laplacian matrix-vector product, local tensor-product solves using fast
diagonalization, and dealiased evaluation of the advection operator on a finer
set of quadrature points.  These kernels 
are sustaining up to 3 TFLOPS FP64 and 5--8 TFLOPS FP32,
per GPU or GCD.  At the strong-scale limit, with MPI overhead, NekRS is
sustaining $\approx$ 1 TFLOPS per rank (i.e., per A100 or GCD) for the
full Navier--Stokes solution~\cite{gb23}.

An important figure of merit is $n_{0.8}$, which is the value of $n/P$ at which
the simulation realizes 80\% parallel efficiency. From the second row, right,
we see that $n_{0.8}=2.5$M for Perlmutter/Crusher and 2M for Polaris (GPUDirect
in green dashed lines). For Polaris without GPUDirect (magenta solid line) we
find $n_{0.8}=4.5$M. The plot on row 3, left, indicates that a remarkably small
$t_{step}$ value of 0.015 seconds per step is realizable on Polaris, albeit at
25\% efficiency.

The plots on the last row, left, of
Figures~\ref{perf10-all-frontier}--\ref{perf170-all-frontier} show that the time
in the advection update strong-scales quite well, as would be expected. The
curves for the single GCD and A100 collapse to nearly the same performance
while the older V100 technology of Summit is about 1.5$\times$ slower.
In the absence of communication, this kernel is sustaining 3--4 TFLOPS FP64 
on these newer architectures, although the graphs here do include the
communication overhead.  By contrast, the last row, right, shows the
performance for the communication-intensive coarse-grid solve, which is
performed using {\tt Hypre} on the host CPUs. Here, both Crusher and Summit show
relatively poor performance at small values of $n/P$ or large values of $P$.
Also, in Figure~\ref{perf10-all-frontier}, lower right, we see that Polaris without
GPUDirect exhibits some level of system noise.

We also observe in Figure~\ref{perf17-all-frontier} that ThetaGPU (dashed blue
lines) is about 1.3--1.5$\times$ slower than Perlmutter and Polaris (GPUDirect).
From the lower left curve we see that the ThetaGPU performance is lower
than the other A100 platforms even for this work-intensive operation.
Moreover, in terms of parallel efficiency, it falls off faster than Polaris
whether Polaris is or is not using GPUDirect communication.


\section{Discussion} \label{sec:disc}

\noindent
In this section we discuss a variety of ``anomalous'' behaviors encountered in
these studies.  By anomalous we mean that they are adverse behaviors that
either appear or disappear with software and hardware updates.  One could argue
that these are passing phenomena not worthy of reporting.  However, users are
actively using these machines, and it is important that everyone understand
potential pitfalls in system performance that might directly impact their own
timing studies or production runtimes.

The behaviors described here include 
   the performance of the large-memory (32GB vs 16GB) nodes on Summit, 
   the use of a nonmultiple of 8 ranks on Crusher, 
   the influence of GPU-direct communication on Polaris,
and 
   the upgrade from  Slingshot 10 to  Slingshot 11 on Perlmutter.

\subsection{Performance on Summit V100 16 GB vs. 32 GB}

Most of the ,4608 nodes on Summit have 16 GB of device memory, which limits how
small one can take $P_0$ in the efficiency definition (\ref{eq:eff1}).  
  few nodes, however, have 32 GB, which allow one to fit more points
onto each V100.  Unfortunately, as seen in
Figures~\ref{perf10-all-frontier}--\ref{perf170-all-frontier}, the Summit 32 GB
curves perform about 10\% slower than their 16 GB counterparts.
The last row of graphs in Figure~\ref{perf17-all-frontier} is particularly
revealing: One can see from the {\tt makef} plot that the V100s perform at the
same rate for both the 16 GB and 32 GB nodes but that the host-based coarse-grid
solve costs differ by almost 1.5$\times$, which indicates either an excessive
device-host communication overhead or some type of interhost communication
slowdown.

\subsection{Performance on Crusher with Rank Dependency}

\begin{figure*}
  \begin{center}
     \includegraphics[width=0.44\textwidth]{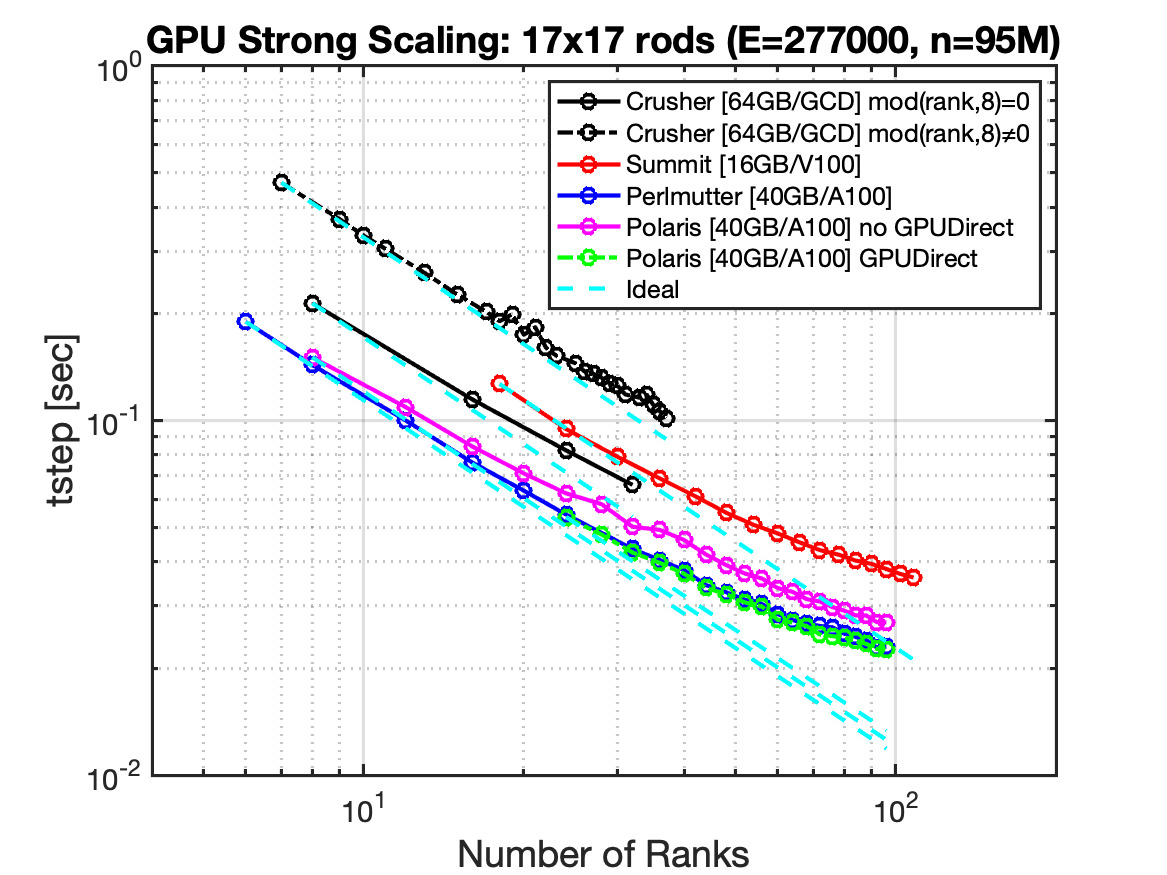 }
     \includegraphics[width=0.44\textwidth]{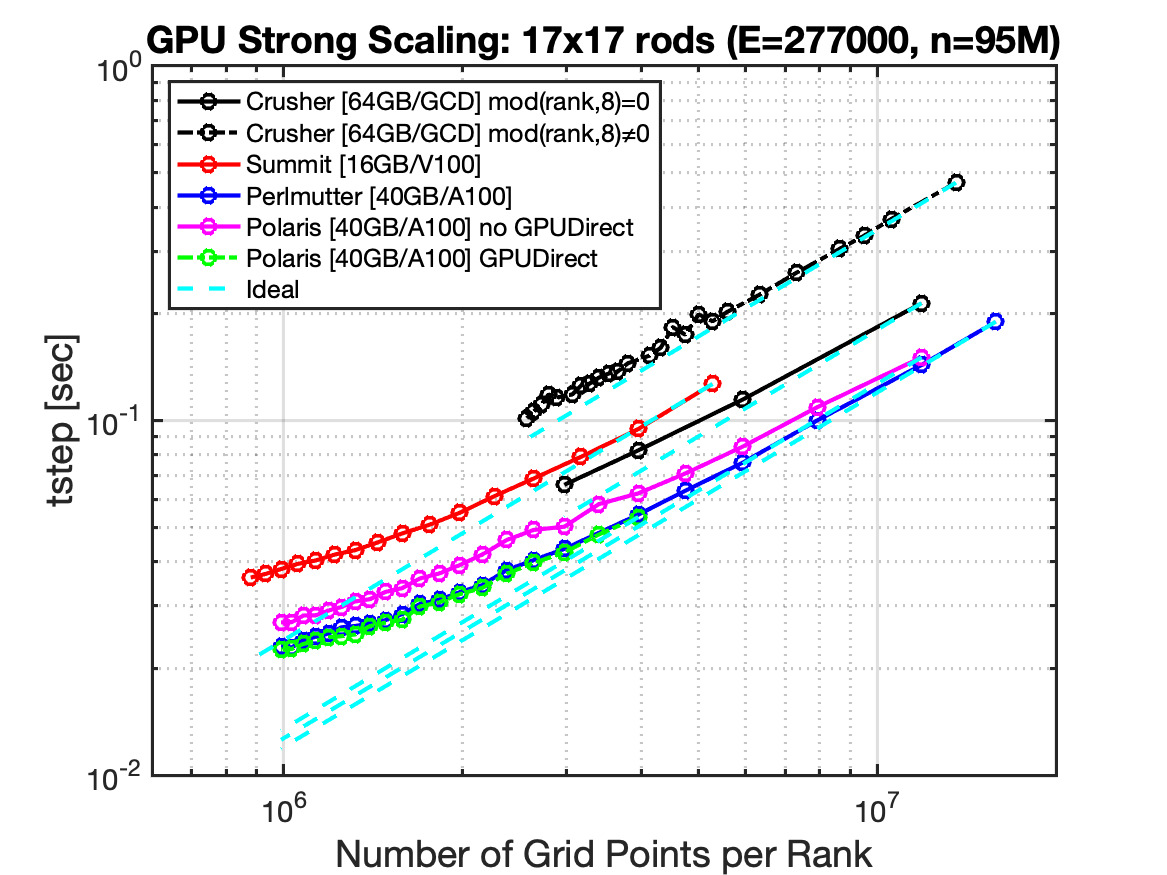 }
     \\
     \includegraphics[width=0.44\textwidth]{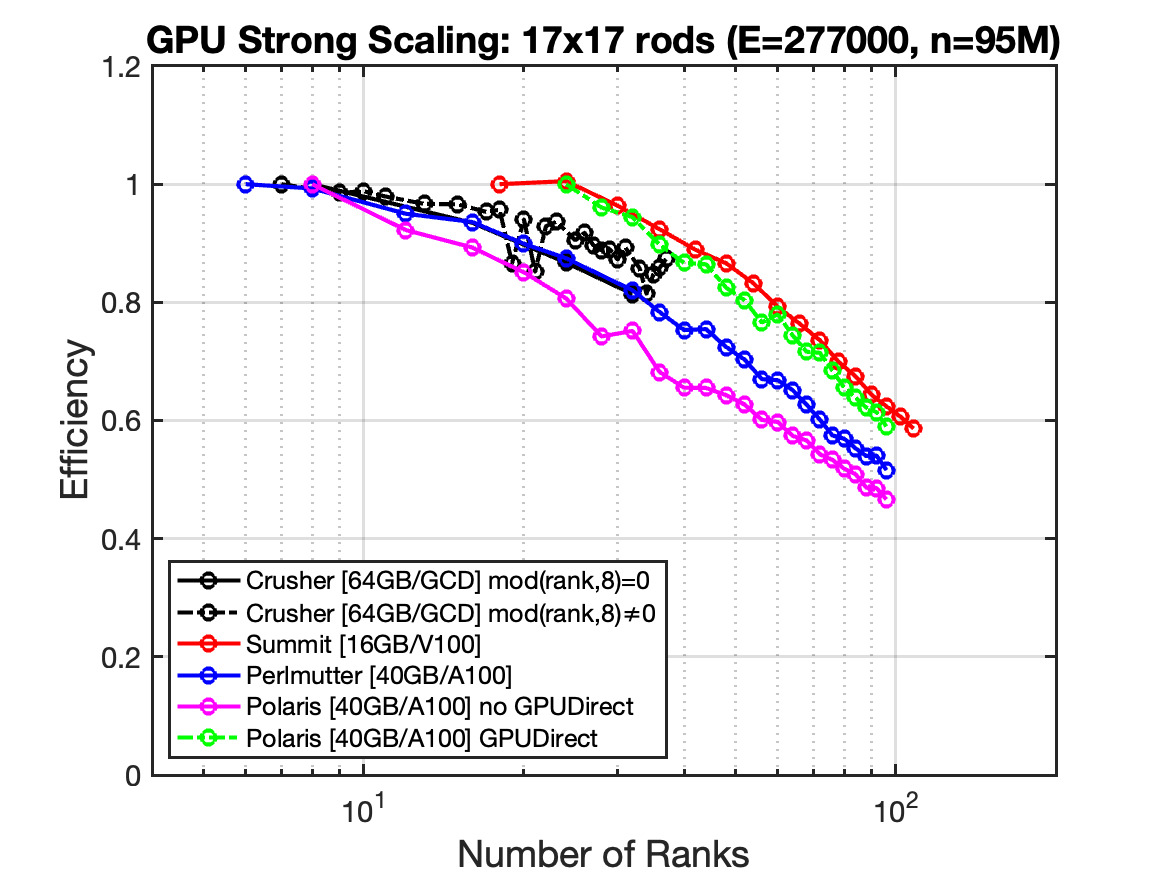 }
     \includegraphics[width=0.44\textwidth]{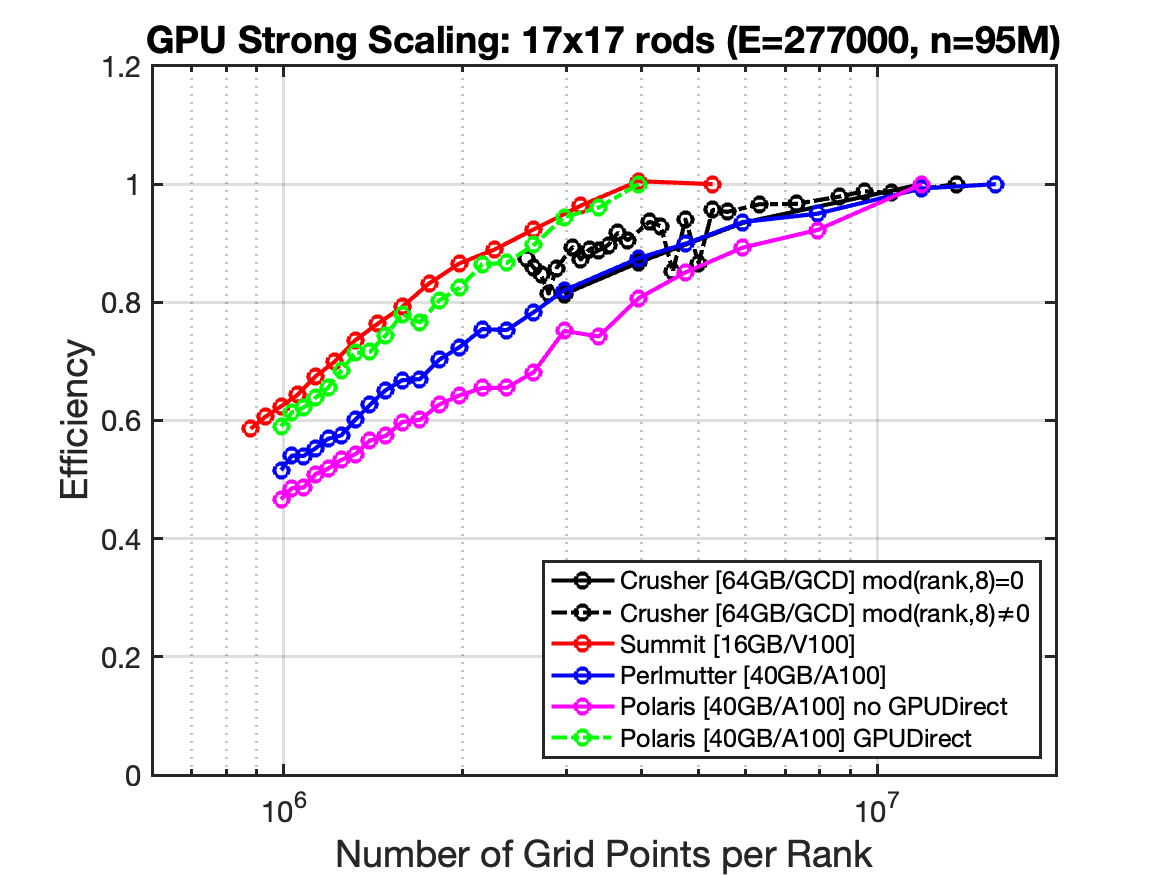 }
     \\
     \includegraphics[width=0.44\textwidth]{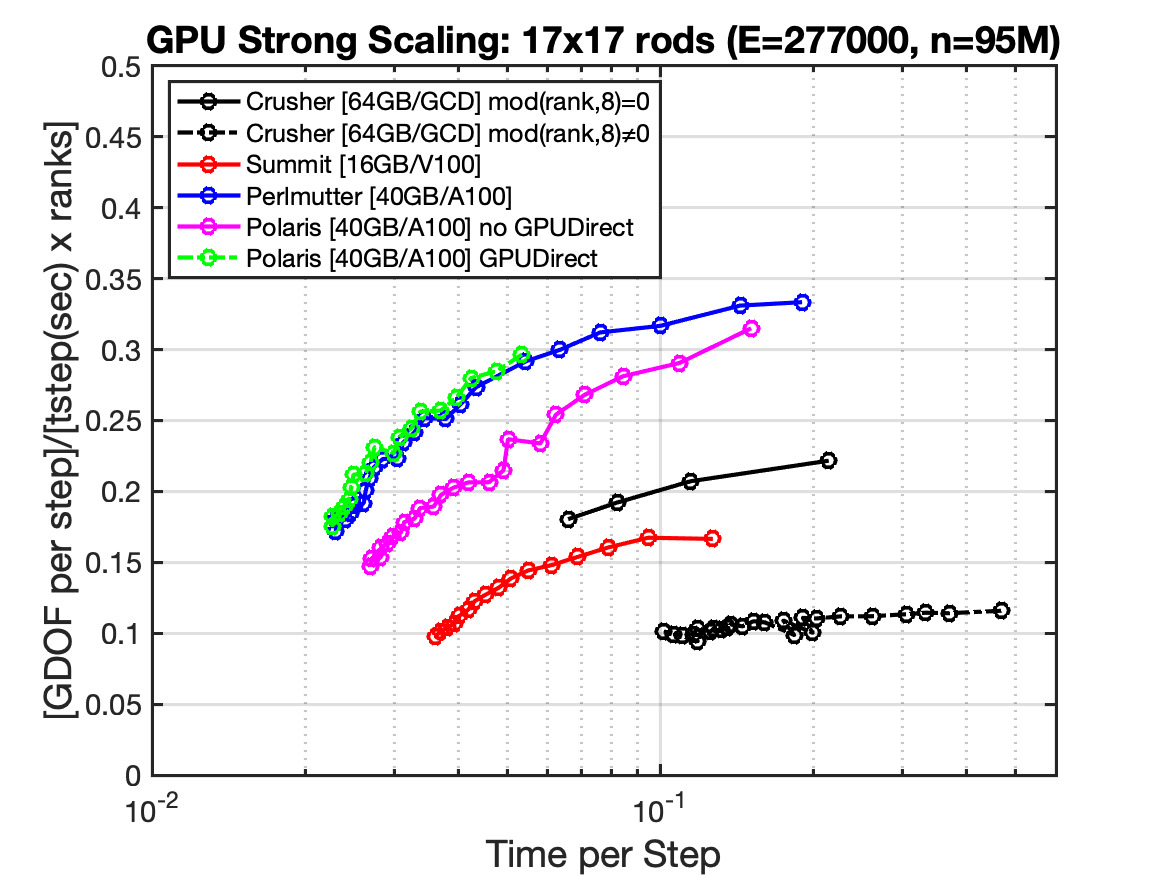 }
     \includegraphics[width=0.44\textwidth]{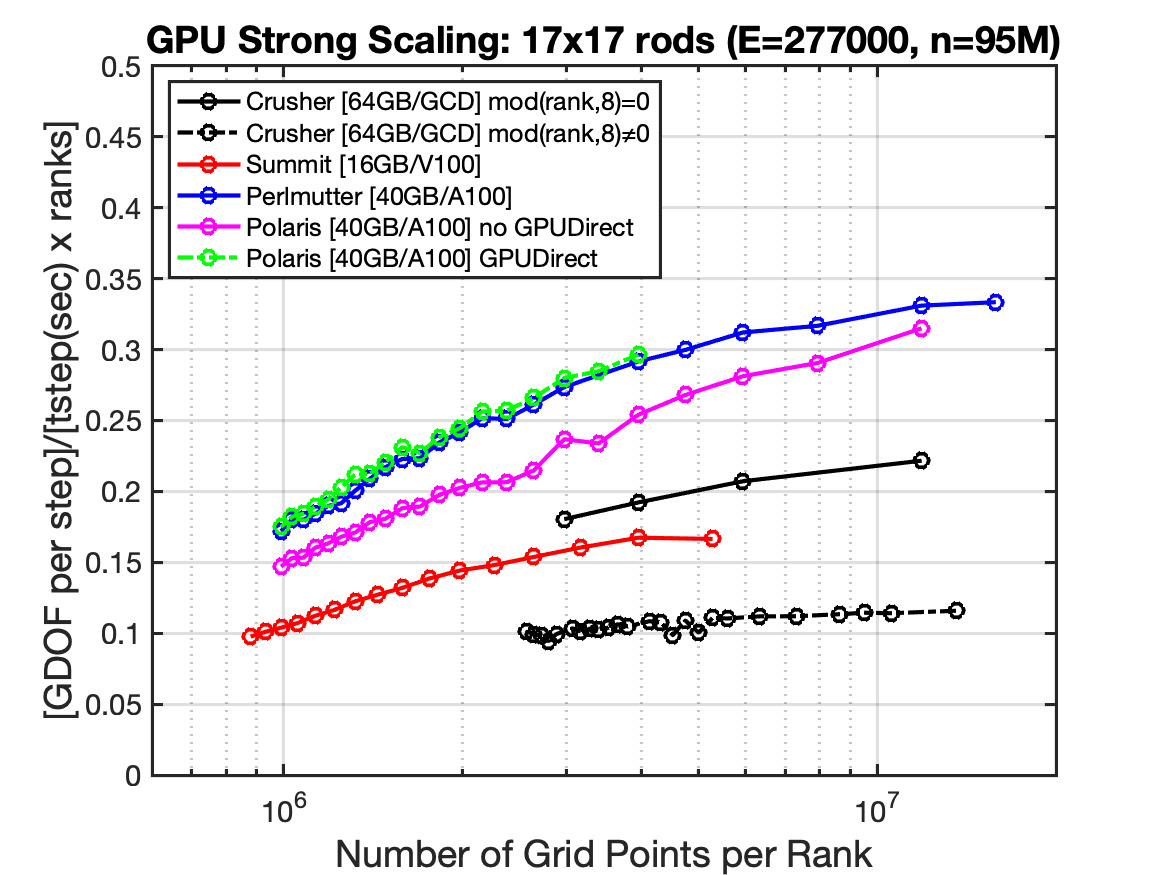 }
     \\
     \includegraphics[width=0.44\textwidth]{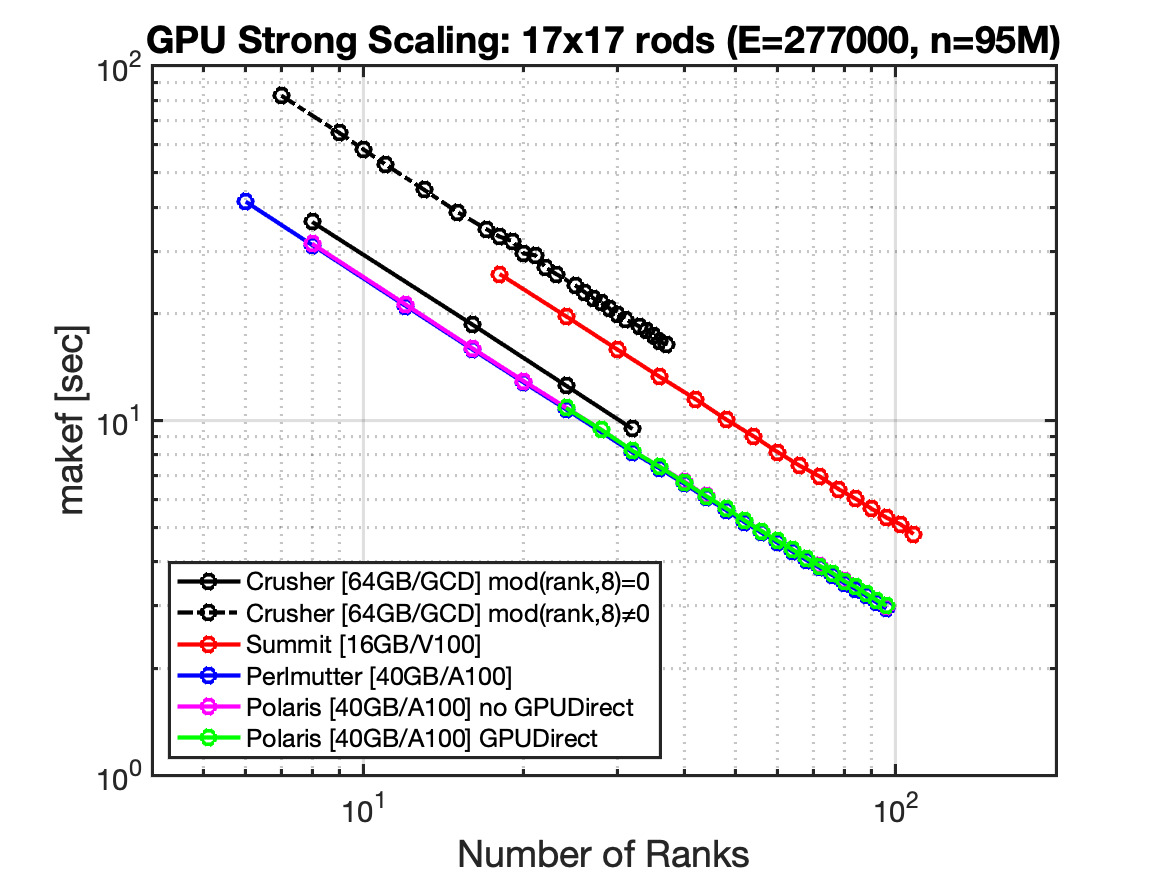 }
     \includegraphics[width=0.44\textwidth]{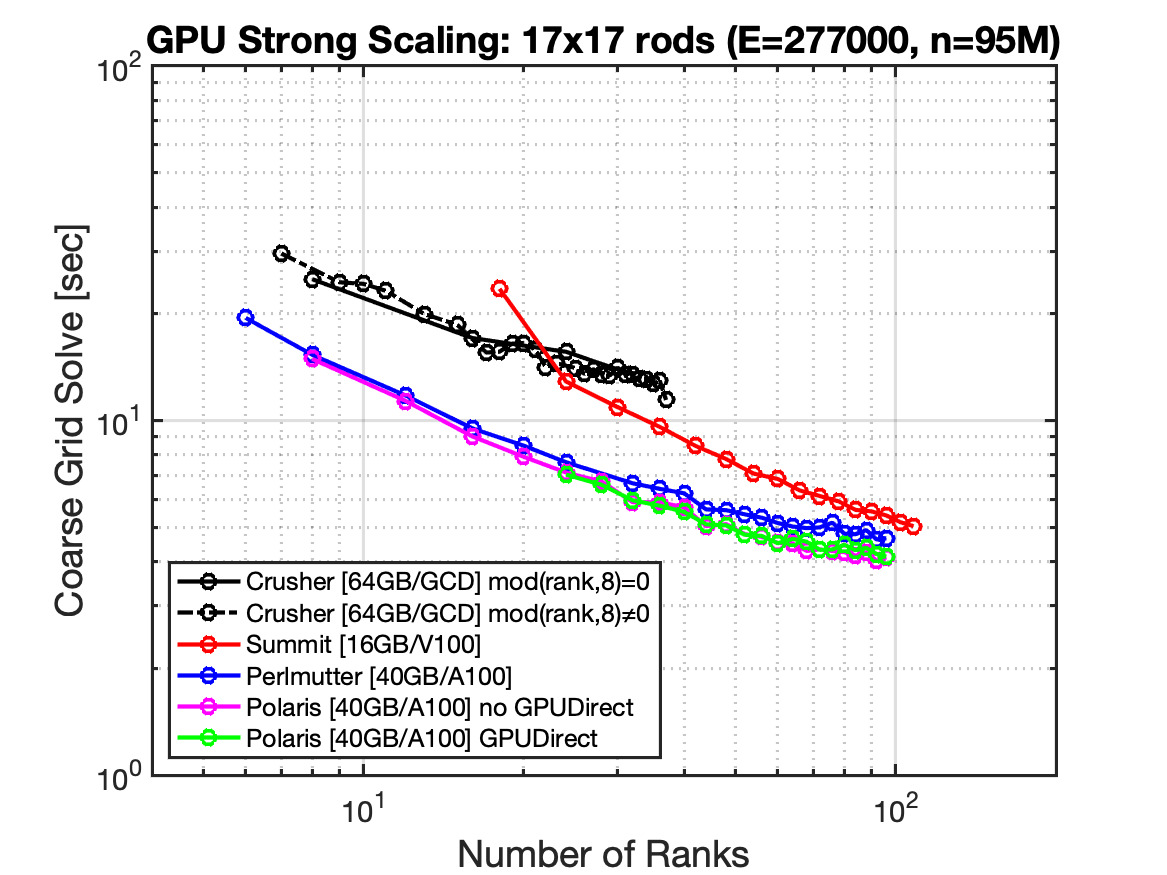 }
     \caption{\label{perf10-rank8} Strong-scaling on Crusher for 17$\times$17 rod bundle with 10 layers.}
  \end{center}
\end{figure*}

\begin{figure*}
  \begin{center}
     \includegraphics[width=0.44\textwidth]{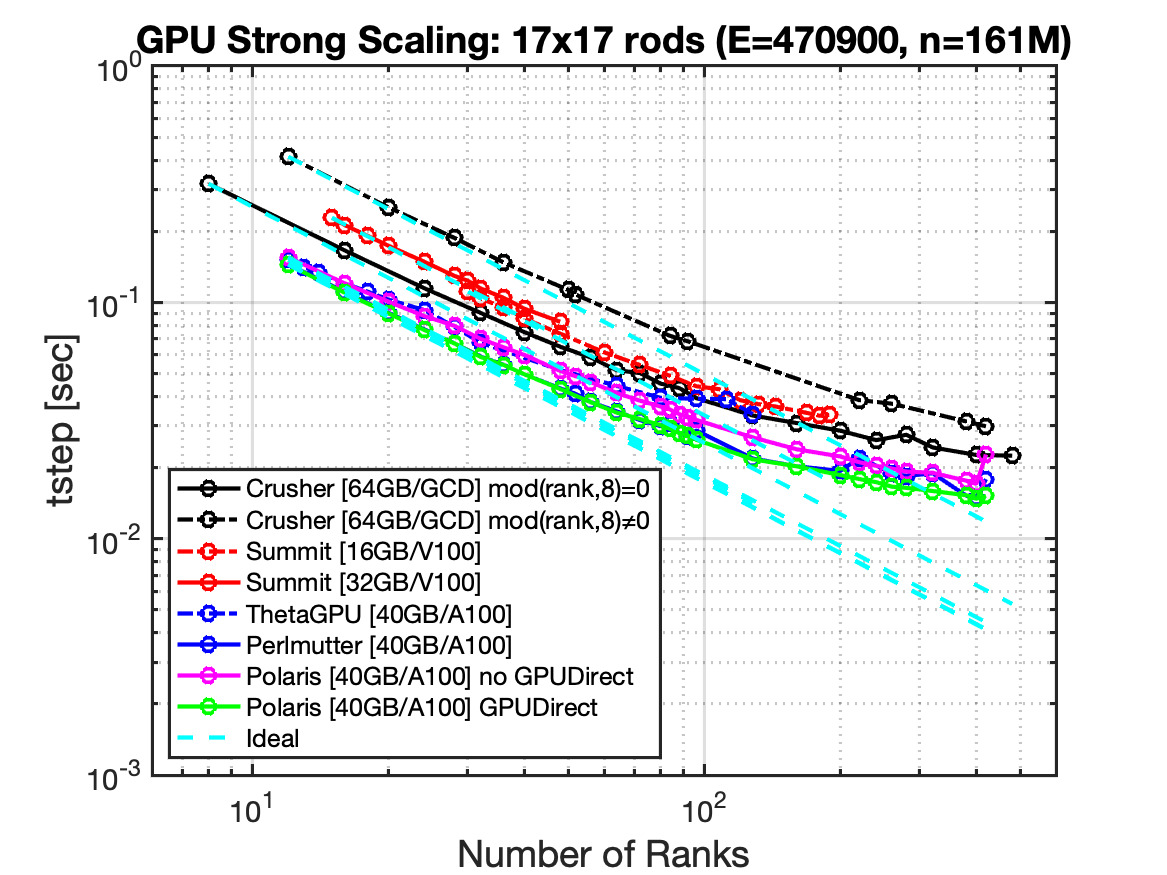 }
     \includegraphics[width=0.44\textwidth]{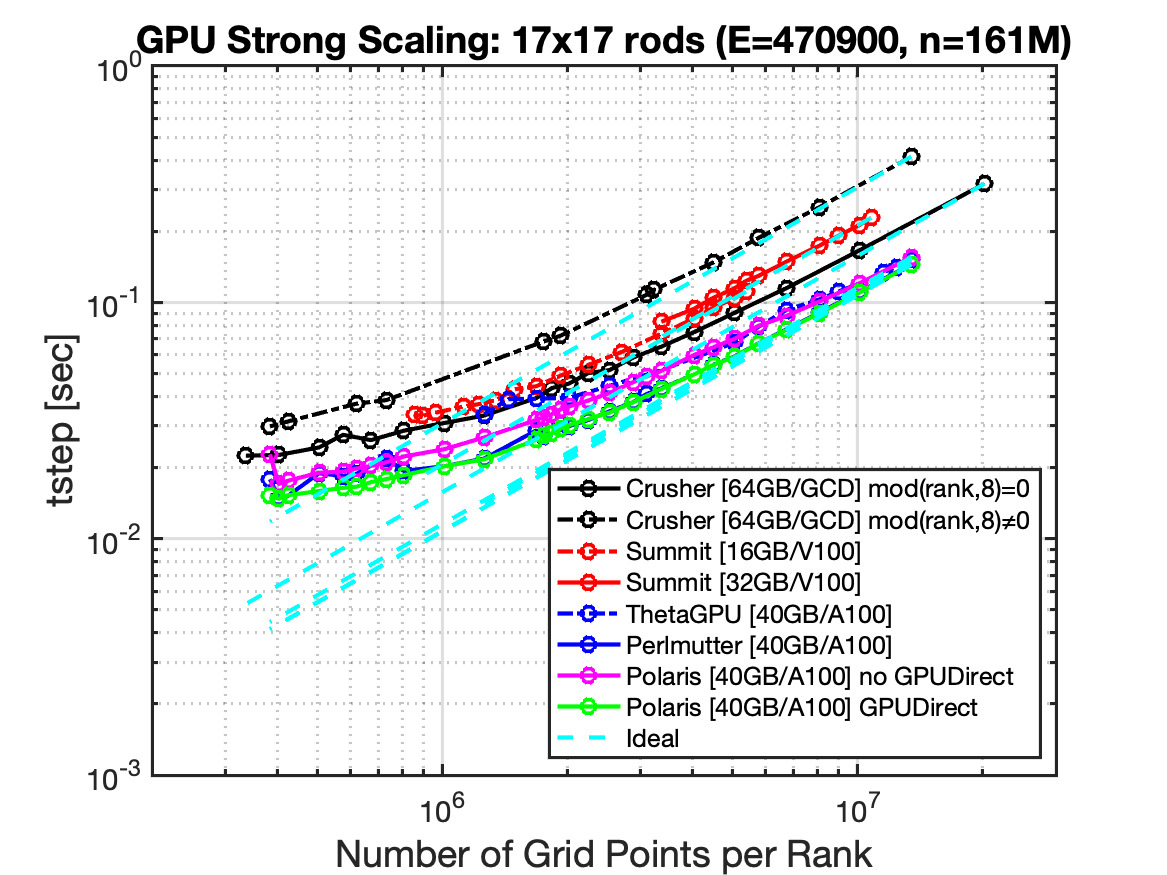 }
     \\
     \includegraphics[width=0.44\textwidth]{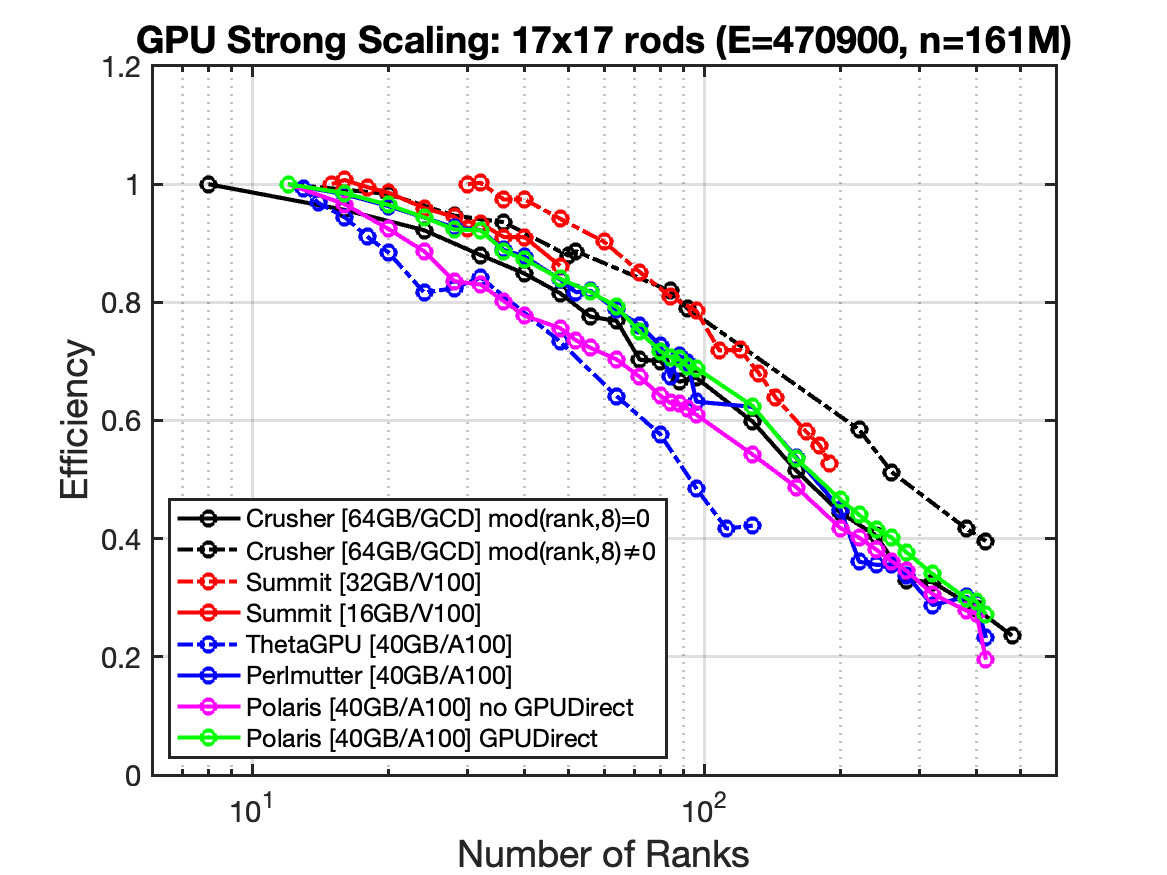 }
     \includegraphics[width=0.44\textwidth]{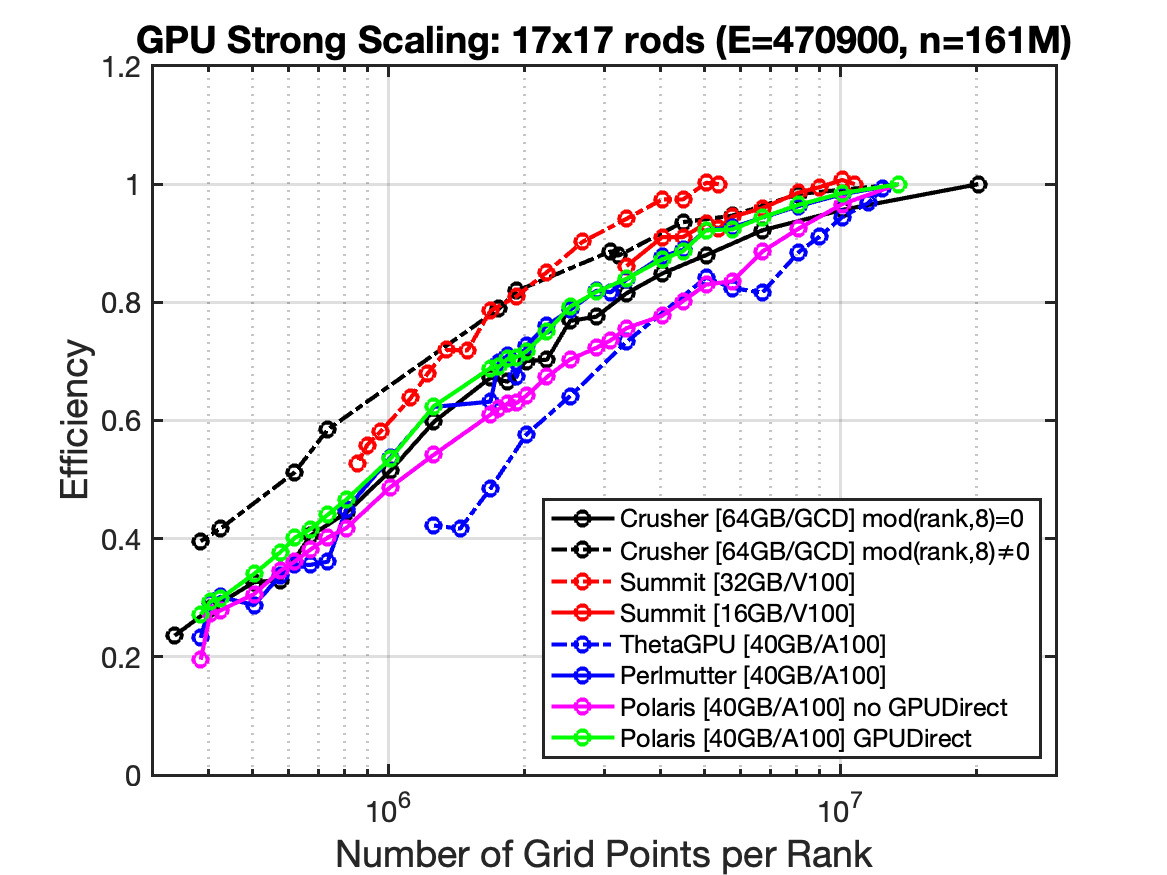 }
     \\
     \includegraphics[width=0.44\textwidth]{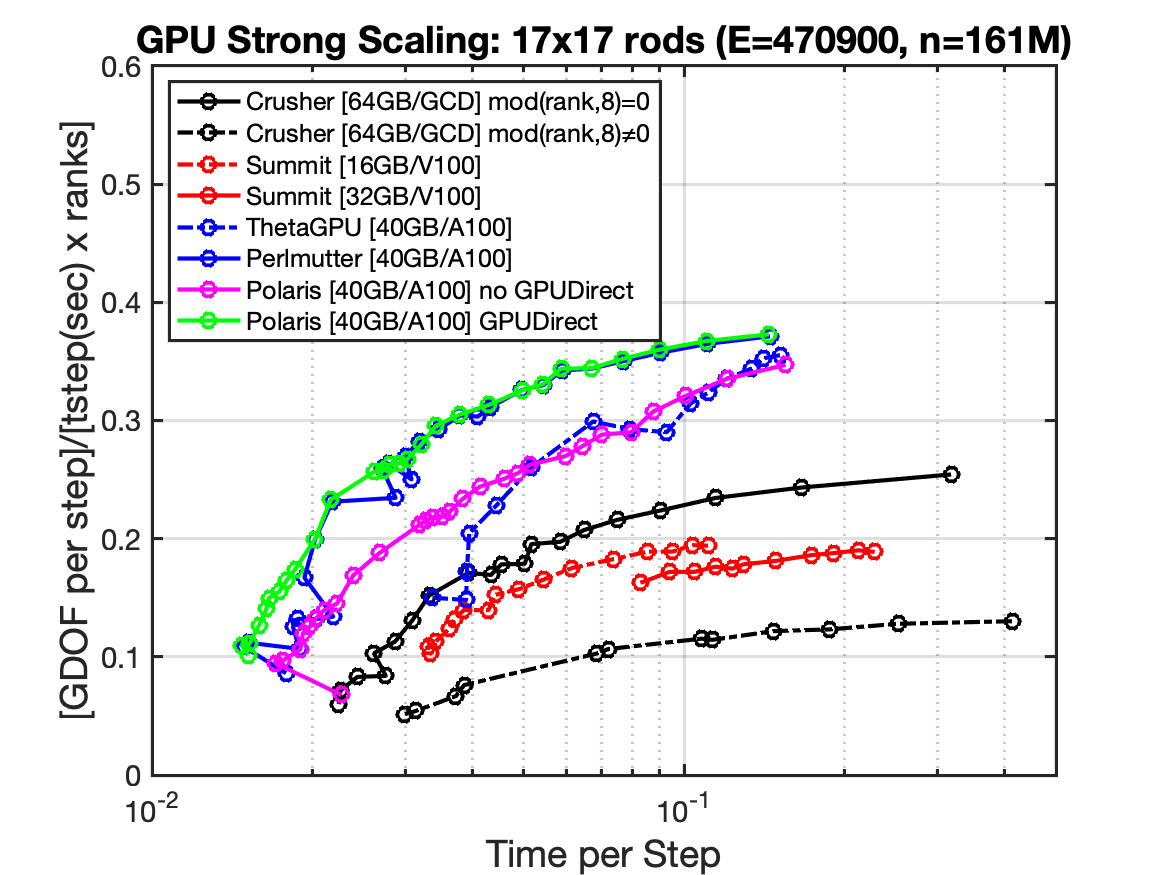 }
     \includegraphics[width=0.44\textwidth]{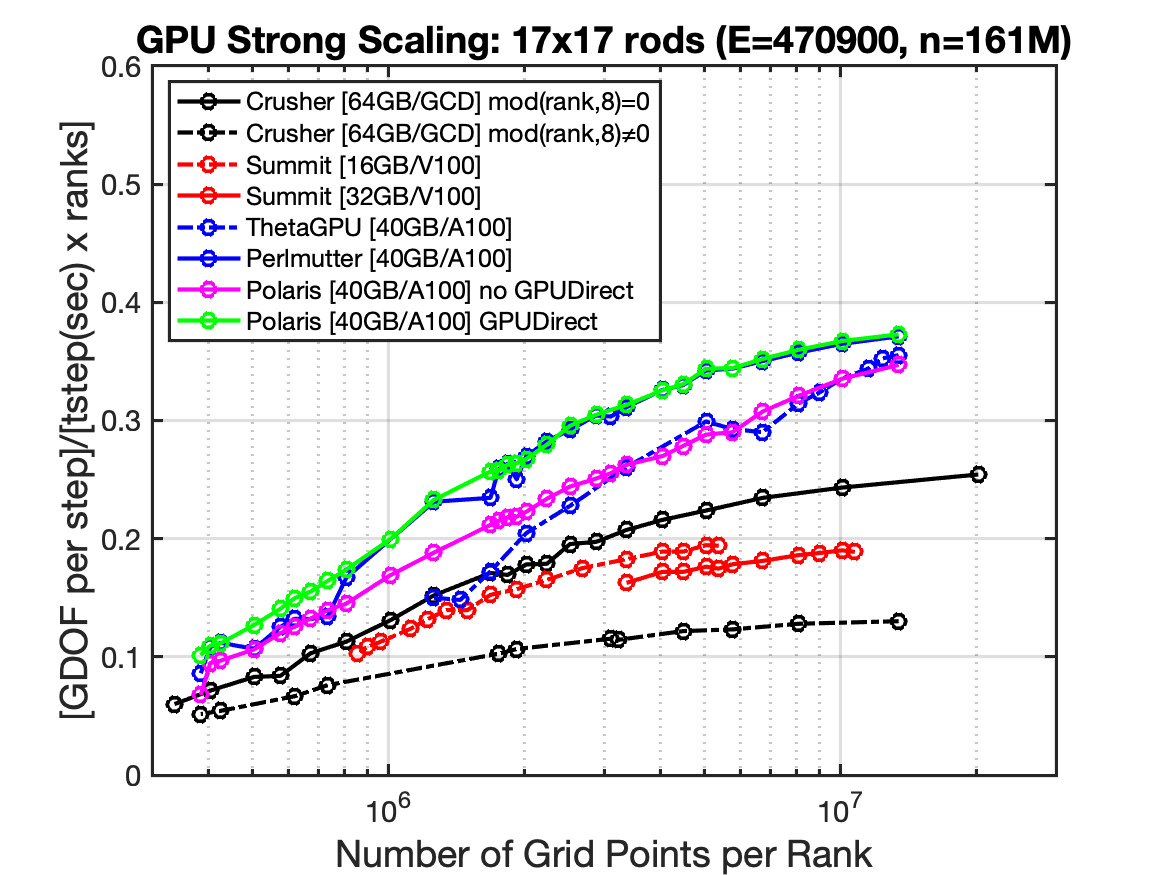 }
     \\
     \includegraphics[width=0.44\textwidth]{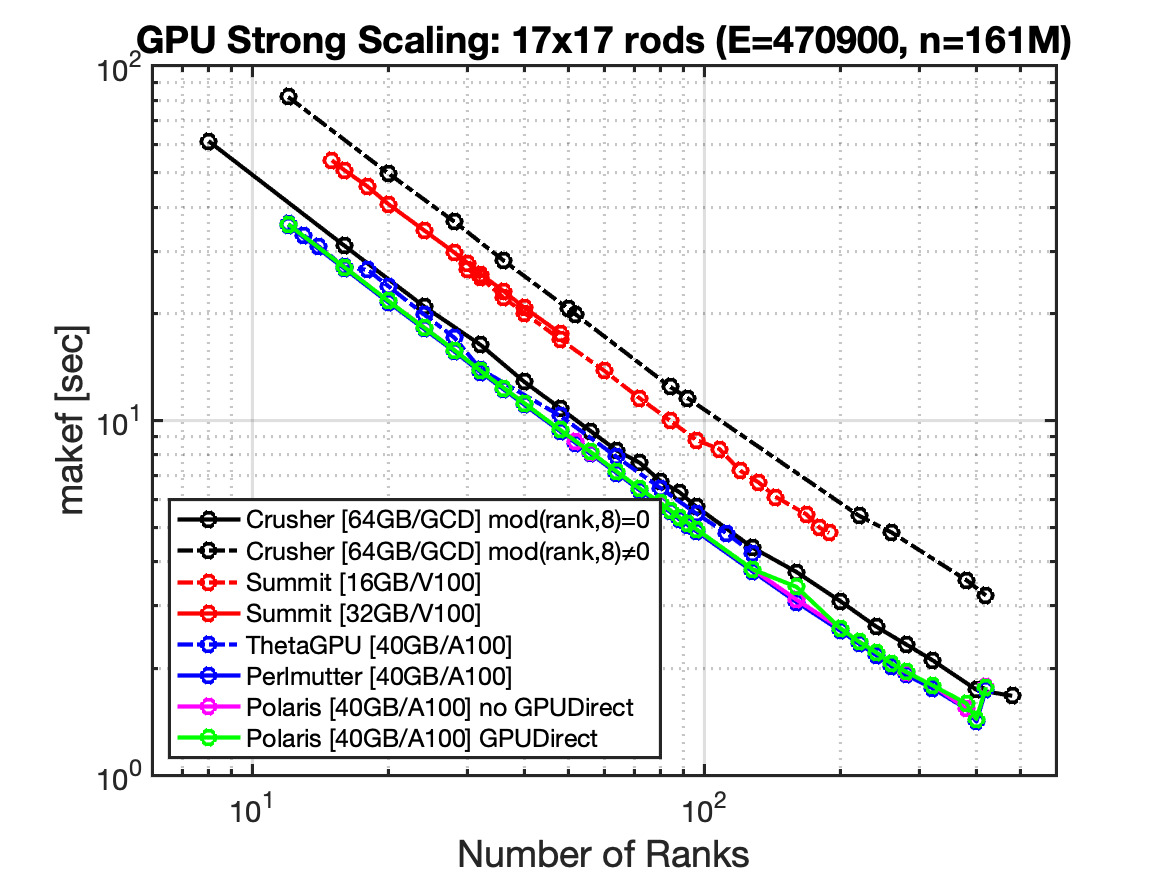 }
     \includegraphics[width=0.44\textwidth]{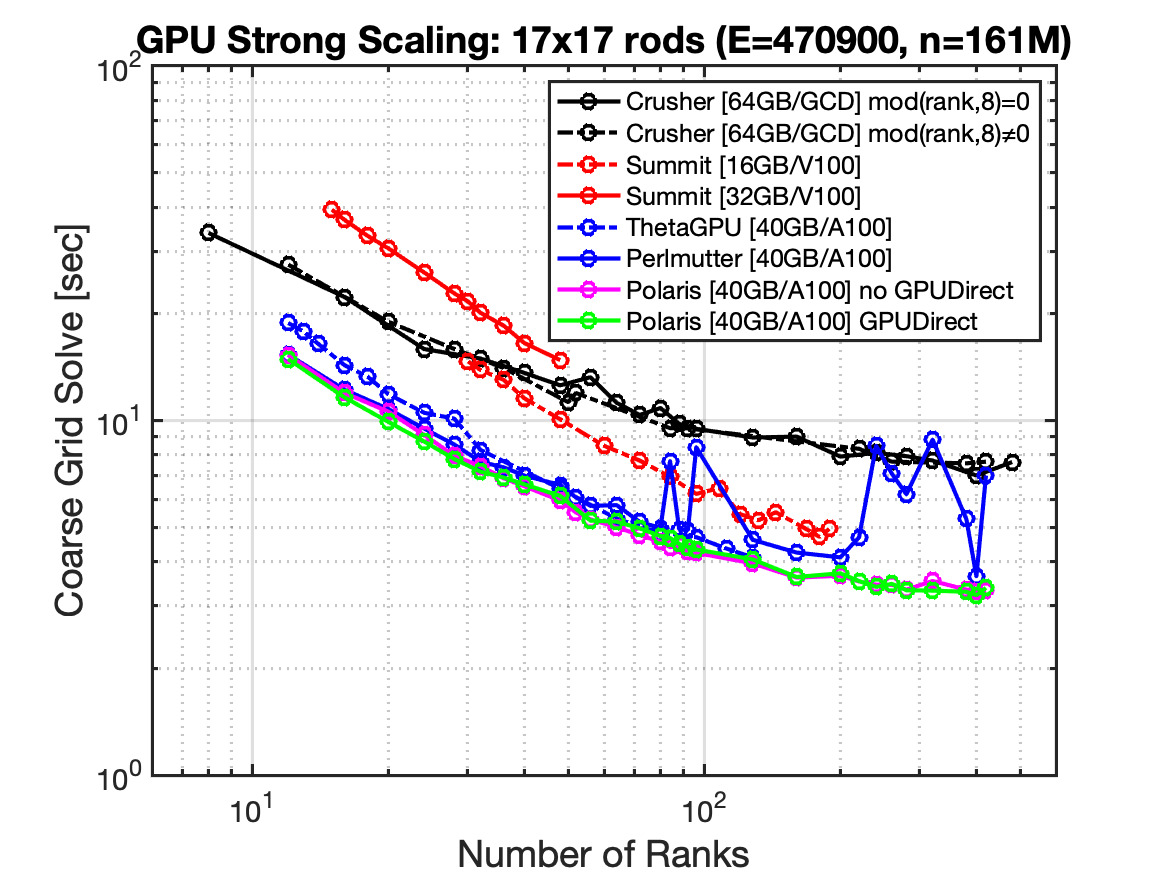 }
     \caption{\label{perf17-rank8} Strong-scaling on Crusher for 17$\times$17 rod bundle with 17 layers.}
  \end{center}
\end{figure*}

\begin{figure*}
  \begin{center}
     \includegraphics[width=0.44\textwidth]{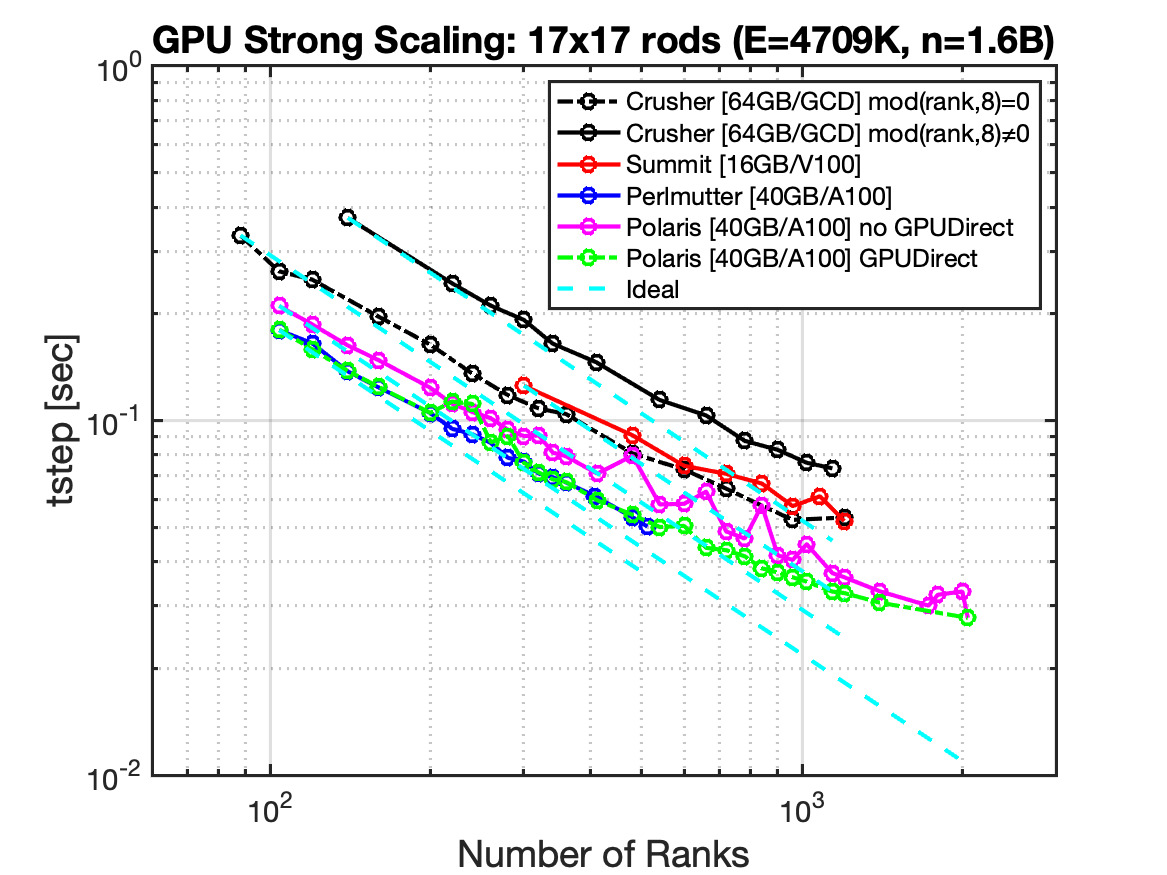 }
     \includegraphics[width=0.44\textwidth]{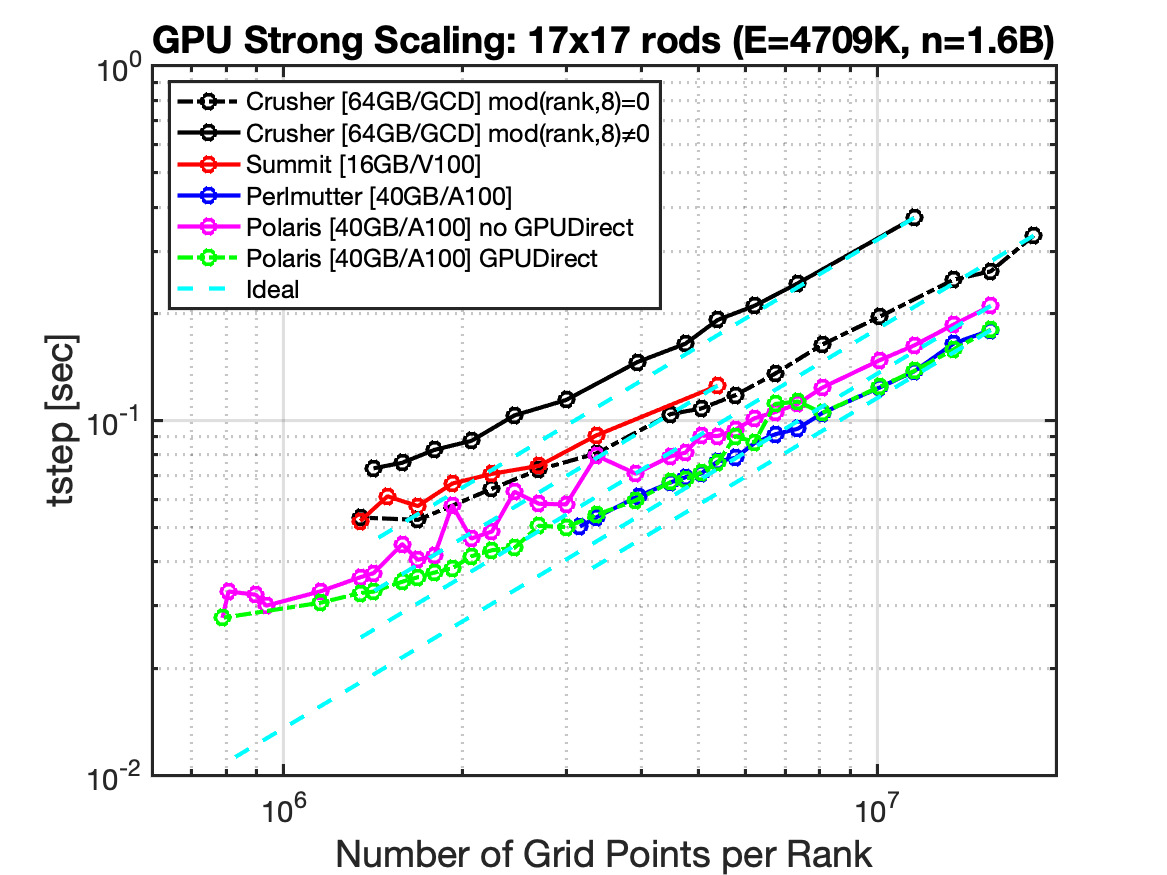 }
     \\
     \includegraphics[width=0.44\textwidth]{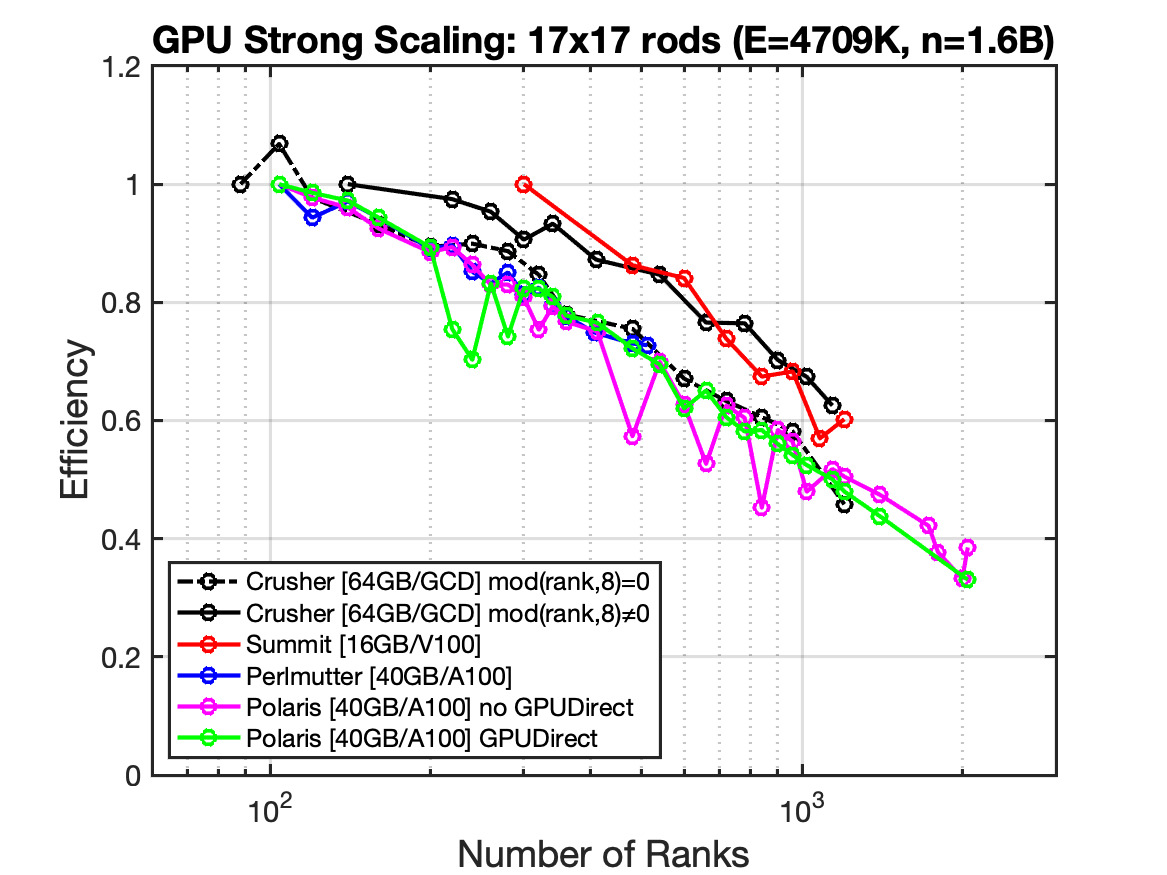 }
     \includegraphics[width=0.44\textwidth]{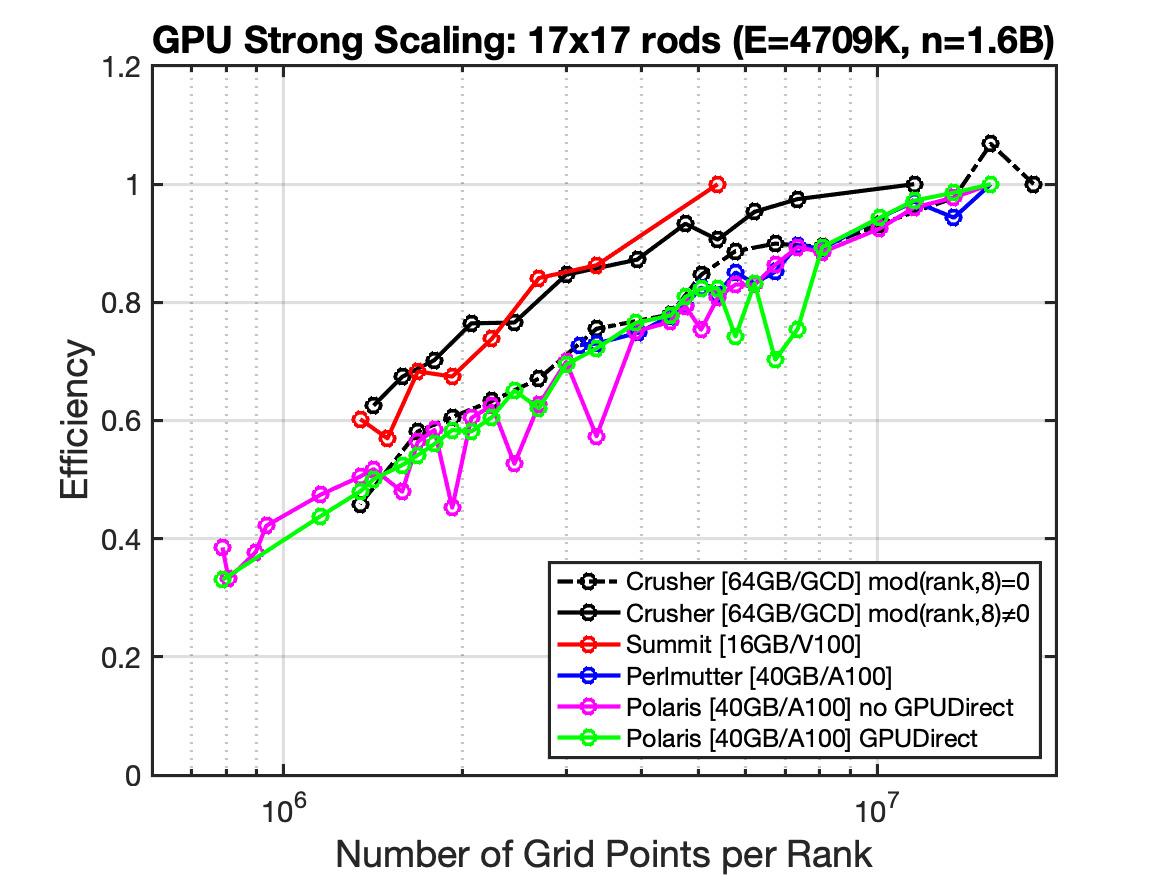 }
     \\
     \includegraphics[width=0.44\textwidth]{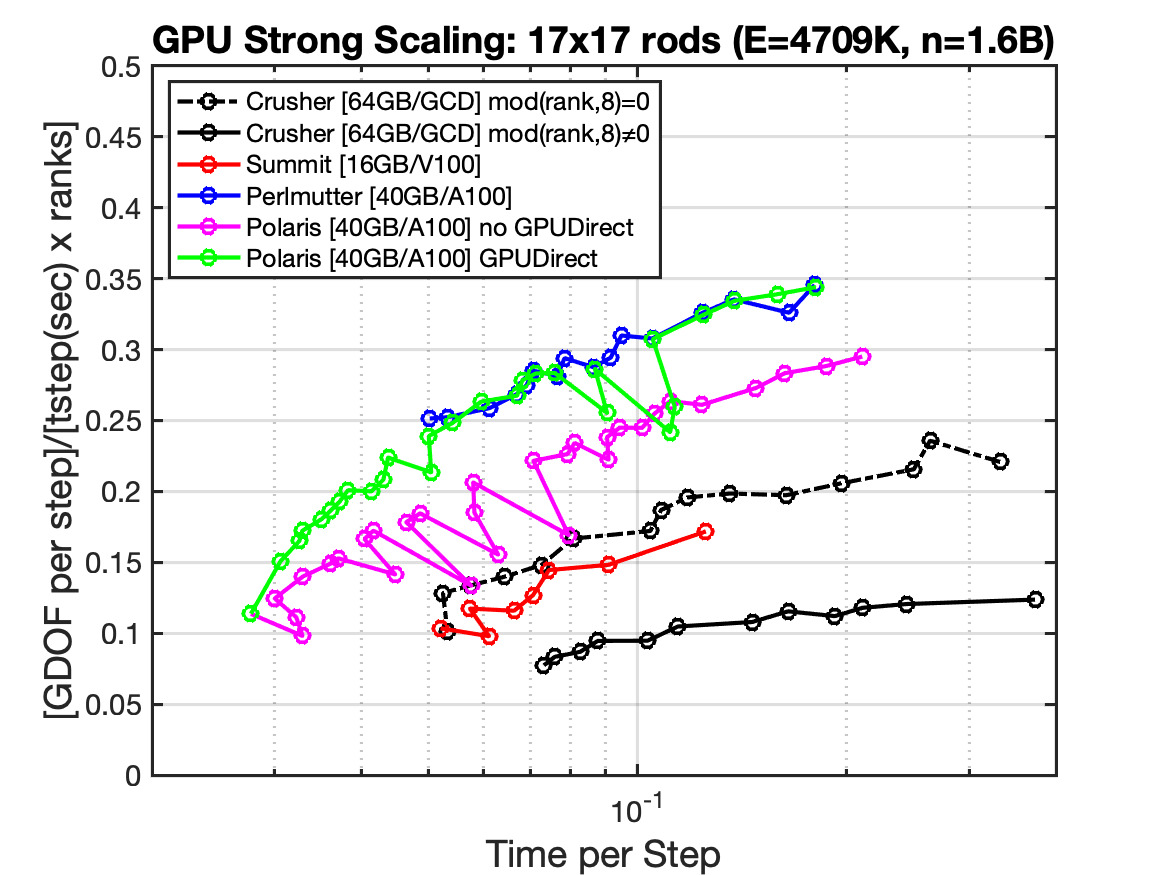 }
     \includegraphics[width=0.44\textwidth]{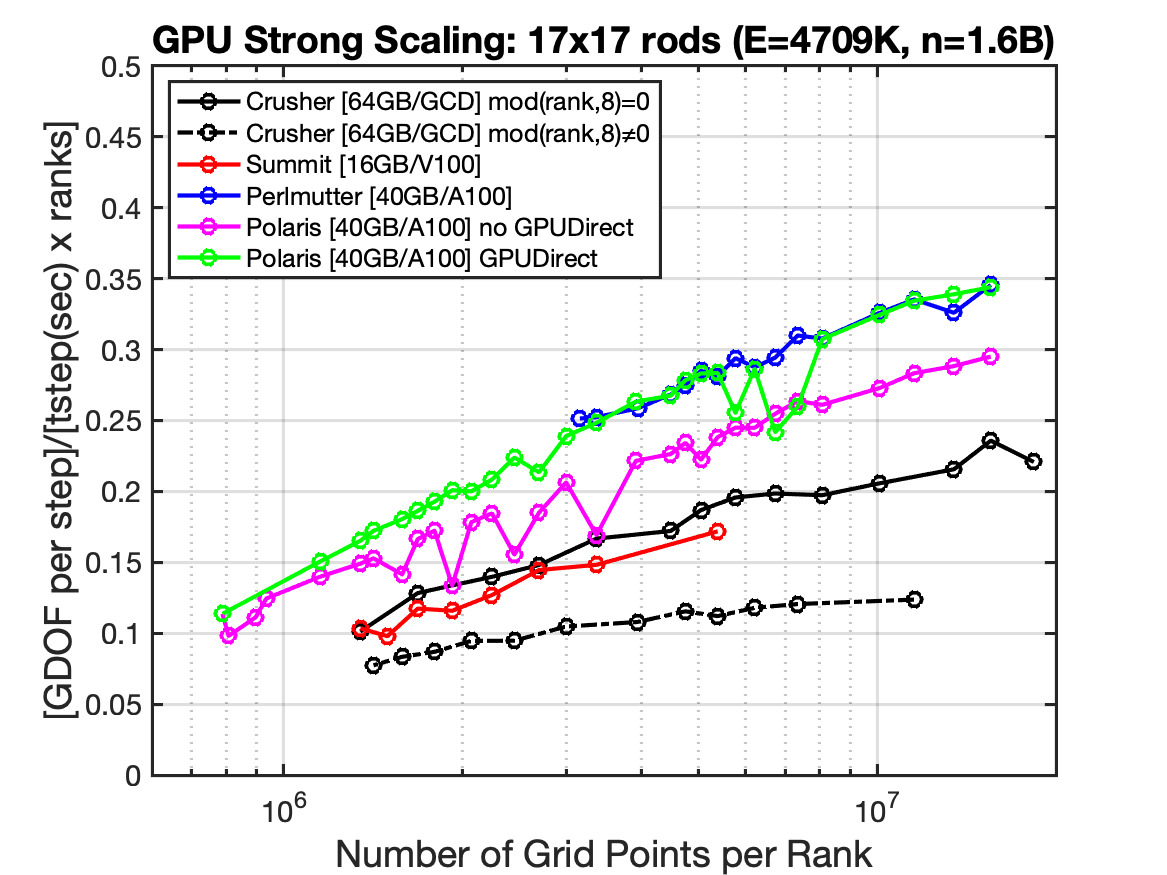 }
     \\
     \includegraphics[width=0.44\textwidth]{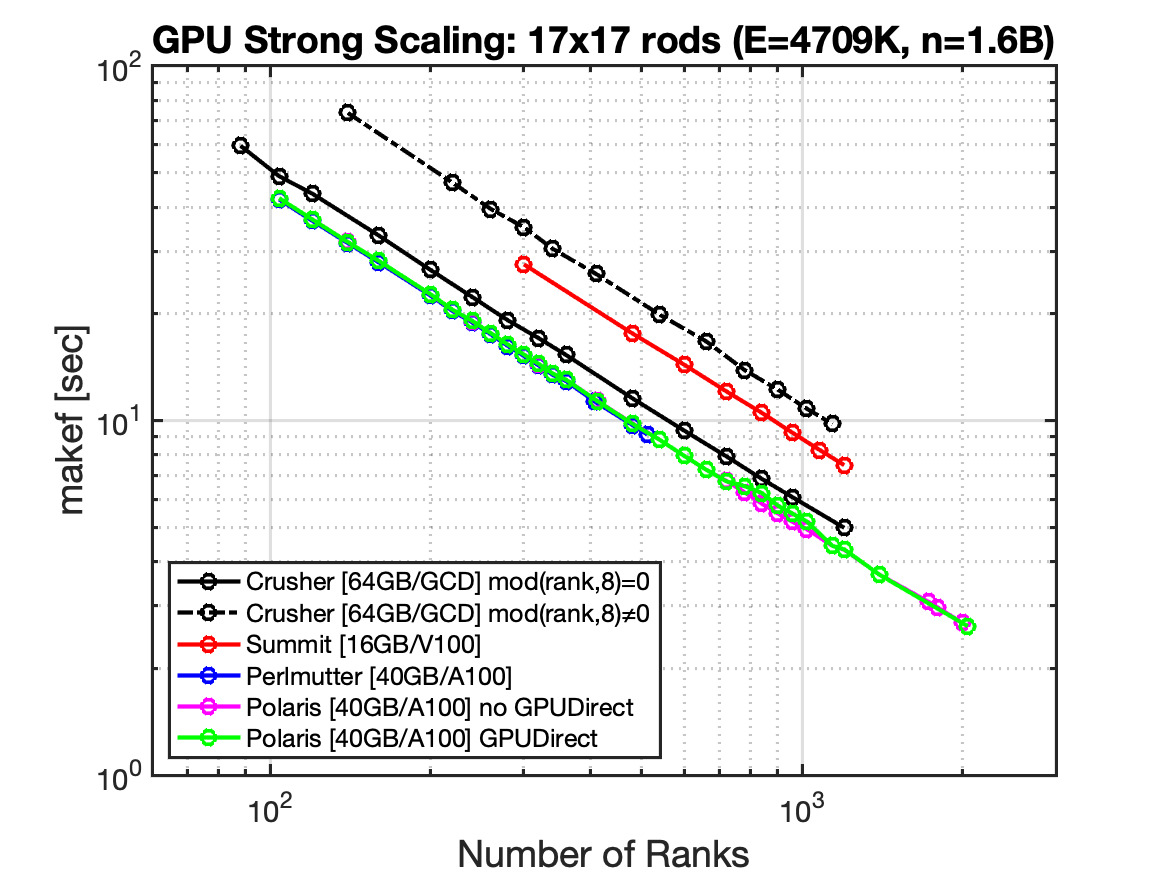 }
     \includegraphics[width=0.44\textwidth]{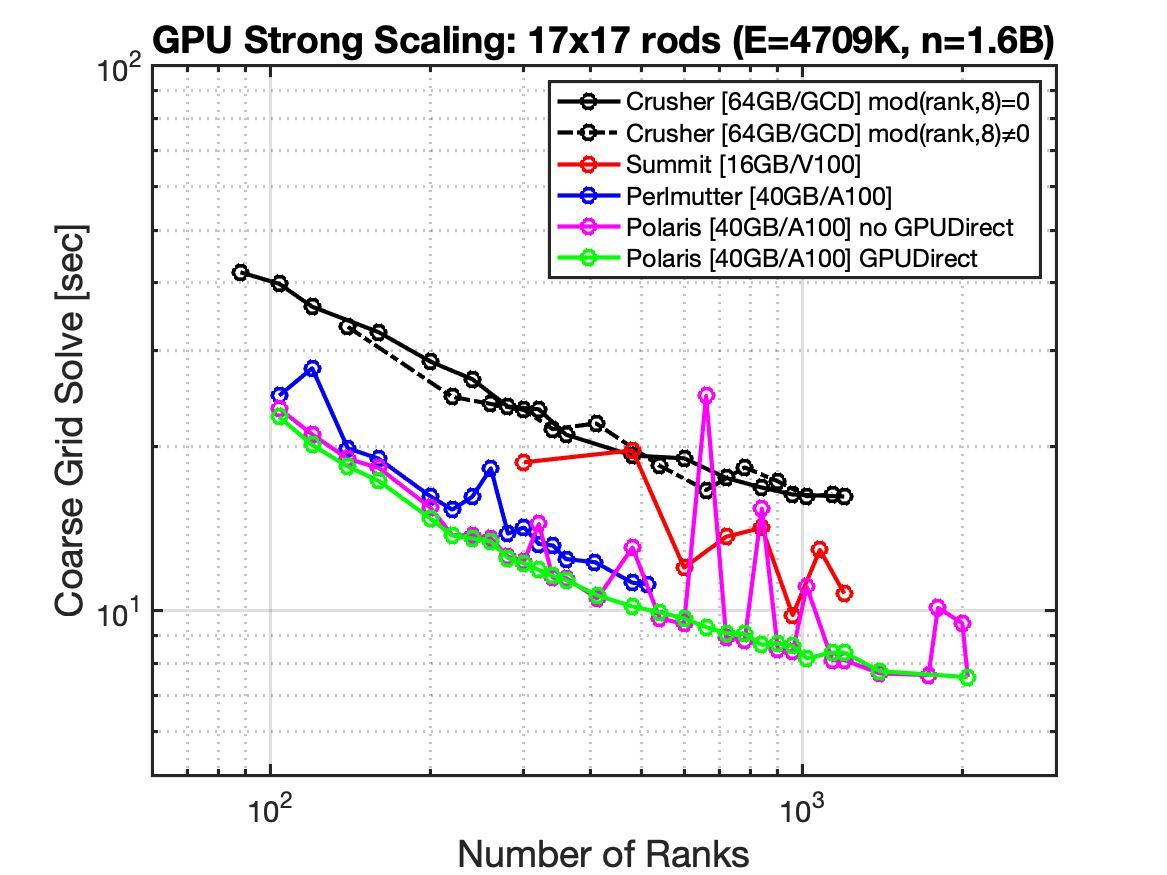 }
     \caption{\label{perf170-rank8} Strong-scaling on Crusher for 17$\times$17 rod bundle with 170 layers.}
  \end{center}
\end{figure*}

\begin{figure*}
  \begin{center}
     \includegraphics[width=0.44\textwidth]{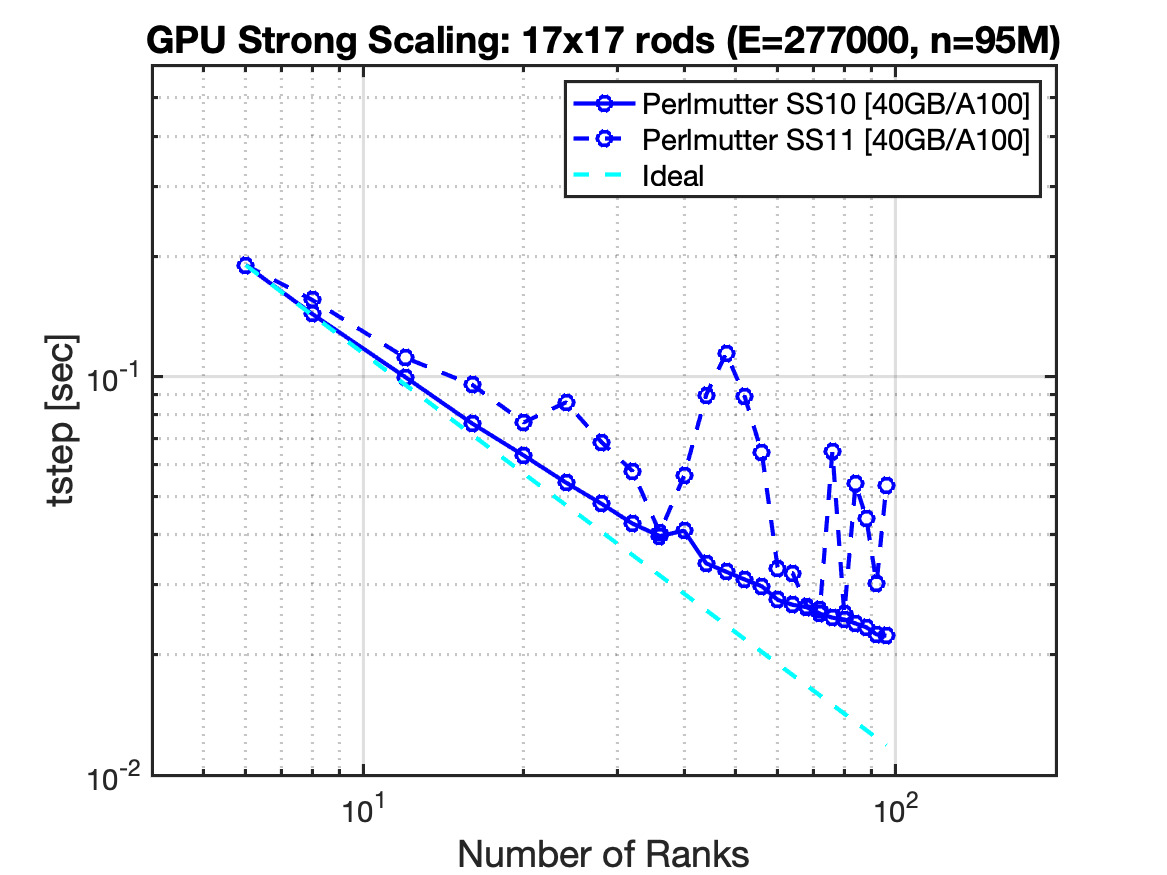 }
     \includegraphics[width=0.44\textwidth]{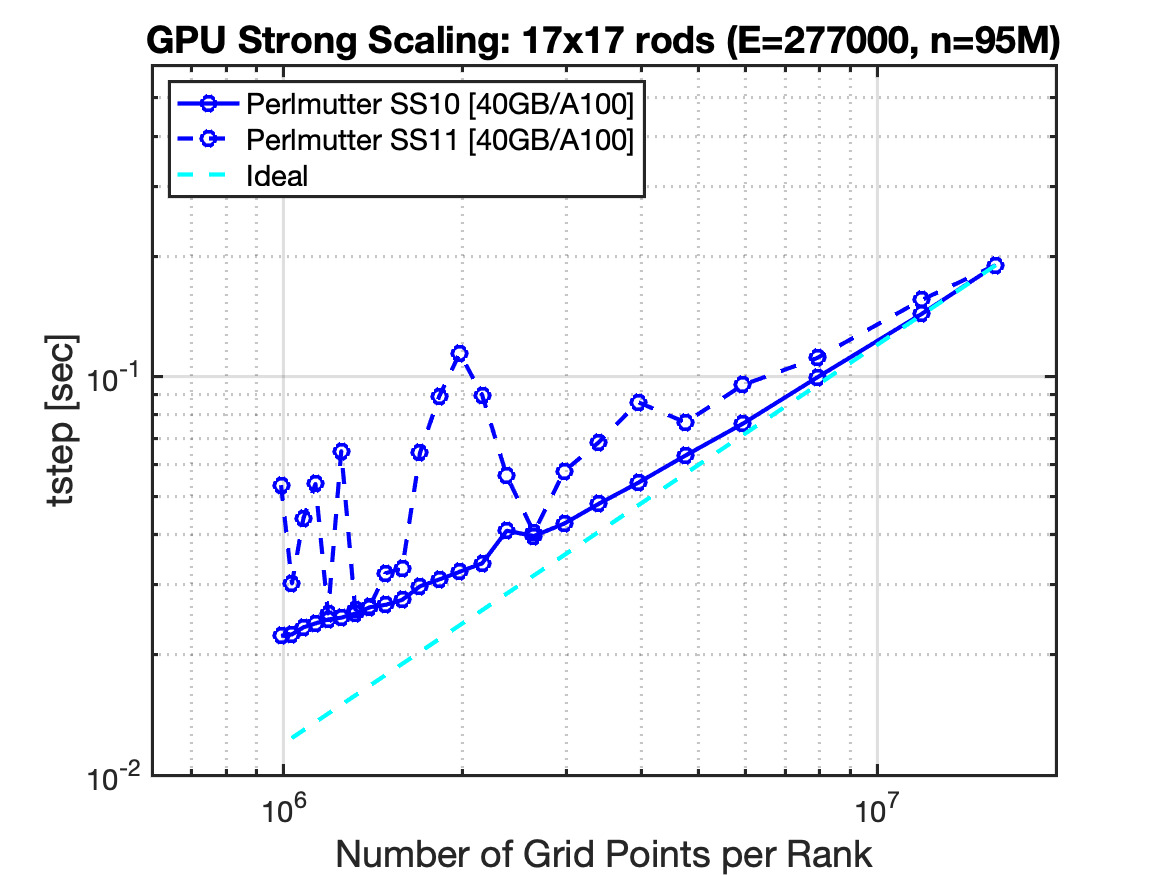 }
     \\
     \includegraphics[width=0.44\textwidth]{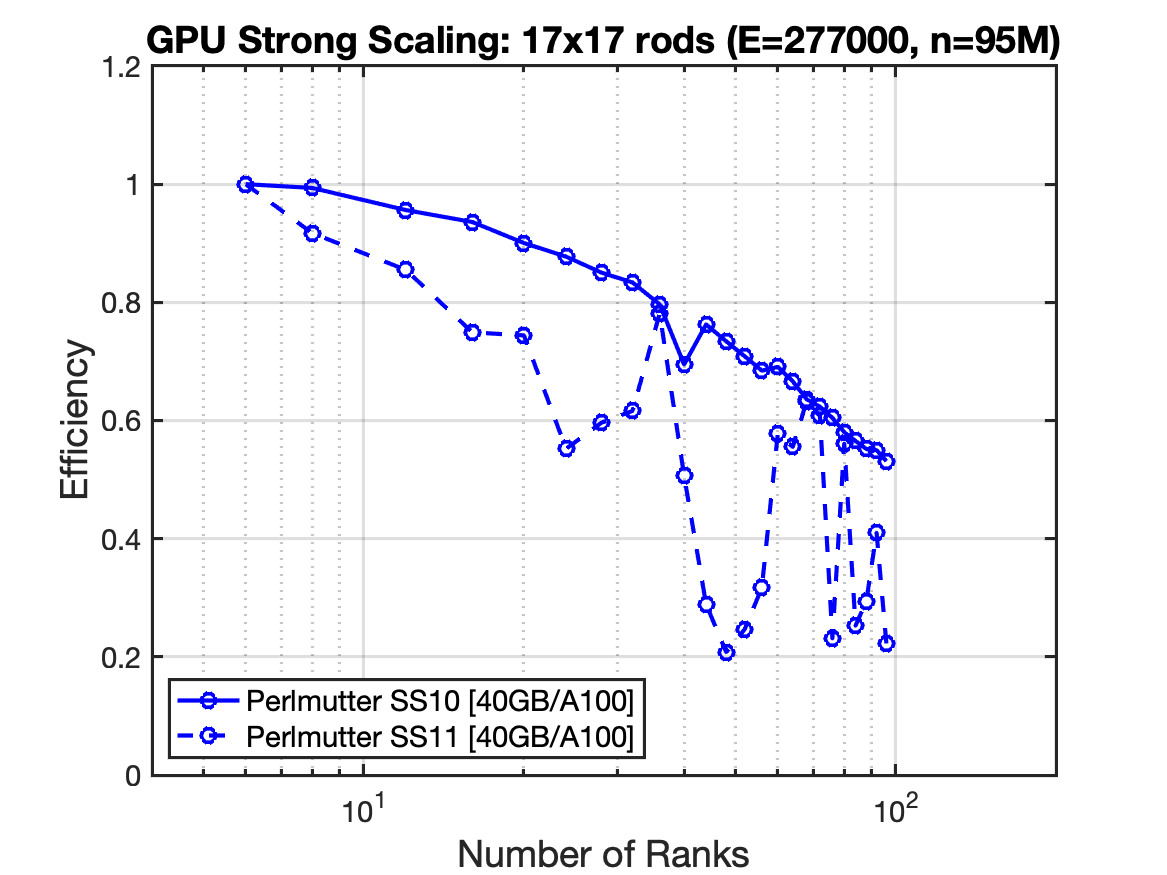 }
     \includegraphics[width=0.44\textwidth]{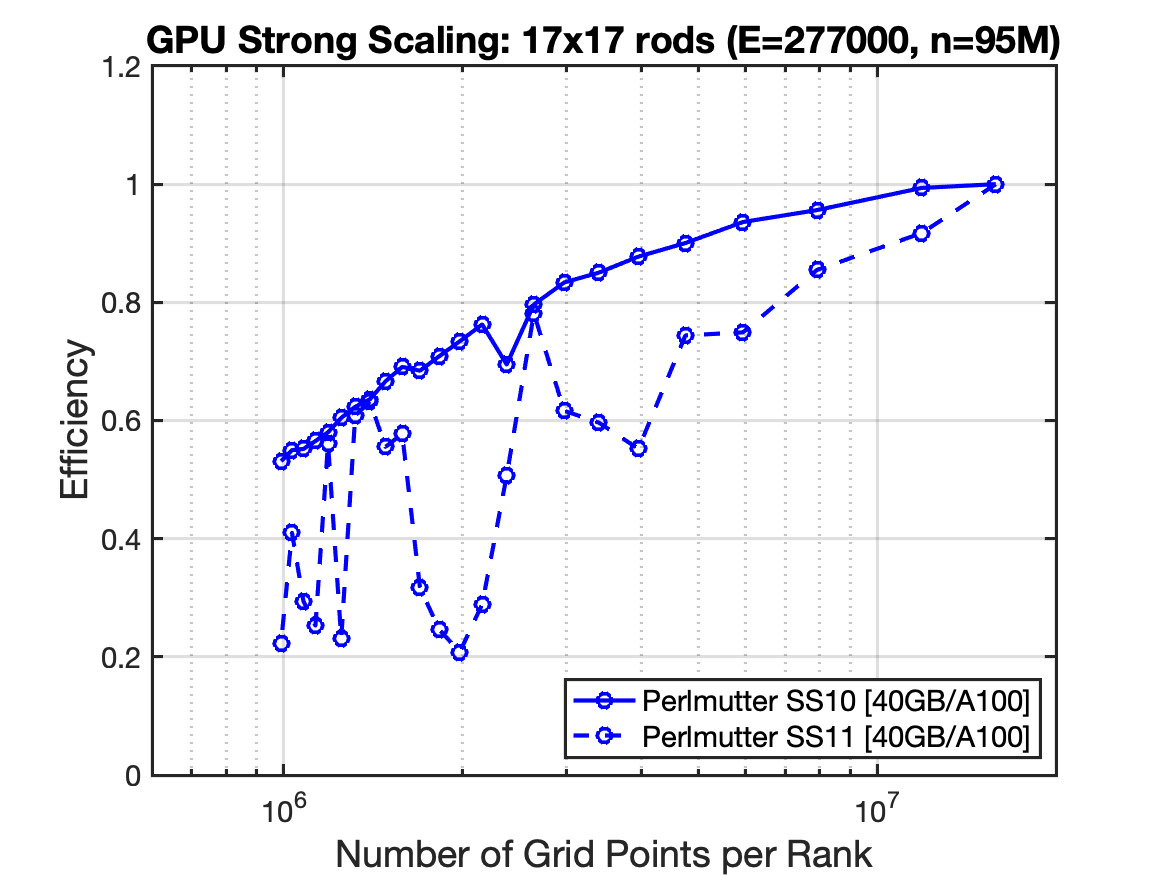 }
     \\
     \includegraphics[width=0.44\textwidth]{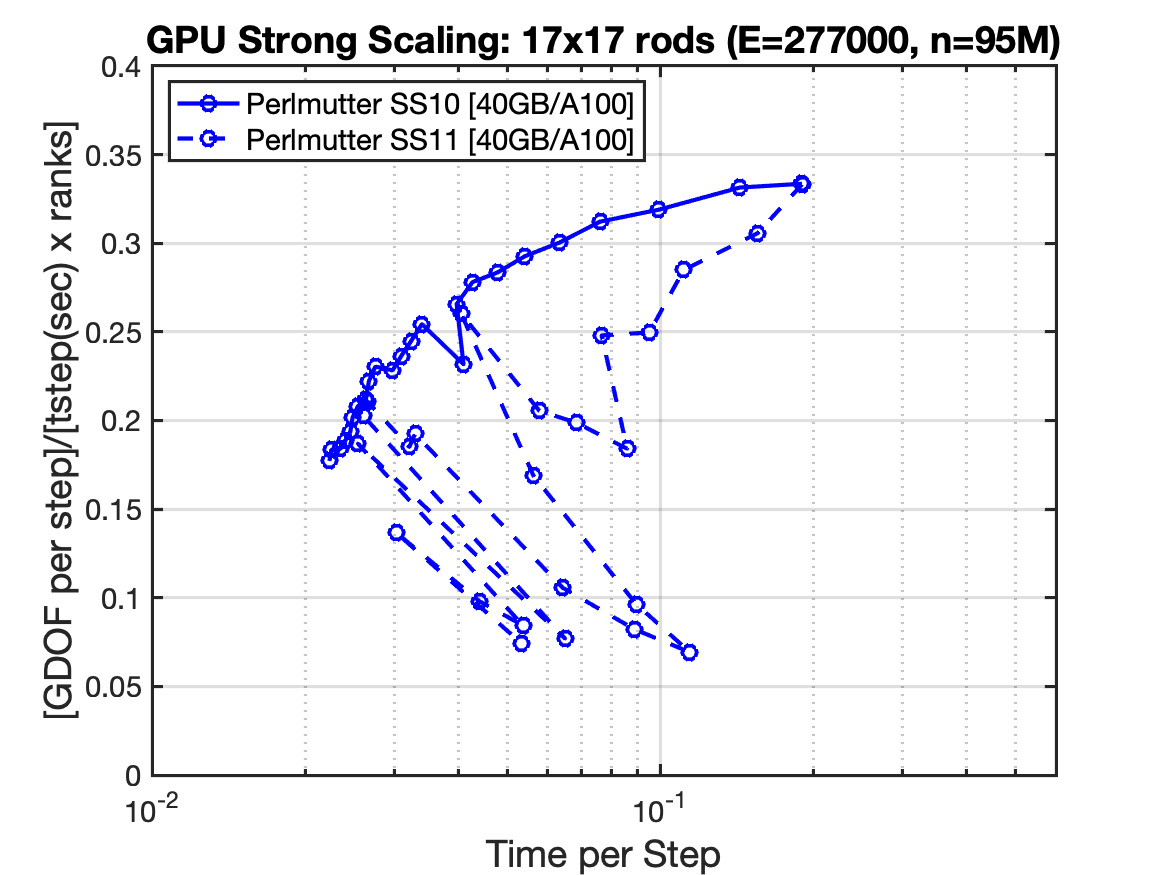 }
     \includegraphics[width=0.44\textwidth]{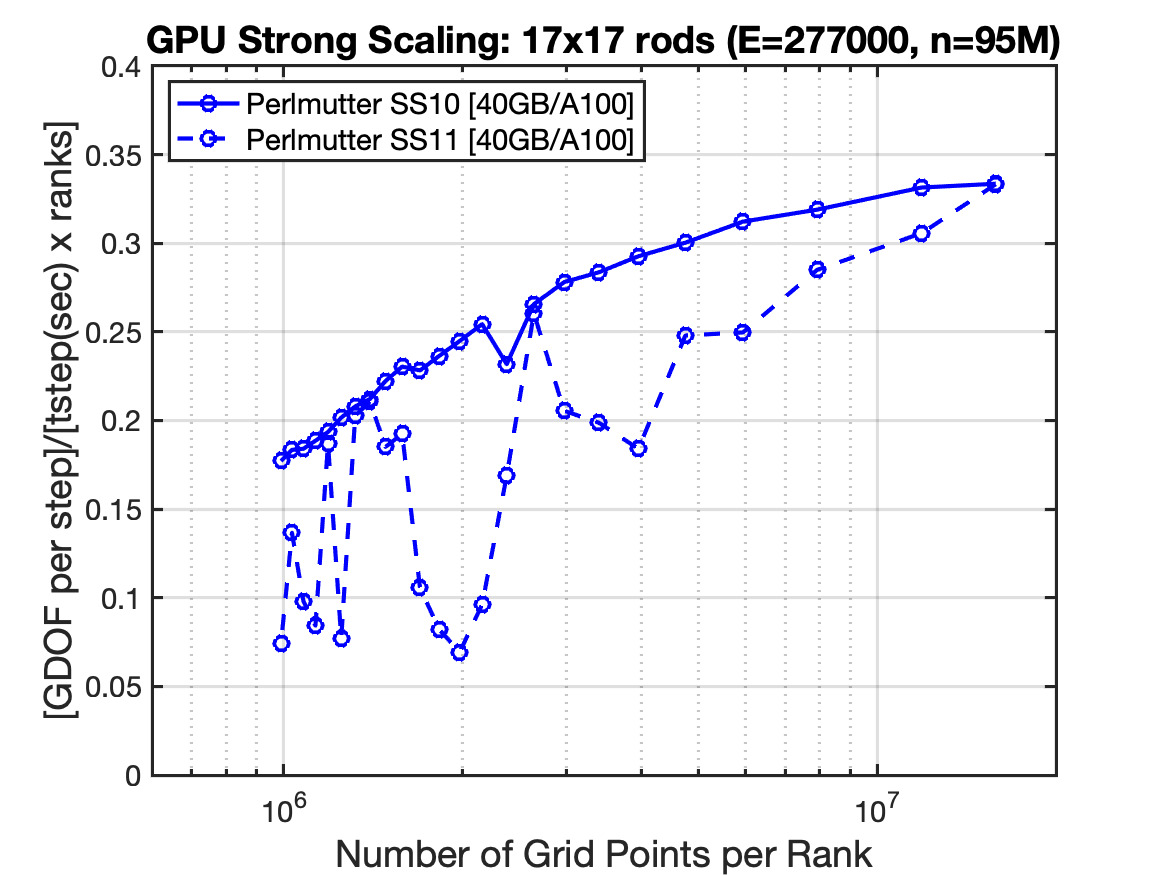 }
     \\
     \includegraphics[width=0.44\textwidth]{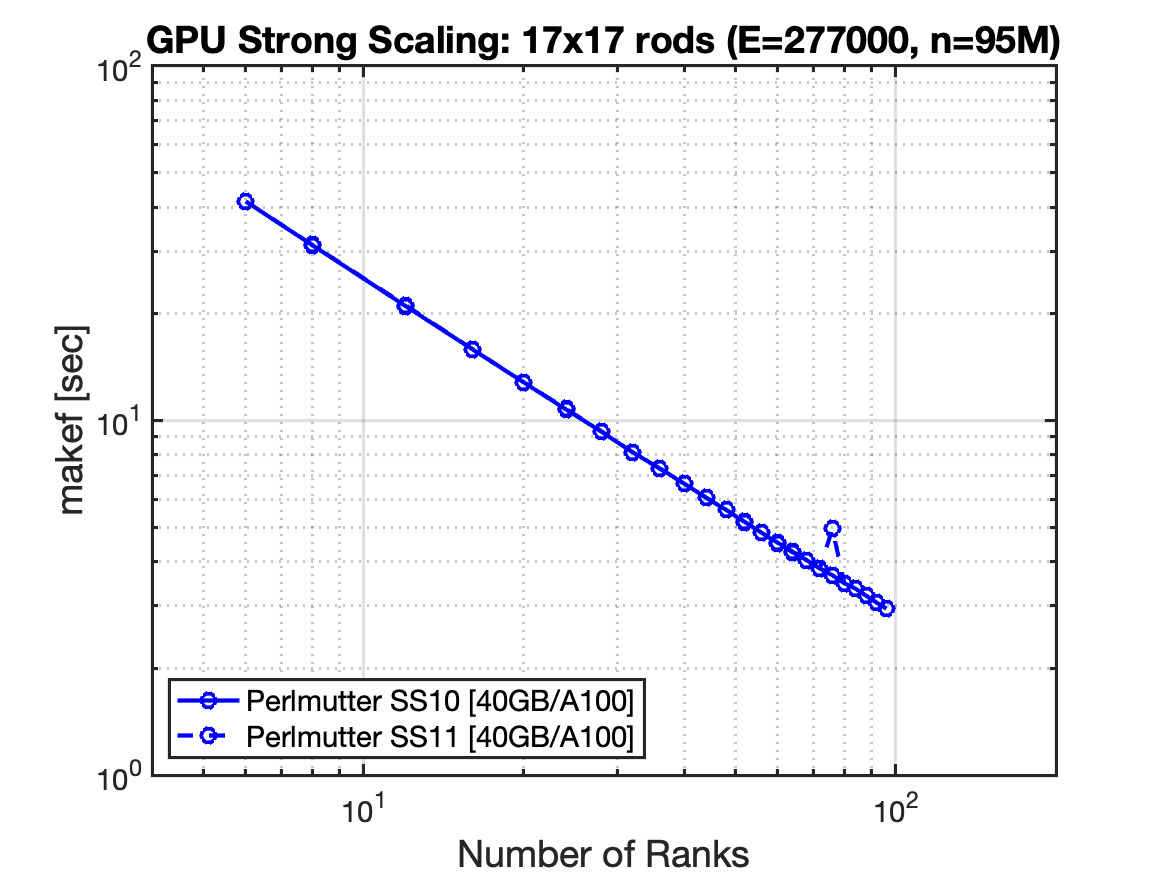 }
     \includegraphics[width=0.44\textwidth]{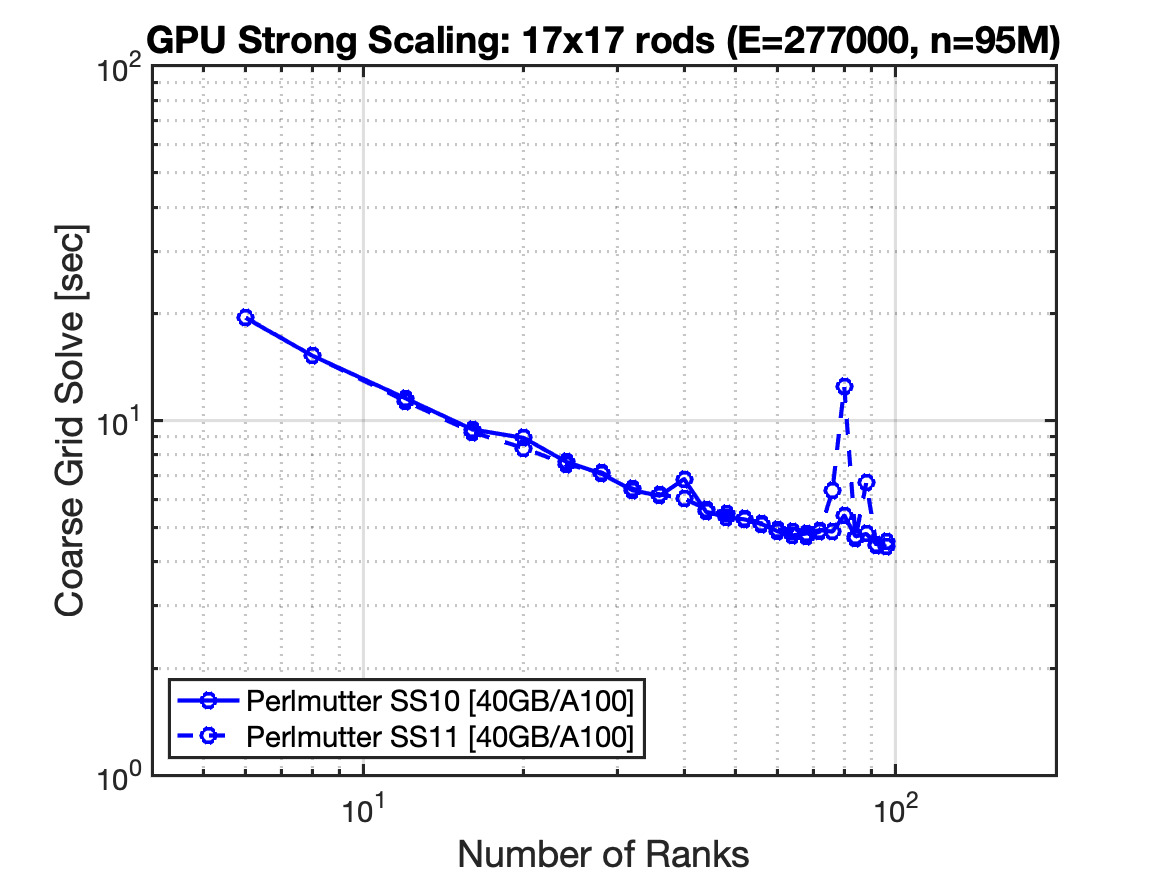 }
   \caption{\label{perf10-perlmutter} Strong-scaling on Perlmutter with Slingshot 10 and 11.}
  \end{center}
\end{figure*}

\noindent
In our timing studies we typically do not require that each node be fully occupied. 
This happens, for example, if one wants to make a device-to-device comparison between
Summit, with six V100 per node, and Crusher, which has 8 GCDs per node.  Our
initial plots of Crusher timing data appeared to be very erratic in its
dependence on $P$.  Closer inspection, however, revealed that the performance
for mod($P$,8)$\neq$0 was about $2\times$ slower than for the case of
mod($P$,8)$=$0.  Figures~\ref{perf10-rank8}--\ref{perf170-rank8} show early
timings taken on Crusher with the results sorted onto two curves corresponding
to multiples and nonmultiples of 8.   Unlike the Summit behavior of the
preceding section, this is clearly a device issue and not a host issue, as can
be seen in the last row in each of
Figures~\ref{perf10-rank8}--\ref{perf170-rank8}.  This anomaly has been resolved
and has not appeared as an issue on Frontier.


\subsection{Performance on Polaris with GPUDirect vs. without GPUDirect}  


As expected, communication overhead is more significant without GPUDirect, and
consequently the {\em no GPUDirect} curves on Polaris show relatively poor
performance as the amount of local work, $n/P$, is reduced.  For example, in
row 2, right, of Figure~\ref{perf17-rank8} the $n_{0.8}$ for Polaris with
GPUDirect is 2.5M, whereas it is 4.5M without GPUDirect.  In the last row of
the same figure we see that neither makef nor the coarse-grid solve are
influenced by the presence or absence of GPUDirect.  For makef, communication
does not matter.  The coarse-grid solve is communication intensive, but all of
that originates from the host.  We can also see in the results of
Figure~\ref{perf170-rank8} that the no GPUDirect results are relatively noisy.

We remark that the number of unknowns in the coarse grid on each rank is
roughly equal to the number of spectral elements on that rank.  For $p=7$ and
$n/P=2$M, we have $E=2$M/(343) $\approx$ 5800 elements per rank, so the local
coarse-grid problem size is 5,800.  If we were running on all $P=$27648 V100s of
Summit, the total coarse-grid size would be $P\times5800=160$M.




\subsection{Performance on Perlmutter with Slingshot 10 vs. Slingshot 11}

One other discovery was a sudden change in the behavior of Perlmutter at NERSC.  
Polaris and Perlmutter have similar
architectures, so it was expected that they would have similar performance, as
is indeed evident in, for example, Figure~\ref{perf17-all-frontier}.
Later in our stuides, however, the Perlmutter interconnect was being upgraded
from Slingshot 10 (SS10) to Slingshot 11 (SS11).   The performance started to vary radically
with $P$, albeit in a highly repeatable fashion.

The strong-scaling plot in Figure~\ref{perf10-perlmutter}, top-left, illustrates
the problem.  For 48 ranks, the Navier--Stokes runtime with SS11 jumps by a 
factor of 3 over that with SS10 and over Polaris, which is still running SS10.
These are repeatable results evidenced by several timings over a period of
several weeks.  Of course, the last row in the figure reveals that this
anomaly is neither a GPU issue, since makef timings are essentially identical,
nor a host-to-host communication issue, since the coarse-grid times are close
to the same.

Inspection of the NekRS logfiles shows that the increase is all focused in
one section of the code, namely, the non-local Schwarz-based smoother in
the $p$-multigrid preconditioner for the pressure Poisson problem.  
{\em That section} of code is running 10$\times$ slower than its SS10
counterpart!    Below, we show the logfile content for the two simulations
with $P=48$ (which is the slowest case).  

\noindent
\begin{minipage}[t]{3.4in}
\noindent {\bf SS10 logfile:}
\footnotesize\begin{verbatim}
name                    time             %  calls
  setup                 3.82904e+01s  0.38  1
    loadKernels         1.03634e+01s  0.27  1
  udfExecuteStep        4.79398e-03s  0.00  2001
  elapsedStepSum        6.13724e+01s  0.62
  solve                 6.12031e+01s  0.61
    min                 2.31879e-02s
    max                 5.51687e-02s
    flop/s              3.36729e+13

    makef               5.59237e+00s  0.09  2000
      udfUEqnSource     3.98969e-02s  0.01  2000
    udfProperties       4.82886e-03s  0.00  2001
    velocitySolve       1.73346e+01s  0.28  2000
      rhs               2.29362e+00s  0.13  2000
    pressureSolve       3.42052e+01s  0.56  2000
      rhs               4.69203e+00s  0.14  2000
      preconditioner    2.26178e+01s  0.66  2470
        pMG smoother    1.51609e+01s  0.67  9880
        coarse grid     5.33568e+00s  0.24  2470
      initial guess     3.18958e+00s  0.09  2000
\end{verbatim} 
\normalsize 
\end{minipage}
\hspace{.3in}
\begin{minipage}[t]{3.0in}
\noindent {\bf SS11 logfile:}
\footnotesize \begin{verbatim}
name                    time             %  calls
  setup                 3.98696e+01s  0.16  1
    loadKernels         8.86541e+00s  0.22  1
  udfExecuteStep        4.79946e-03s  0.00  2001
  elapsedStepSum        2.06042e+02s  0.84
  solve                 2.05867e+02s  0.84
    min                 5.50540e-02s
    max                 3.32500e-01s
    flop/s              1.00108e+13

    makef               5.57575e+00s  0.03  2000
      udfUEqnSource     3.99624e-02s  0.01  2000
    udfProperties       4.88246e-03s  0.00  2001
    velocitySolve       1.72489e+01s  0.08  2000
      rhs               2.29522e+00s  0.13  2000
    pressureSolve       1.79243e+02s  0.87  2000
      rhs               4.48683e+00s  0.03  2000
      preconditioner    1.67813e+02s  0.94  2470
        pMG smoother    1.49445e+02s  0.89  9880
        coarse grid     5.53950e+00s  0.03  2470
      initial guess     3.20173e+00s  0.02  2000
\end{verbatim} 
\normalsize 
\end{minipage}

\vspace{.2in}
\noindent
We also note that SS11 delivers higher bandwidth than does SS10 in all of the set-up
tests done by NekRS when it is making a runtime selection of the best
communication algorithm for each subproblem.  We list the logfile output for
those below.

\noindent
\begin{minipage}[t]{3.4in}
\noindent {\bf SS10 logfile:}
\footnotesize \begin{verbatim}
pw+device (MPI: 7.37e-05s / bi-bw:  54.5GB/s/rank)
pw+device (MPI: 1.75e-04s / bi-bw:  23.0GB/s/rank)
pw+device (MPI: 1.77e-04s / bi-bw:  22.7GB/s/rank)
pw+device (MPI: 1.76e-04s / bi-bw:  22.8GB/s/rank)
pw+device (MPI: 7.29e-05s / bi-bw:  55.2GB/s/rank)
pw+device (MPI: 7.29e-05s / bi-bw:  55.1GB/s/rank)
pw+device (MPI: 5.50e-05s / bi-bw:  73.1GB/s/rank)
pw+device (MPI: 5.48e-05s / bi-bw:  73.4GB/s/rank)
pw+device (MPI: 5.37e-05s / bi-bw:  96.3GB/s/rank)
pw+device (MPI: 5.16e-05s / bi-bw: 100.2GB/s/rank)
pw+device (MPI: 4.64e-05s / bi-bw:  16.3GB/s/rank)
pw+device (MPI: 4.90e-05s / bi-bw:  15.4GB/s/rank)
pw+device (MPI: 3.84e-05s / bi-bw:  33.6GB/s/rank)
pw+host   (MPI: 2.46e-05s / bi-bw:   3.6GB/s/rank)
\end{verbatim} 
\normalsize 
\end{minipage}
\hspace{.3in}
\begin{minipage}[t]{3.4in}
\noindent {\bf SS11 logfile:}
\footnotesize \begin{verbatim}
pw+device (MPI: 4.38e-05s / bi-bw:  91.8GB/s/rank)
pw+device (MPI: 8.45e-05s / bi-bw:  47.6GB/s/rank)
pw+device (MPI: 8.45e-05s / bi-bw:  47.6GB/s/rank)
pw+device (MPI: 8.58e-05s / bi-bw:  46.9GB/s/rank)
pw+device (MPI: 4.52e-05s / bi-bw:  89.0GB/s/rank)
pw+device (MPI: 4.48e-05s / bi-bw:  89.8GB/s/rank)
pw+device (MPI: 4.07e-05s / bi-bw:  98.7GB/s/rank)
pw+device (MPI: 3.97e-05s / bi-bw: 101.2GB/s/rank)
pw+device (MPI: 3.52e-05s / bi-bw: 146.7GB/s/rank)
pw+device (MPI: 3.47e-05s / bi-bw: 148.8GB/s/rank)
pw+device (MPI: 3.75e-05s / bi-bw:  20.1GB/s/rank)
pw+device (MPI: 3.58e-05s / bi-bw:  21.1GB/s/rank)
pw+device (MPI: 2.74e-05s / bi-bw:  47.2GB/s/rank)
pw+host   (MPI: 1.66e-05s / bi-bw:   5.4GB/s/rank)
\end{verbatim} 
\normalsize 
\end{minipage}

\vspace{.2in}
\noindent
For completeness, we also include the kernel performance numbers
as reported in the NekRS logfiles: \\

\noindent
\begin{minipage}[t]{3.4in}
\noindent {\bf SS10 logfile:}
\footnotesize \begin{verbatim}
 Ax: N=7 FP64 GDOF/s=13.2 GB/s=1260 GFLOPS=2184 kv0
 Ax: N=7 FP64 GDOF/s=13.2 GB/s=1260 GFLOPS=2183 kv0
 Ax: N=3 FP64 GDOF/s=12.6 GB/s=1913 GFLOPS=1883 kv5

 Ax: N=7 FP32 GDOF/s=25.0 GB/s=1194 GFLOPS=4145 kv4
 Ax: N=3 FP32 GDOF/s=18.0 GB/s=1368 GFLOPS=2693 kv2

fdm: N=9 FP32 GDOF/s=44.9 GB/s= 812 GFLOPS=7452 kv4
fdm: N=5 FP32 GDOF/s=34.1 GB/s= 825 GFLOPS=4301 kv1

flop/s  3.36729e+13  (701 GFLOPS/rank)
\end{verbatim} 
\normalsize 
\end{minipage}
\hspace{.3in}
\begin{minipage}[t]{3.4in}
\noindent {\bf SS11 logfile:}
\footnotesize \begin{verbatim}
 Ax: N=7 FP64 GDOF/s=13.2 GB/s=1256 GFLOPS=2179 kv0
 Ax: N=7 FP64 GDOF/s=13.2 GB/s=1257 GFLOPS=2180 kv0
 Ax: N=3 FP64 GDOF/s=12.6 GB/s=1912 GFLOPS=1882 kv5

 Ax: N=7 FP32 GDOF/s=25.0 GB/s=1194 GFLOPS=4144 kv5
 Ax: N=3 FP32 GDOF/s=18.1 GB/s=1369 GFLOPS=2696 kv2

fdm: N=9 FP32 GDOF/s=44.9 GB/s= 812 GFLOPS=7444 kv4
fdm: N=5 FP32 GDOF/s=34.1 GB/s= 825 GFLOPS=4303 kv1

flop/s  1.00108e+13  (208 GFLOPS/rank)
\end{verbatim} 
\normalsize 
\end{minipage}

\vspace{.2in}
\noindent
In the table above, {\tt kv} reflects the particular kernel version chosen out
of the suite of available kernels in NekRS for the particular operation.  We
see that the 64-bit {\tt Ax} kernels (the matrix-vector product with the
Laplace operator for spectral element order $N$) realize $\approx 2$ TFLOPS per
device, while their 32-bit counterparts realize 3--4 TFLOPS   (32-bit
arithmetic is used in the preconditioner only). The fast-diagonalization method
(fdm) implements local tensor-product solves of the form $\uz = S\Lambda^{-1}
S^T \ur$, where $S=(S_t \otimes S_s \otimes S_r)$ is the orthogonal matrix of
eigenvectors for the overlapped Poisson problem in local ($r-s-t$) coordinates
in the reference element, $\Oh=[-1,1]^3$ \cite{dfm02}.  This is a fast
operation and can be seen to sustain $>$ 7 GFLOPS (FP32).  Note that, as
expected, the kernel performance, which does not include any MPI overhead, is
not dependent on the Slingshot version.  NekRS also reports the total observed
GFLOPS, which is seen to by 701 GFLOPS/rank for SS10 and 208 GFLOPS/rank for
SS11.  We note that, for this particular case, $P=48$ and $n=96790120$, for
which $n/P = 2.01$M, which is close to (but below) the value of $n_{0.8}$.
We can anticipate that the saturated floating-point performance for SS10
would thus be about $701/0.8=876$ GFLOPS.

\vspace{.2in}
\noindent
At present we do not know why the $p$-multigrid smoother time is an
order of magnitude larger on SS11 than SS10.  That question is under
investigation but has been hampered by downtime of Perlmutter.

%
%
%

\section{Conclusions}

We explored the performance of a highly tuned incompressible flow code,
NekRS, on current-generation HPC architectures featuring accelerator-based
nodes.   The principal accelerators under consideration are the NVIDIA V100,
NVIDIA A100, and MI250X with eight GCDs per node.  We found that the raw
performance of a single GCD is about 85\% of a single A100.  We also found that
the AMD-based Crusher platform had slower host-based communication than its
Polaris/Perlmutter counterpart, as witnessed by the relatively poor performance
of the Hypre-based coarse-grid solver, which runs on the host.  This situation
is significantly improved on Frontier.

What is critical to end-user performance is the ratio $n_{0.8}/P$, which
governs the time-to-solution (\ref{eq:t08}).  Despite the pessismistic results
in \cite{ceed_bp_paper_2020}, where it appeared that GPU-based platforms might
be $3\times$ slower than ANL's IBM BG/Q, Mira, we are finding that NekRS is in
$3\times$ {\em faster} than Nek5000 on Mira.  This gain can be attributed to
careful attention to reducing $n_{0.8}$, to improved pressure preconditioners
tuned specifically for accelerator-based nodes, to the use of 32-bit precision
in the preconditioners, and to extensive use of overlapped communication and
computation.

\section*{Funding} 
This material is based upon work supported by the U.S. Department of Energy,
Office of Science, under contract DE-AC02-06CH11357 and by the Exascale Computing
Project (17-SC-20-SC). 
The research used resources of the Oak Ridge Leadership Computing Facility at Oak Ridge National Laboratory, 
which is supported by the Office of Science of the U.S. Department of Energy under Contract DE-AC05-00OR22725.
This research also used resources of
the Argonne Leadership Computing Facility at Argonne National Laboratory,
which is a DOE Office of Science User Facility supported under Contract DE-AC02-06CH11357.
This research also used resources of the National Energy Research Scientific Computing Center (NERSC), a U.S. Department 
of Energy Office of Science User Facility located at Lawrence Berkeley National Laboratory, 
operated under Contract No. DE-AC02-05CH11231





\bibliographystyle{./Frontiers-Harvard}   
\bibliography{abl.bib,ananias.bib,emmd.bib}



%

\end{document}